\documentclass[%
reprint,
superscriptaddress,
nofootinbib,
amsmath,amssymb,
aps,
]{revtex4-2}

\usepackage{color}
\usepackage{graphicx}
\usepackage{dcolumn}
\usepackage{bm}
\usepackage{textcomp}
\usepackage{rotating} 
\usepackage[T1]{fontenc}

\begin {document}

\title{Anomalous statistics in the Langevin equation with fluctuating diffusivity: from Brownian yet non-Gaussian diffusion to anomalous diffusion and ergodicity breaking 
 }

\author{Takuma Akimoto}
\email{takuma@rs.tus.ac.jp}
\affiliation{%
  Department of Physics and Astronomy, Tokyo University of Science, Noda, Chiba 278-8510, Japan
}%

\author{Jae-Hyung Jeon}
\affiliation{%
Department of Physics, Pohang University of Science and Technology, Pohang, 37673, Republic of Korea
}
\affiliation{%
  Asia Pacific Centre for Theoretical Physics, Pohang, 37673, Republic of Korea
}

\author{Ralf Metzler}
\affiliation{%
  Institute of Physics \& Astronomy, University of Potsdam, 14476 Potsdam-Golm, Germany
}%
\affiliation{%
  Asia Pacific Centre for Theoretical Physics, Pohang, 37673, Republic of Korea
}

\author{Tomoshige Miyaguchi}
\affiliation{%
  Department of Systems Engineering, Wakayama University, 930 Sakaedani, Wakayama, 640-8510, Japan
}%

\author{Takashi Uneyama}
\affiliation{%
  Department of Materials Physics, Nagoya University, Nagoya 464-8603, Japan
}%

\author{Eiji Yamamoto}
\affiliation{%
  Department of System Design Engineering, Keio University, Yokohama, Kanagawa 223-8522, Japan
}%



\date{\today}

\begin{abstract}
Diffusive motion is a fundamental transport mechanism in physical and biological systems, governing dynamics across a wide range of scales---from molecular transport to animal foraging. In many complex systems, however, diffusion deviates from classical Brownian behaviour, exhibiting striking phenomena such as  Brownian yet non-Gaussian diffusion (BYNGD) and anomalous diffusion.
BYNGD describes a frequently observed statistical feature characterised by the coexistence of linear mean-square displacement (MSD) and non-Gaussian displacement distributions. Anomalous diffusion, in contrast, involves a nonlinear time dependence of the MSD and often reflects mechanisms such as trapping, viscoelasticity, heterogeneity, or active processes. Both phenomena challenge the conventional framework based on constant diffusivity and Gaussian statistics.
This review focuses the theoretical modelling of such behaviour via the Langevin equation with fluctuating diffusivity (LEFD)—a flexible stochastic framework that captures essential features of diffusion in heterogeneous media. LEFD not only accounts for BYNGD but also naturally encompasses a wide range of anomalous transport phenomena, including subdiffusion, ageing, and weak ergodicity breaking. Ergodicity is discussed in terms of  the correspondence between time and ensemble averages, as well as the trajectory-to-trajectory variability of time-averaged observables. 
The review further  highlights the empirical relevance of LEFD and related models in explaining diverse experimental observations and underscores their value to uncovering the physical mechanisms governing transport in complex systems.
\end{abstract}

\maketitle

\tableofcontents

\clearpage

\begin{table}[b]
\begin{tabular}{|l|l|}
\hline
$d$ & Dimension of the embedding space\\
$D$ & Diffusion coefficient\\
$T$ & Temperature\\
$\gamma$ & Drag coefficient\\
$\tau_R$ & Relaxation time\\
$P({\bm r},t)$ & Positional probability density function (PDF)\\
$\alpha$ & Subdiffusion exponent/power-law exponent in the sojourn-time PDF\\
$\langle \bm{r}^2(t)\rangle$ & Mean squared displacement (ensemble average)\\
$\Delta$ & Lag time in the displacement\\
$\overline{\delta^2(\Delta;t)}$ & Time-averaged squared displacement
(TSD)\\m
$\Sigma^2(t;\Delta)$ & Relative standard deviation (RSD) of the TSD\\
$\tau_D$ & Characteristic time of $D(t)$\\
$ \psi_1(t)$ & Correlation function of $D(t)$\\
$\tau_c$ & Crossover time of the RSD\\
$\rho(\tau)$ & Sojourn-time PDF\\
$\mu$ & Mean sojourn time \\
$G_d({\bm r},t;D)$ & Green's function for the diffusion equation with
diffusion coefficient $D$ \\
$D_\tau$ & Instantaneous diffusion coefficient given that the sojourn time
is $\tau$ in the annealed transit time model\\
$\tau_d$ & Disengagement time in the reptation model\\
$\chi$ & Mass ratio in the binary system\\
\hline
\end{tabular}
\caption{Notations used in this review.}
\end{table}


\clearpage

\section{Introduction}

Diffusive motion is a cornerstone physical transport process in nature,
governing the dynamics in systems ranging from cellular environments to
glass-forming liquids. In many complex systems, however, diffusion often
deviates from its classical behaviour and exhibits rich phenomena. One such deviation is anomalous diffusion, where the mean squared displacement (MSD) follows a power-law time dependence rather than growing linearly, This behaviour has been widely observed across disciplines and often reflects complex mechanisms such as trapping, viscoelasticity, or spatial heterogeneity 
\cite{bouchaud90,metzler00,goychuk09}.
Another notable deviation is what is often referred to as ``Brownian yet non-Gaussian diffusion" (BYNGD). 
While many different terms have been coined for this phenomenon, BYNGD is widely used to denote a recurrent statistical feature observed in numerous experimental and computational studies---namely, the coexistence of a linear MSD and non-Gaussian displacement probability density functions (PDFs).
This behaviour has become a central topic in statistical physics and soft matter science, challenging classical assumptions of constant diffusivity and Gaussian displacement statistics. Instead, these observations point to a central role played by a ``fluctuating diffusivity," which reflects the temporal or spatial heterogeneity inherent in many real-world systems. Building upon the foundational work of Einstein and Langevin on thermally driven diffusion, the concept of fluctuating diffusivity offers a deeper framework for understanding stochastic dynamics in heterogeneous media. This review provides a comprehensive overview of fluctuating diffusivity and its statistical modelling with extensions to anomalous diffusion.


Brownian motion---the famed random jittery motion of microscopic particles
suspended in a fluid---has been a cornerstone in the development of
nonequilibrium statistical physics and stochastic modelling since its
discovery \cite{livi,van1992stochastic}. In 1827, the botanist Robert Brown
observed this erratic motion while studying microscopic granules ejected
by pollen grains in water, providing the first documented description of
what is now known as Brownian motion \cite{brown1828xxvii}. Decades later,
Einstein and Smoluchowski independently developed theoretical frameworks
linking Brownian motion to molecular collisions, providing a quantitative
basis for its stochastic nature and revolutionising our understanding of
matter \cite{Einstein1905,smoluchowski1906kinetischen}. These groundbreaking
theories were experimentally confirmed by Perrin, further cementing Brownian
motion as a fundamental concept in modern physics \cite{Perrin1909}.


Einstein's theoretical work culminated in the formulation of the diffusion
equation, which describes the time evolution of the probability density
function (PDF) $\rho(x,t)$ for the position $x$ of a diffusing particle in
the course of time,\footnote{We here develop the framework in embedding
dimension $d=1$, generalisation to higher dimensions is straightforward.}
\begin{equation}
\label{diffeq}
\frac{\partial\rho(x,t)}{\partial t}=D\frac{\partial^2\rho(x,t)}{\partial x^2}, 
\end{equation} 
where $D$ is the diffusion coefficient. Formally, this equation
is equivalent to Fick's second law \cite{fick} and Fourier's law of
heat conduction \cite{fourier}. The solution of the diffusion equation
(\ref{diffeq}) for $\delta$-function initial condition at $x=0$ and natural boundary
conditions $\lim_{|x|\to\infty}\rho(x,t)=0$, a Gaussian PDF with mean $0$
and variance $2Dt$, reflects the hallmark of normal diffusion: a linear
growth of the MSD with time, $\langle x(t)^2\rangle=2Dt$. Smoluchowski's
contributions further reinforced this understanding, emphasising the role of
molecular collisions in shaping the dynamics of diffusing particles. The
linear time dependence of the MSD and the Gaussian PDF form the foundation of
classical diffusion theory. However, many experimental systems deviate
 from these classical behaviours, leading to phenomena such as
anomalous diffusion and non-Gaussian diffusion---topics that are the focus of this review.

At the microscopic level, the random zigzag motion of a diffusing particle---beautifully recorded by Perrin \cite{Perrin1909}---arises from incessant collisions
with surrounding molecules, which generate thermal forces and friction. These
interactions establish a balance between thermal fluctuations, which drive the
motion, and friction, which resists it. This balance is quantitatively described by the
Einstein-Smoluchowski relation, linking the diffusion coefficient $D$
to the temperature $T$ and the drag coefficient $\gamma$:
\begin{equation}
D=\frac{k_{\rm B}T}{\gamma},
\label{eq: ER}
\end{equation}
where $k_{\rm B}$ is the Boltzmann constant.  For spherical particles,
Stokes' law relates the drag force to the particle's velocity relative to
the surrounding medium as $F=-\gamma v$, where $\gamma=6 \pi\eta R$ is the
drag coefficient, $R$ the particle radius, and $\eta$ the viscosity of the
ambient liquid \cite{Stokes}. The Stokes law and the Einstein-Smoluchowski
relation (\ref{eq: ER}) lead to the Stokes-Einstein relation, which connects
the diffusion coefficient $D$ to the particle size and the medium properties:
\begin{equation}
D=\frac{k_{\rm B}T}{6\pi\eta R}.
\label{eq: SK relation}
\end{equation}
The Stokes-Einstein relation highlights how the diffusion of a particle depends
on both its physical characteristics and the properties of the surrounding
medium, establishing a foundational link between microscopic interactions and
macroscopic transport.

With the introduction of the concept of fluctuating forces, Langevin laid a
fundamental concept for nonequilibrium statistical physics \cite{kubo2012statistical,brenig,
van1992stochastic,livi,Langevin1908,coffey}. Extending Newton's equation of
motion with a noise term to mimic these fluctuating forces, the stochastic
Langevin equation provided a groundbreaking framework for describing the
time evolution of the velocity $v(t)$ of a Brownian particle
\cite{Langevin1908},\footnote{Brownian motion is sometimes used synonymously with overdamped dynamics, in which inertial effects are negligible. Langevin's original formulation, however, includes inertia and describes the velocity evolution explicitly.}
\begin{equation}
\label{LE}
m\dot{v}(t)=-\gamma v(t)+\zeta (t).
\end{equation}
Here $m$ is the mass of the Brownian particle, $\gamma$ is the friction
coefficient, and $\zeta (t)$ represents the random thermal force arising from
collisions with molecules in the surrounding medium. The first and second
terms on the right-hand side of Eq.~(\ref{LE}), i.e., the frictional and
random forces, respectively, balance each other to maintain thermal
equilibrium. Invoking the central limit theorem, the random force $\zeta(t)$
is modelled as Gaussian white noise, characterised by zero mean and an
infinitesimally short correlation time, implying that the thermal noise is
uncorrelated in time. The fluctuation-dissipation theorem provides a quantitative link
between $\gamma$ and $\zeta(t)$, in order to ensure thermal equilibrium
\cite{Callen1951,kubo1957statistical,kubo1966fluctuation,kubo2012statistical},
\begin{equation}
\label{FDT}
\langle\zeta(t)\zeta(t')\rangle=2\gamma k_{\rm B}T\delta(t-t'),
\end{equation}
where $\langle\cdot\rangle$ represents ensemble averaging. This relationship
highlights how thermal fluctuations, in the form of random forces, are
intrinsically coupled to dissipation due to friction. From the Langevin
equation (\ref{LE}), the MSD of a Brownian particle with initial equilibrium
distribution of the velocity \cite{risken} evolves as 
\begin{equation}
\langle[x(t)-x(0)]^2\rangle=\frac{2k_{\rm B}T\tau_R}{m}\left\{t+\tau_R\exp
\left(-\frac{t}{\tau_R}\right)-\tau_R\right\},
\label{MSD LE}
\end{equation}
where $\tau_R=m/\gamma$ is the relaxation timescale, characterising how rapidly the particle's velocity relaxes due to friction.
In the short-time limit, the MSD scales ballistically,
\begin{equation}
\langle[x(t)-x(0)]^2\rangle\sim\frac{k_{\rm B}T}{m}t^2\quad(t\ll\tau_R),
\end{equation}
indicating that motion is dominated by inertia. In the long-time limit, the MSD simplifies to
\begin{equation}
\label{MSD LE ltl}
\langle[x(t)-x(0)]^2\rangle\sim\frac{2k_{\rm B}T}{\gamma}t\quad(t\to\infty),
\end{equation}
which is consistent with linear growth of the MSD and yields the diffusion coefficient
 $D=\langle[x(t)-x(0)]^2\rangle/(2t)$ in the long-time limit. This recovers the
Einstein-Smoluchowski relation (\ref{eq: ER}). The Langevin equation thus offers a comprehensive framework for describing 
normal diffusion across both inertial and diffusive timescales.

As previously mentioned, Brownian motion exhibits two hallmark characteristics. First, the
MSD increases linearly with time, as shown in expression (\ref{MSD LE ltl}) for
the long time limit. Second, the displacement PDF, or the propagator, follows a
Gaussian distribution. This arises because the thermal noise in the Langevin equation is modelled as a Gaussian process, which leads to Gaussian statistics for particle displacements. 
In the overdamped limit ($t\gg\tau_R$), the Langevin equation reduces to 
\begin{equation}
\label{brownian motion}
\dot{x}(t)=\sqrt{2D}\xi(t),
\end{equation}
where $\xi(t)$ is a Gaussian white noise with zero mean and unit variance, and
$D$ is the diffusion coefficient given by expression (\ref{eq: ER}). The PDF
of a Gaussian process is fully defined by its first two moments, and accordingly, it satisfies the diffusion equation
(\ref{diffeq}), i.e., the partial differential equation describing the time evolution of the PDF $P(x,t)$. 
Solving this equation with the initial condition $P(x,0)=\delta(x)$ and the boundary condition $\lim_{|x|\to\infty}P(x,t)=0$, we obtain
\begin{equation}
\label{gaussian propagator}
P(x,t)=\frac{1}{\sqrt{4\pi Dt}}\exp\left(-\frac{x^2}{4Dt}\right). 
\end{equation}
This Gaussian propagator reflects the normal diffusion, characterised by a linear MSD,
 $\langle x(t)^2\rangle=2Dt$. Notably, both the Gaussian PDF (\ref{gaussian propagator}) and the linear form of the
MSD can also be derived from random walk models \cite{hughes,igor2011}.

Classical Brownian motion is characterised by the linear growth in time of the
MSD and the Gaussian propagator, however, many real-world systems deviate from
this behaviour, exhibiting what is known as ``anomalous diffusion" \cite{pt,Metzler2014}. In anomalous diffusion, the MSD follows a power-law relationship with time,
$\langle x(t)^2\rangle\propto t^\alpha$, where the anomalous diffusion
exponent $\alpha\neq1 $. This behaviour, commonly observed in systems with
strong disorder \cite{Scher1975,bouchaud90,metzler00} or long-range dependency
\cite{mandelbrot,Metzler2014}, challenges the classical framework of ordinary
diffusion. Anomalous diffusion has been extensively observed in experiments,
see, inter alia, \cite{Solomon1993,Wong2004,Tolic-Norrelykke2004,
banks2005anomalous,Golding2006,Szymanski2009,Weber2010,Weigel2011,Jeon2011,
Skaug2013,Tabei2013,akselrod2014visualization,Manzo2015,rajyaguru2024diffusion,rajyaguru2025quantifying}, dynamical systems
\cite{Geisel1984,ishizaki1991anomalous,Dr2000,Barkai2003,Akimoto2010,
Akimoto2012,Sato2019}, and molecular dynamics simulations
\cite{muller1992computational,Neusius2008,Akimoto2011,Jeon2012,
yamamoto2014origin,Jeon2016,jeremy,Tamm2015}, highlighting its ubiquity
in nature \cite{klafter2005anomalous,pt}. Unlike normal diffusion, anomalous
diffusion is often, but not always, associated with non-Gaussian propagators,
deviating from the Gaussian characteristic of Brownian motion \cite{metzler00}.
This dual departure---from linear MSD growth and Gaussian PDFs---underscores
the fundamental differences between anomalous and classical diffusion.

More recently, the phenomenon referred to as BYNGD has garnered significant
interest in statistical physics, soft matter, and biological sciences
\cite{wang2009anomalous,wang2012brownian,He2013,Bhattacharya2013,Guan2014,
kwon2014dynamics,Chubynsky2014,Jain2016,Chechkin2017,tyagi2017non,
sposini2018random,Miotto2021,Rusciano2022,Alexandre2023,sposini2023glassy,
Sposini2024PRL,Sposini2024PRE}. A defining feature of BYNGD is the coexistence
of a non-Gaussian PDF with a normal MSD, that grows linearly with time. Both
experimental and theoretical studies have significantly advanced our understanding of the underlying mechanisms behind this behaviour.
 Experimental techniques such as single-particle tracking (SPT), fluorescence microscopy,
and advanced imaging methods have played a crucial role in detecting and characterising the non-Gaussian nature of diffusion in complex environments 
\cite{wang2009anomalous,wang2012brownian,He2013,Bhattacharya2013,Guan2014,
Rusciano2022,Alexandre2023}. These experiments reveal deviations from the
classical Gaussian description, showcasing features such as Laplace-like
(i.e., exponential) displacement  distributions and signatures of heterogeneity in the local diffusion environment.
Exploring BYNGD across diverse physical
systems enhances our understanding of the underlying mechanisms
driving these complex dynamics. Moreover, these studies have broad implications
for fields ranging from biophysics to materials science, offering new
perspectives on diffusion in heterogeneous and non-equilibrium systems.
Interestingly, Laplace-like displacement distributions are also commonly observed in financial markets and geophysical systems such as hydrology. In these contexts, they are typically modelled using variance-gamma processes \cite{madan1998variance, madan1990variance,kotz2001classical} and their fractional generalisations \cite{meerschaert2004fractional,gajda2017generalized,flm},
highlighting the mathematical and phenomenological parallels across disciplines.

Extensive theoretical studies on BYNGD have highlighted the significance of
environmental heterogeneity and therefore distributed diffusivities in shaping
non-Gaussian propagators \cite{Chubynsky2014,Jain2016,Chechkin2017,tyagi2017non,
sposini2018random}. In particular, Chubynsky and Slater introduced the concept
of diffusing diffusivity to account for this unique diffusion behaviour
\cite{Chubynsky2014}. In their framework, the diffusion coefficient varies over
time according to an advection-diffusion process, giving rise to time-dependent
instantaneous diffusivity. These fluctuations are linked to temporal
changes in a particle's shape and its local environment---such as
temperature and viscosity---via the Stokes-Einstein relation [see 
Eq.~(\ref{eq: SK relation})]. Temporal fluctuations in diffusivity
are commonly observed in glass-forming liquids \cite{Yamamoto-Onuki-1998,
Yamamoto-Onuki-1998a,Richert-2002,sposini2022detecting} and living cells
\cite{javer2013short,parry2014bacterial,Manzo2015,lampo2017cytoplasmic,
gu2018transcription,shi2018interphase}. These fluctuations contribute to the emergence of non-Gaussian
PDFs, anomalous diffusion, and weak ergodicity breaking
\cite{Yamamoto-Onuki-1998,Yamamoto-Onuki-1998a,Richert-2002,javer2013short,
parry2014bacterial,Manzo2015,lampo2017cytoplasmic,gu2018transcription,
shi2018interphase}.  Moreover,  recent molecular dynamics simulations of a protein have
further demonstrated that fluctuations in the protein's conformations directly
effect a fluctuating diffusivity \cite{Yamamoto2021}. Consequently, instantaneous
fluctuating diffusivities are inherent in natural phenomena. 
Understanding their impact on global diffusion characteristics---including the emergence of non-Gaussian statistics, global diffusivity fluctuations, and the efficiency of target search processes---is essential. These insights pave the way toward uncovering the mechanisms that drive complex diffusion dynamics in heterogeneous and biological environments.

The objective of this review is to explore the intriguing and non-trivial
phenomena arising from processes with a stochastic diffusivity, a key driver of
BYNGD. The paper is organised as follows: Section II reviews experimental
observations and molecular dynamics simulations that reveal the significance of
a fluctuating diffusivity in diverse systems. Section III introduces the
fundamental theories underpinning the Langevin equation with a fluctuating
diffusivity (LEFD), providing a robust framework for modelling these dynamics.
Sections IV to VII present a detailed analysis of key diffusion characteristics,
including the MSD, non-Gaussian propagators, and diffusivity fluctuations across
several stochastic models. Section VII highlights the broad applicability of the
LEFD framework in physics, chemistry, biology, and molecular dynamics
simulations. Finally, Section VIII summarises the findings and discusses
future directions for research in this rapidly evolving field.

\section{Fluctuating diffusivity revealed by experiments and simulations}

The concept of fluctuating diffusivities of a tracer particle is based on what
by-now turns out to be a widespread phenomenon observed in diverse systems,
inter alia, including complex fluids, living biological cells and their
membranes, glass-forming liquids, and molecular systems. These observations
can be categorised into processes with local or global diffusivity fluctuations,
respectively, corresponding to short- and long-timescale variations in 
diffusivity. SPT is a powerful experimental technique for capturing both types
of fluctuations \cite{wang2009anomalous,wang2012brownian,Bhattacharya2013,
Guan2014,Golding2006,Weber2010,Weigel2011,Jeon2011,he2016dynamic}.
Concurrently, molecular dynamics simulations provide single-particle
trajectories, allowing the analysis of fluctuating diffusivities at high
temporal and spatial resolutions \cite{yamamoto2017dynamic,Yamamoto2021,
Jeon2016,amanda}. A fundamental tool to estimate the diffusivity from a
single-particle trajectory is the time-averaged squared displacement
(TSD) defined as 
\begin{equation}
  \overline{\delta^2 (\Delta; t)} \equiv \frac{1}{t-\Delta} \int_0^{t-\Delta} | {\bm r} (t'+\Delta) -  {\bm r} (t') |^2 dt',
  \label{eq: TSD}
\end{equation}
where ${\bm r} (t')$ is the test particle's position at time $t'$, $\Delta$ is the lag time, and $t$ is a measurement time.
This expression is written for a general \( d \)-dimensional position vector.  
In one dimension, it simplifies to\begin{equation}
\label{eq: TSD-1D}
\overline{\delta^2(\Delta;t)}\equiv\frac{1}{t-\Delta}\int_0^{t-\Delta}\Big(
x(t'+\Delta)-x(t')\Big)^2dt',
\end{equation}
where $x(t')$ is the particle's position at time $t'$. The TSD, computed from a single
trajectory, is analysed as a function of the lag time $\Delta$, 
which corresponds to the width of the window slid along the trajectory.  If
the TSD increases linearly with  $\Delta$, the diffusion coefficient can be
directly extracted from its slope. For anomalous diffusion processes, the
TSD may instead exhibit a power-law scaling $\Delta^{\alpha}$; or, it may
exhibit a linear scaling in $\Delta$, depending on the precise physical
process, see below \cite{pt,Metzler2014,bouchaud1992,barkai2005,staspnas}. In
the following, we examine local and global diffusivity fluctuations observed
in experiments and molecular dynamics simulations, highlighting their
significance in understanding diffusion in heterogeneous environments.

\subsection{Global diffusivity fluctuations} 

In traditional statistical physics, ergodicity ensures the equivalence between
time-averaged and ensemble-averaged observables \cite{staspnas,khinchin,pt,
Metzler2014,bouchaud1992,barkai2005}. In an ergodic system, TSDs calculated from
different single-particle trajectories converge to the same value in the
long-time limit, matching the ensemble average. This property guarantees
reproducibility in experiments: repeated measurements under
identical conditions yield consistent results. However, recent SPT experiments
revealed distinct deviations from ergodicity. TSDs obtained from different
trajectories and time windows often exhibit significant variations in their
apparent diffusion coefficients, leading to trajectory-to-trajectory fluctuations
\cite{Golding2006,Weber2010,Weigel2011,Jeon2011,Manzo2015,amanda,beta,beta1,
stefanie}. These findings challenge the conventional assumption of ergodicity
in diffusion processes and highlight the need to consider nonergodic effects in
complex systems. In what follows, we highlight a selection of these experiments and simulations.

An experimental study was conducted to track the random motion of fluorescently labelled mRNA molecules inside live {\it E. coli\/} cells
\cite{Golding2006}. The analysis revealed that the TSDs grow as $\overline{
\delta^2(\Delta;t)}\sim D_\alpha\Delta^\alpha$ with $\alpha\approx0.7$ for large $\Delta$,
indicating subdiffusive behaviour. Furthermore, the generalised diffusion coefficient $D_\alpha$
obtained from different single-particle trajectories and time windows exhibited
large fluctuations, even for long measurement times, pointing to global
diffusivity fluctuations. Notably, in this experiment, the relative standard
deviation of the generalised diffusion coefficient exceeded the value of 3,
based on analysis of 70 trajectories across 3 experiments with varying
time windows. These results suggest that the observed subdiffusion and the
substantial trajectory-to-trajectory fluctuations are indicative of a
superdense and heterogeneous cytoplasmic medium, where local variations in
the intracellular environment significantly impact molecular transport
\cite{Golding2006}.

Compelling evidence for anomalous diffusion, ergodicity
breaking, and significant trajectory-to-trajectory fluctuations of TSDs was provided through a combination of high-resolution
SPT microscopy data from endogenous lipid granules in
living {\it Schizosaccharomyces pombe\/} ({\it S. pombe\/}) yeast cells and
analytical modelling \cite{Jeon2011}. This experiment utilised an optical
tweezers setup to capture the short-term motion of lipid granules in {\it
S. pombe\/} cells through high-precision SPT microscopy. Analysis of the TSDs revealed a crossover: 
initially, the TSD exhibited linear growth (from 0.1 to around 1 msec), followed by a power-law
scaling $\overline{\delta^2(\Delta;t)}\propto\Delta^\beta$, 
consistent with the weakly non-ergodic behaviour predicted for continuous time random walks 
(CTRWs) with heavy-tailed waiting time distributions.\footnote{We use the symbol $\sim$ to denote asymptotic equivalence, meaning that the ratio of two quantities tends to one in a specified limit (e.g., $\Delta \to \infty$). In contrast, we write $\propto$ to indicate that two quantities exhibit the same scaling behaviour, possibly up to a multiplicative constant, and without requiring an explicit asymptotic limit.} In particular, for CTRWs in which the PDF of waiting times 
 follows $\psi(\tau)\propto \tau^{-1-\alpha}$ with $0<\alpha<1$, the mean
waiting time diverges. In confined environments, this leads to a crossover from $\beta=1$ to $\beta
=1-\alpha$ in the TSDs; experimental values indicate $\alpha\approx0.80, \ldots, 0.85$, as 
theoretically expected \cite{staspnas}, see also the discussion in
\cite{pt,pt1,Metzler2014,He2008,Lubelski2008,Neusius2009,Miyaguchi2011a,
Miyaguchi2013}. At even longer times, monitored by video particle tracking, the
motion of the granules follows ergodic viscoelastic diffusion with an anomalous exponent $\alpha
\approx0.8$, consistent with the short-time data \cite{Jeon2011}. Notably, some TSDs
eventually transitioned to normal diffusion at timescales around 1 sec,  but still exhibited substantial 
trajectory-to-trajectory fluctuations, highlighting the persistent heterogeneity even at long observation times.

SPT experiments were conducted to monitor the motion of dendritic cell-specific intercellular adhesion molecule 3-grabbing nonintegrin (DC-SIGN) on living-cell membranes \cite{Manzo2015}. They employed an automated algorithm to track
the positions of quantum dots with nanometre precision, generating over
600 trajectories consisting of more than 200 frames, with some extending
up to 2000 frames. The data were captured at a camera rate of 60 Hz 
 to facilitate the evaluation of the ergodicity of the TSD and
the ensemble-averaged analysis. To investigate DC-SIGN dynamics, they
conducted video microscopy of quantum-dot-labelled DC-SIGN expressed in
Chinese hamster ovary cells under an epi-illumination configuration. The
full receptor is referred to as wild-type DC-SIGN (WT DC-SIGN) to distinguish
it from its mutated forms. The TSDs of WT DC-SIGN displayed normal diffusion,
indicated by linear growth with lag time.  However, these TSDs displayed considerable
trajectory-to-trajectory fluctuations and the  diffusion
coefficients extracted from individual trajectories spanned several orders of magnitude,
reflecting strong heterogeneity. In contrast, the ensemble-averaged
analysis of the MSD revealed subdiffusion with an exponent of $\alpha \approx
0.84$, providing clear evidence of ergodicity breaking. 
The authors further explored the role of protein structure by analysing mutant forms of DC-SIGN and confirmed that ergodicity breaking and global diffusivity fluctuations of the TSDs persisted.
Unlike the diffusion behaviour observed for lipid granules \cite{Jeon2011}, these results did not align
with CTRW characteristics, suggesting a different underlying mechanism for the observed nonergodic behaviour.

The experimental observations of ergodicity breaking and global diffusivity
fluctuations in TSDs challenge the assumption of ergodicity in living cells.
These findings indicate the presence of significant trajectory-to-trajectory
fluctuations, suggesting that conventional models, such as the CTRW
\cite{pt,pt1,staspnas,He2008,Lubelski2008,Neusius2009,Miyaguchi2011a,
Miyaguchi2013}, may not fully capture the complexity of intracellular
diffusion. This highlights the need for further theoretical exploration to
develop models that better describe nonergodic dynamics and global diffusive
fluctuations in biological systems.

\subsection{Local diffusivity fluctuations}

Local diffusivity fluctuations may occur even when the TSDs obtained from
long-time trajectories do not exhibit large trajectory-to-trajectory
fluctuations. Several methods are used to evaluate local diffusivity
fluctuations \cite{MontielCangYang2006,KooMochrie2016,Akimoto2017,samu,
stefanie,yael,amanda}. A key point in detecting local diffusivity fluctuations
is to identify transition points of diffusive states from a single trajectory
\cite{granik2019single,munoz2020single,seckler2022bayesian,gorka,henrik_jpp,janczura,henrik_rev,gorka1,janczura,persson,doris}.

Local diffusivity fluctuations arise when the instantaneous mobility of a tracer, $D(t)$, varies over time due to changes in its immediate environment or in its own internal state. Such variations can have several distinct physical origins. First, the surrounding medium may possess environmental structural or viscoelastic heterogeneity that is quasi-static on experimental timescales, so the tracer experiences different mobilities as it moves between microenvironments. Second, internal dynamics of the tracer itself, such as conformational changes in proteins, can modulate its hydrodynamic drag and thus its diffusivity. Third,  {\color{black}even without spatial displacement, the local environment may evolve dynamically in time---for instance due to dynamical heterogeneity or time-dependent external fields---causing temporal modulations of mobility.} These mechanisms, while physically different, all produce temporally fluctuating diffusivities that can be captured within the LEFD framework and are key to understanding BYNGD and other hallmarks of heterogeneous dynamics.

As discussed in the previous section,
global diffusivity fluctuations were observed in living-cell membranes \cite{Manzo2015}, where 
local diffusivity fluctuations were also investigated using the same experimental
framework. By plotting the displacement $x(t+\Delta)-x(t)$ over time for
WT DC-SIGN trajectories, they analysed temporal variations in the diffusivity.
A likelihood-based algorithm was employed to detect time-dependent changes in
the diffusivity. This approach involved computing maximum-likelihood estimators
to determine local diffusion coefficients and applying a likelihood-ratio test to
pinpoint transitions between diffusive states \cite{MontielCangYang2006,
KooMochrie2016}. Their results revealed that while DC-SIGN trajectories
exhibited Brownian motion within certain time intervals, significant
diffusivity changes occurred between these intervals. By identifying transition
points, they extracted diffusion coefficients from the TSDs, demonstrating
the presence of temporal fluctuations in local diffusivity. These findings
suggest that a fluctuating diffusivity contributes to anomalous diffusion,
ergodicity breaking, and trajectory-to-trajectory fluctuations observed in
living-cell membranes.

Molecular dynamics (MD) simulations by Yamamoto et al.~\cite{Yamamoto2021}
provided further insight into local diffusivity fluctuations in single
proteins. They conducted five independent all-atom MD simulations of the
protein super Chignolin, each lasting 40 microseconds, to explore how protein
conformational fluctuations influence the diffusivity. To quantify
local diffusivity changes, they employed a local time-averaged diffusivity
analysis, as used in previous research \cite{Akimoto2017}. A key discovery was
the correlation between the instantaneous diffusivity $D_I$ and the protein's
radius of gyration $R_g$, leading to the modified Stokes-Einstein relation
\begin{equation}
D_I\propto\frac{1}{R_g+R_0},
\end{equation}
where $R_0$ represents a constant associated with the hydration layer
surrounding the protein. This extension of the classical Stokes-Einstein
relation demonstrates that conformational fluctuations play a key role in
modulating local diffusivity. Their findings \cite{Yamamoto2021} highlight
the intricate interplay between protein conformational dynamics and their
diffusive behaviour, reinforcing the importance of local diffusivity
fluctuations in biological systems.

The motion of lipid molecules in bilayer membranes crowded with membrane
proteins was studied by MD simulations, revealing pronounced non-Gaussian
displacement PDFs and intermittent diffusivity fluctuations \cite{Jeon2016}.
These fluctuations are due to the transient trapping of lipids between
clusters of crowding proteins as well as transient surface attachment to
individual protein molecules. The resulting motion deviates from the purely
Gaussian, viscoelastic motion of lipid molecules in simple bilayer membranes
\cite{Jeon2012}. Instead, a multifractal, spatiotemporally heterogeneous
anomalous lateral diffusion emerges. The observed non-Gaussianity fitted
to stretched Gaussian functional forms $P(r,t)\propto\exp\left(-[r/(ct^{\alpha/2})]^{\delta}\right)$ with stretching exponents $\delta\approx1.3
\ldots1.7$ of the radial component is therefore attributed to the stochastically
changing diffusivity within individual trajectories \cite{Jeon2016}.

MD simulations of supercooled liquids and glass-forming liquids have provided
compelling evidence of dynamic heterogeneity, characterised by fluctuations in
local diffusivity \cite{Kob1997,Yamamoto-Onuki-1998,Yamamoto-Onuki-1998a,
kawasaki2007correlation,Berthier2011,hachiya2019unveiling,Miotto2021,
sposini2023glassy}. Dynamic heterogeneity refers to spatial and temporal
variations in the local diffusivity, driven by cage effects and
cooperative particle motions \cite{Donati1998}. In MD simulations, these
local diffusivity fluctuations manifest through the displacements of
particles and the structural relaxation time \cite{Yamamoto-Onuki-1998,
Yamamoto-Onuki-1998a,kawasaki2007correlation,Berthier2011,
hachiya2019unveiling}, providing a visual and quantitative characterisation of dynamic heterogeneity
 This complexity gives rise to anomalous behaviours
such as non-Gaussian displacement PDFs, slow relaxation, and a non-monotonic
increase in the MSD \cite{Yamamoto-Onuki-1998,Yamamoto-Onuki-1998a,
kawasaki2007correlation,Hurtado2007,Berthier2011,Kim2013,hachiya2019unveiling}.
Local diffusivity fluctuations in supercooled and glass-forming liquids stem
from multiple mechanisms. As a particle diffuses in a heterogeneous environment,
it experiences time-dependent variations in the local diffusivity, further
influenced by structural rearrangements in the surrounding medium. The
interplay between diffusing particles and evolving heterogeneous structures is
critical for the understanding of local diffusivity fluctuations and their
impact on materials properties.

The phenomenon of BYNGD has attracted much interest due to its ubiquity
in nature \cite{wang2009anomalous,wang2012brownian,Bhattacharya2013,
He2013,Guan2014,kwon2014dynamics,yael}. Local diffusivity fluctuations play
a pivotal role in BYNGD \cite{wang2012brownian}. In a pioneering SPT
experiment, the diffusion of colloidal beads along linear phospholipid bilayer tubes was observed, revealing characteristic features of BYNGD
 \cite{wang2009anomalous, wang2012brownian}. While
the MSD across hundreds of trajectories appeared normal-diffusive, the
short-time propagator deviated significantly from a Gaussian PDF, instead
exhibiting a Laplace PDF of the form $P(x) \propto\exp(-c|x|)$. This deviation was attributed
to diffusivity variations within the system \cite{wang2012brownian}. 
The non-Gaussian PDF observed in BYNGD can result from local diffusivity fluctuations. Alternative mechanisms, such as impulsive stochastic processes (e.g., shot-noise driven systems) \cite{Thiffeault2015}, can also give rise to similar statistical features. 
When particles
diffuse in a heterogeneous environment, in which the diffusivity varies over
time and space, the PDF reflects an average over multiple diffusion
coefficients rather than a single value. This phenomenon can be formally
described using a superstatistical approach \cite{Chechkin2017,
beck2003superstatistics}, in which the total propagator is obtained by
integration over the PDF $P(D)$ of diffusion coefficients,
\begin{equation}
\label{eq: superstatistics approach0}
P({\bm r},t)=\int_0^\infty G_d({\bm r},t;D)P(D)dD,
\end{equation}
where $G_d({\bm r},t; D)$ is the Green's function for the diffusion equation
with diffusion coefficient $D$,
\begin{equation}
G_d({\bm r},t;D)=\frac{1}{\sqrt{(4\pi Dt)^d}}\exp\left(-\frac{{\bm r}^2}{4Dt}
\right).
\end{equation}
The PDF derived from Eq.~(\ref{eq: superstatistics approach0}) typically
exhibits a non-Gaussian distribution, highlighting how a varying diffusivity
effects a non-Gaussian PDF. To investigate these diffusive fluctuations,
an iterative algorithm was employed to infer the underlying distribution of local diffusivities from displacement data
\cite{wang2012brownian, lucy1974iterative}. While this method does not directly track
instantaneous diffusivity fluctuations, it provides compelling indirect evidence
of their presence. These results significantly contribute to theoretical
studies on the impact of fluctuating diffusivity in complex environments
\cite{Chubynsky2014,Jain2016,Chechkin2017,tyagi2017non}.

\section{Fundamental properties of Langevin equation with fluctuating diffusivity}
\label{sec: Fundamental properties of LEFD}

In this section, we introduce the stochastic model of the Langevin equation
with fluctuating diffusivity (LEFD) and provide its general theoretical
framework for analysing moments of displacement and the relative standard
deviation of the TSDs  in LEFD systems. The general form of the LEFD is given by 
\begin{equation}
\frac{d{\bm r}(t)}{dt}={\bm B}(t)\cdot{\bm \xi}(t),
\label{LEFD_general}
\end{equation}
where ${\bm r}(t)$ is the $d$-dimensional position of a Brownian particle at
time $t$, ${\bm B}(t)$ is the noise coefficient matrix, and ${\bm \xi}(t)$ is
a $d$-dimensional Gaussian white noise, i.e., 
\begin{equation}
\langle{\bm \xi}(t)\rangle=0,\quad\langle\xi_i (t)\xi_j(t')\rangle=\delta_{ij}
\delta(t-t'),
\end{equation} 
where $\xi_i(t)$ is the $i$th component of ${\bm \xi}(t)$. We assume that ${\bm
B}(t)$ follows a stochastic process that is independent of the position ${\bm
r}(t)$ and the noise ${\bm \xi}(t)$. As a result, the LEFD model multiplicatively
couples two independent stochastic processes, ${\bm B}(t)$ and  ${\bm \xi}(t)$,
as formulated in Eq.~(\ref{LEFD_general}).
This form represents an annealed version of a more general Langevin equation in which the noise coefficient matrix depends explicitly on space and time, ${\bm B}({\bm r}, t)$. In such cases, care must be taken in specifying the stochastic interpretation (e.g., It\^o, Stratonovich, and H\"anggi-Klimontovich), as it affects the resulting dynamics and probability density evolution \cite{van1992stochastic,igor2010}. 
In particular, multiplicative noise gives rise to a spurious drift, which modifies the drift term in the Fokker-Planck equation depending on the stochastic interpretation. 
Here, by treating ${\bm B}(t)$ as independent of position, the LEFD simplifies the description of heterogeneous dynamics by focusing on temporal fluctuations in diffusivity alone.

When the noise coefficient matrix is isotropic, i.e., 
\begin{equation}
{\bm B}(t)=\sqrt{2D(t)}{\bm 1},
\end{equation} 
where ${\bm 1}$ is the $d$-dimensional unit matrix, the LEFD simplifies to 
\begin{equation}
\frac{d{\bm r}(t)}{dt}=\sqrt{2D(t)}{\bm\xi}(t).
\label{LEFD}
\end{equation}
While we primarily focus on the isotropic form of the LEFD in this review, the general tensorial form of ${\bm B}(t)$ is also introduced and exemplified through specific models, with a comprehensive analysis of its statistical consequences reserved for future investigations.
 In a loose sense, the isotropic model with time-dependent diffusivity corresponds to an ``annealed" system, where the properties
of each local region change upon each visit by the tracer particle. This contrasts with
 ``quenched" disorder, in which the local properties remain fixed throughout the observation period---meaning that whenever the tracer revisits a particular site, it experiences the same local properties (e.g., fixed waiting times as in a quenched trap model \cite{bouchaud90}, in contrast to the annealed waiting times in a standard CTRW framework).
In non-isotropic cases, where ${\bm B}(t)$ is a full tensor that encodes direction-dependent fluctuations, the model captures anisotropic diffusion arising from structural or dynamical anisotropy in the environment---such as in complex fluids, active matter, or entangled polymers \cite{aporvari2020anisotropic,doi1978dynamics,Doi-Edwards-book,degennes,rubinstein,Miyaguchi2017}.

\subsection{Langevin equation models with fluctuating diffusivity}

Diffusion with fluctuating diffusivity is a ubiquitous phenomenon, occurring
in diverse systems across physics, chemistry, and biology. In this section,  we present
several representative models, in which the diffusion dynamics are
governed by spatially or temporally fluctuating diffusivity.

\subsubsection{Diffusion in a heterogeneous environment}

In heterogeneous environments, the Langevin equation in the overdamped limit
takes the form 
\begin{equation}
\frac{d{\bm r}(t)}{dt}=\sqrt{2D({\bm r}(t))}{\bm \xi}(t),
\label{eq: LEFD hetero}
\end{equation}
where the instantaneous diffusivity depends on the particle position ${\bm
r}(t)$. According to the Stokes-Einstein relation [Eq.~(\ref{eq: SK relation})],
the local diffusivity is determined by the ambient temperature, viscosity,
and the particle radius. Notably,  when a temperature gradient is present, 
Eq.~(\ref{eq: LEFD hetero}) serves as a model for thermally driven
diffusion, relevant to the description of the Soret effect \cite{Soret,
Giglio1977,Iacopini2003,duhr2006molecules,dominguez2011soret}.

Theoretical models of diffusion in heterogeneous environments exhibit
anomalous statistical properties, such as anomalous diffusion, non-Gaussian
propagators, and/or ergodicity breaking \cite{Cherstvy2014,Leibovich2019,
adrian}. However, because the noise is multiplicatively coupled to the
${\bm r}(t)$-dependent noise strength in Eq.~(\ref{eq: LEFD hetero}), this
equation differs from the LEFD framework, in which $D(t)$
evolves independently of ${\bm r}(t)$ (see above). In particular, the
stochastic interpretation of Eq.~(\ref{eq: LEFD hetero}) depends explicitly
on the physically meaningful stochastic interpretation used to evaluate the
Langevin equation---such as It\^o, Stratonovich, or H{\"a}nggi-Klimontovich
formalisms \cite{Ito1944,Stratonovich1966,van1992stochastic,Gardiner,
adrian,lubensky,oded}.

When the heterogeneous environment is dynamic and fluctuating, an annealed 
description becomes appropriate. In such cases, the LEFD [Eq.~(\ref{LEFD})] 
provides an effective modelling framework. For instance, the annealed
transit time model (ATTM) describes diffusion in a spatially heterogeneous
environment and can be formulated as an LEFD model with a stochastic
diffusivity process $D(t)$ \cite{Massignan2014, Manzo2015}. Similarly, the
CTRW with a purely time-dependent waiting time PDF also represents an annealed approach
\cite{bouchaud90}. Diffusion in glassy and supercooled liquids provides another example of fluctuating diffusivity, where the mobility of particles varies in both space and time due to evolving environmental fluctuations \cite{Hurley1995,
Yamamoto-Onuki-1998,Yamamoto-Onuki-1998a,sillescu1999heterogeneity,
Richert-2002,Berthier2011}. In this context, Eq.~(\ref{eq: LEFD hetero})
captures local (quenched) variations in diffusivity, whereas the LEFD represents the annealed counterpart, effectively describing the influence of temporally fluctuating environments.

\subsubsection{Center-of-mass motion in the reptation model}

The reptation model \cite{doi1978dynamics,Doi-Edwards-book,degennes,rubinstein}
describes the motion of entangled polymers, whose centre-of-mass dynamics
follow an LEFD-like equation. The noise coefficient matrix is given by 
\begin{equation}
{\bm B}(t)=\sqrt{\frac{3D_{\rm com}}{\langle {\bm R}_{e}^2\rangle}} \frac{{\bm R}_e(t){\bm R}_e(t)}{|{\bm R}_e(t)|},
\label{eq: noise matrix in reptation}
\end{equation} 
where $D_{\rm com}$ is the diffusivity and ${\bm R}_e(t)$ is the end-to-end
vector of the polymer. Since the end-to-end vector ${\bm R}_e(t)$ is independent
of the position ${\bm r}(t)$, the reptation model can be formulated within the
LEFD framework using Eq.~(\ref{eq: noise matrix in reptation}). Here, the fluctuating
diffusivity is attributed to the fluctuating shape of a polymer. The LEFD
formalism is therefore applicable  not only to entangled polymers but also to 
other systems characterised by fluctuating particle shapes, such as
 proteins undergoing conformational changes
\cite{Yamamoto2021}. The centre-of-mass motion of these deformable particles exhibits 
time-dependent diffusivity, reinforcing the utility of the LEFD approach for describing dynamics 
 in both biological and polymeric systems. 

\subsubsection{Diffusion in media with fluctuating viscosity}

In some cases, diffusivity fluctuations arise due to environmental viscosity
fluctuations \cite{rozenfeld1998brownian,luczka2000diffusion}. The underdamped
Langevin equation with fluctuating viscosity takes the form:
\begin{equation}
m\dot{v}(t)=-\gamma(t)v(t)+\sqrt{2\gamma(t)k_{\rm B}T}\xi(t),
\label{eq: LE fluctuating viscosity}
\end{equation}
where $\gamma(t)$ is a stochastic process representing a fluctuation of the
environment, assumed to be independent of the position $x(t)$,
velocity $v(t)$, and noise $\xi(t)$. In the overdamped limit, this model
reduces to an LEFD form, whose instantaneous diffusivity is given by $D(t)=
2k_{\rm B}T/\gamma(t)$. This description captures diffusion in
dynamically changing environments such as viscoelastic fluids, colloidal
suspensions, and biological media, where fluctuations in local viscosity 
modulate the particle motion.

\subsection{Brownian yet non-Gaussian diffusion}
\label{sec: III BYNGD}

The phenomenon of BYNGD describes systems in which the MSD grows linearly
with time (indicating normal diffusion, also called ``Fickian"), while the 
propagator exhibits non-Gaussian characteristics at short times.
Over longer timescales, the propagator typically crosses over to
 a Gaussian PDF, once the correlation time of the diffusivity fluctuations is exceeded. 
 This behaviour naturally emerges when the instantaneous
diffusivity $D(t)$ follows a stationary stochastic process.

To illustrate this, consider the squared displacement
\begin{equation}
|\bm{r}(t) - \bm{r}(0)|^2 = \int_0^\infty \! \int_0^\infty 
K(t,t')\, \bm{\xi}(t) \cdot \bm{\xi}(t') \, dt\, dt',
\end{equation}
where $ K(t,t') = \sqrt{2D(t)}\sqrt{2D(t')} $.
Taking the average, we obtain the MSD
\begin{equation}
\langle |{\bm r}(t)-{\bm r}(0)|^2\rangle_{\rm st}=2d\langle D\rangle_{\rm st}t,
\label{eq: normal diffusion in LEFD}
\end{equation}
where $\langle\cdot\rangle_{\rm st}$ denotes an ensemble average under
stationary initial conditions. The stationary mean diffusivity is given by
\begin{equation}
\langle D\rangle_{\rm st}=\int_0^\infty DP_{\rm st}(D)dD ,
\end{equation}
where $P_{\rm st}(D)$ represents the stationary PDF of the diffusivity $D$.
This result confirms that when  $D(t)$ is stationary, the MSD follows normal
diffusion with effective diffusivity $\langle D\rangle_{\mathrm{st}}$ at all times,
despite the local variations in the diffusivity. However, if a stationary
distribution for $D(t)$ does not exist, the MSD may exhibit anomalous scaling. 
In such cases, the MSD follows 
\begin{equation}
\langle |{\bm r}(t)-{\bm r}(0)|^2 \rangle=2d \int_0^t \langle D(t')\rangle dt', 
\end{equation}
where $ \langle D(t)\rangle$ is the
time-dependent mean diffusivity. When  $\langle D(t)\rangle$ varies with time,
the MSD deviates from linearity, signalling anomalous diffusion.  
A detailed discussion of non-stationary initial conditions for the diffusivity is provided in Section~\ref{subsec: two-state model: MSD}
and in Refs.~\cite{Miyaguchi2016,sposini2018random}. We also note that the Brownian diffusing-diffusivity
can be formulated using subordination techniques  and  bi-variate
Fokker-Planck equation \cite{Chechkin2017,Sposini2024PRL,Sposini2024PRE}.

\subsubsection{Non-Gaussian propagator in a short-time limit} 

The short-time behaviour of the PDF can be effectively analysed using the
superstatistical framework \cite{beck2003superstatistics,Chechkin2017}. 
In this regime, the instantaneous diffusivity $D(t)$ can be treated as a
constant over short timescales. Specifically, for $t\ll\tau_D$, where
$\tau_D$ is a characteristic correlation time of $D(t)$, $D(t)$ can be
approximated by $D(t)\approx D(0)$. Under these conditions, the PDF $P({\bm
r},t)$ is described by an average over the diffusivity PDF, i.e.,
\begin{equation}
P({\bm r},t)\approx\int_0^\infty G_d({\bm r},t;D)P(D)dD,
\label{eq: superstatistics approach}
\end{equation}
where $G_d({\bm r},t;D)$ represents the Green's function for normal diffusion
with the fixed diffusion coefficient $D$. This formulation highlights the
superstatistical nature of the process, in which the propagator emerges as
an ensemble average over random diffusivities. 

In Fourier space, the PDF transforms into
\begin{equation}
\tilde{P}({\bm k},t)=\int_0^\infty\exp(-{\bm k}^2Dt)P(D)dD.
\end{equation}
This representation provides a direct means to relate the moments of the
displacement to the diffusivity PDF $P(D)$. Using the general relation
between the $n$th moment of the displacement and the PDF in Fourier space,
\begin{equation}
\langle{\bm r}(t)^n\rangle=i^n\left.\frac{\partial^n\tilde{P}({\bm k},t)}{
\partial{\bm k}^n}\right|_{{\bm k}=0},
\label{relation between moments and F}
\end{equation}
we obtain explicit expressions for the second and fourth moments as
\begin{equation}
\langle{\bm r}(t)^2\rangle_{\rm st}=2d\langle D\rangle_{\rm st}t
\label{2nd moment of ri}
\end{equation}
and 
\begin{equation}
\langle{\bm r}(t)^4\rangle_{\rm st}=4d(d+2)\langle D^2\rangle_{\rm st}t^2, 
\label{4th moment of ri}
\end{equation}
where $\langle D\rangle_{\rm st}$ and
$\langle D^2\rangle_{\rm st}$ represent the stationary averages of $D$
and $D^2$, respectively. To quantify the non-Gaussianity of the propagator,
we introduce  the non-Gaussian parameter (NGP) \cite{Rahman1964}
\begin{equation}
A(t)\equiv\frac{d\langle{\bm r}(t)^4\rangle}{(d+2)\langle{\bm r}(t)^2\rangle^2}
-1.
\label{eq: NGP def}
\end{equation}
Substituting  Eqs.~(\ref{2nd moment of ri}) and
(\ref{4th moment of ri}) yields
\begin{equation}
A(t)\approx\frac{\langle D^2\rangle_{\rm st}-\langle D\rangle_{\rm st}^2}{\langle
D\rangle_{\rm st}^2}>0.
\label{eq: NGP short-time}
\end{equation}
Since  $A(t)$ remains a positive constant for $t\ll\tau_D$, this result
demonstrates that even though the MSD follows normal diffusion, $\langle {\bm
r}(t)^2\rangle_{\rm st}=2d\langle D\rangle_{\rm st}t$, the propagator  remains non-Gaussian at
short times due to the presence of the fluctuating diffusivity. This phenomenon
is a key feature of BYNGD, in which normal MSD growth coexists with a non-Gaussian displacement distribution.  
Note that a zero value for the NGP does not necessarily imply
that the underlying PDF is Gaussian. For instance, when the centre of the PDF is
Gaussian and has a crossover to far-out Laplace tails, the NGP converges to
zero rather quickly \cite{flm}.

\subsubsection{Gaussian propagator in the long-time limit} 

To analyse the propagator in the long-time limit, we employ the subordination
approach for the LEFD \cite{bochner,Fogedby1994,Chechkin2017}. In this sense,
the LEFD equation (\ref{LEFD}) can be recast in terms of the two coupled
stochastic equations,
\begin{equation}
\frac{d{\bm r}(\tau)}{d\tau}=\sqrt{2}\xi(\tau)
\end{equation}
and
\begin{equation}
\frac{d\tau(t)}{dt}=D(t).
\end{equation}
Since $D(t)$ is a stochastic process, the operational time $\tau(t)$, defined
as
\begin{equation}
\tau(t)=\int_0^tD(t')dt',
\end{equation}
also follows a stochastic process. The PDF of ${\bm r}(\tau)$, with initial
condition ${\bm r}(0)=0$, is given by the Green's function
\begin{equation}
G_d({\bm r},\tau;1)=\frac{1}{\sqrt{(4\pi\tau)^d}}\exp\left(-\frac{r^2}{4\tau}
\right),
\end{equation}
where $r=|{\bm r}|$. To obtain the PDF of ${\bm r}(t)$, we integrate out $\tau$
by averaging over its distribution:
\begin{equation}
P({\bm r},t)=\int_0^\infty G_d({\bm r},\tau;1)\rho(\tau,t)d\tau,
\end{equation}
where $\rho(\tau,t)$ is the PDF of $\tau$ at time $t$. Taking the Fourier
transform $\tilde{P}({\bm k},t)$ of $P({\bm r},t)$, we  obtain
\begin{equation}
\tilde{P}({\bm k},t)=\int_0^\infty\tilde{G}_d({\bm k},\tau;1)\rho(\tau,t)d\tau.
\end{equation}
Using the Fourier transform of the Green's function, this expression reduces
to the Laplace transform of $\rho(\tau,t)$ with respect to $\tau$,
\begin{equation}
\tilde{P}({\bm k},t)=\int_0^\infty e^{-k^2\tau}\rho(\tau,t)d\tau,
\end{equation}
where the Laplace variable is $k^2$.

To determine the asymptotic behaviour of the PDF, we assume that $D(t)$ is
a stationary process. By the central limit theorem, the PDF of $\tau(t)$
converges to a Gaussian distribution with mean $\mu_t$  and variance $\sigma_t^2$,
given by $\mu_t\equiv\langle D\rangle_{\rm st}t$ and $\sigma_t^2\equiv c_{D}t$.
The constant $c_D$ represents the correlation function of $D(t)$
\cite{bouchaud90} (see also \cite{Chechkin2017}),
\begin{equation}
c_D=2\int_0^{\infty}C(t)dt,
\end{equation}
where $C(t)=\langle D(0)D(t)\rangle_{\rm st} - \langle D\rangle_{\rm
st}^2$. In the long-time limit, the Fourier-transformed PDF simplifies to
\begin{equation}
\tilde{P}({\bm k},t)\sim\int_0^\infty e^{-k^2\tau}\frac{1}{\sqrt{2\sigma_t^2}}
\exp\left(-\frac{(\tau-\mu_t)^2}{2\sigma_t^2}\right)d\tau.
\label{Fourier asymptotic}
\end{equation}
Taking $t\to\infty$ and $k\to0$, we obtain
\begin{equation}
\tilde{P}({\bm k},t)\sim e^{-\mu_t k^2}. 
\end{equation}
Applying the inverse Fourier transform, the PDF in real space becomes
\begin{equation}
P({\bm r},t)=\frac{1}{(4\pi\langle D\rangle_{\rm st}t)^{d/2}}\exp\left(-
\frac{|{\bm r}|^2}{4\langle D\rangle_{\rm st}t}\right).
\label{eq: prop LEFD long-time}
\end{equation}
Thus, in the long-time limit, the PDF indeed converges to a Gaussian PDF,
indicating that diffusion crosses over to ordinary Brownian motion at
long times, with an effective diffusivity.

\begin{figure*}
\includegraphics[width=.9\linewidth, angle=0]{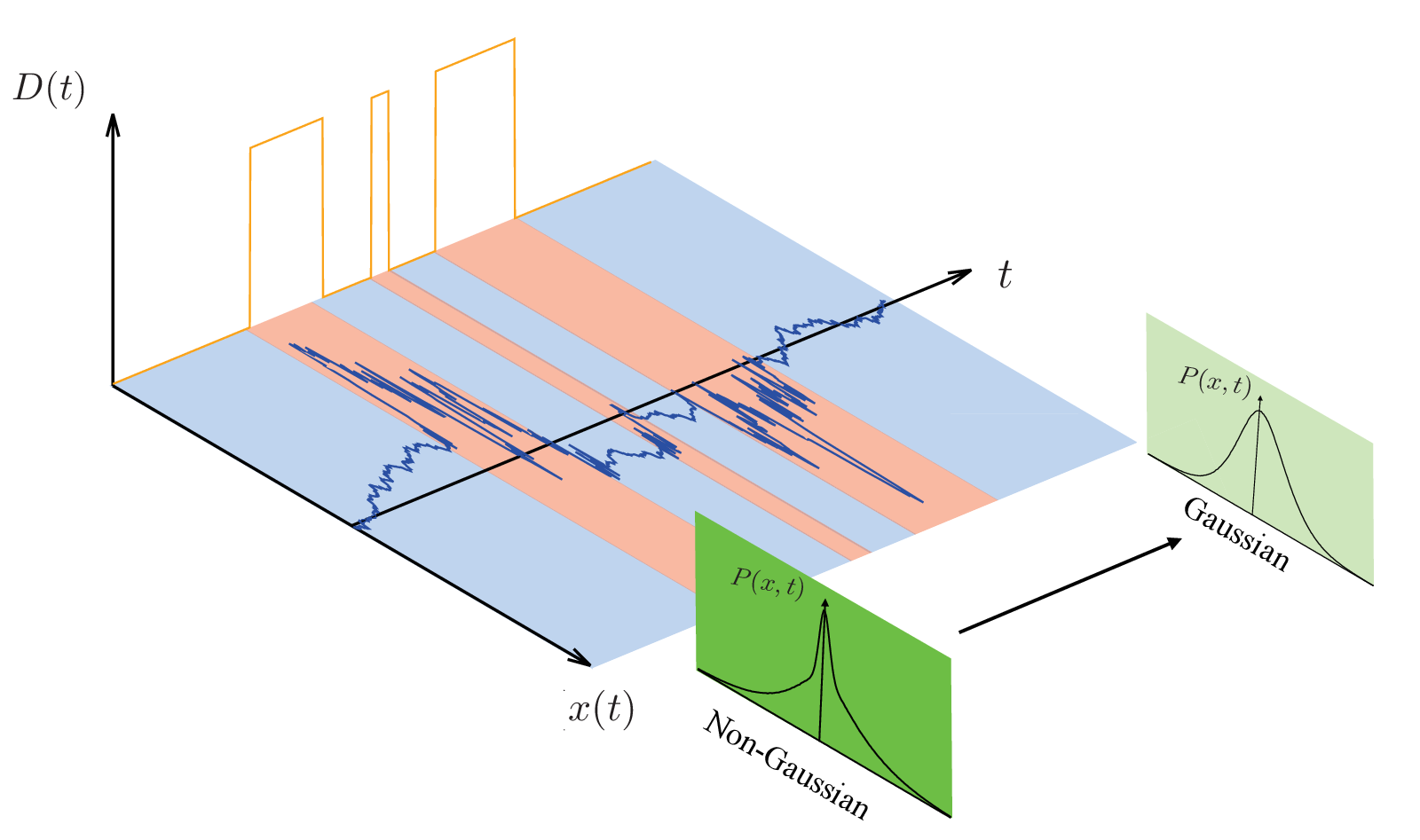}
\caption{Statistical signatures of fluctuating diffusivity in LEFD. The figure illustrates a representative trajectory and the corresponding instantaneous diffusivity in the LEFD, alongside the evolution of the propagator from a non-Gaussian to a Gaussian form over time. This transition reflects the underlying fluctuating diffusivity, which causes deviations from Gaussian statistics at short times. These features are hallmarks of heterogeneous diffusion dynamics.}
\label{fig: main LEFD}
\end{figure*}

We estimate the asymptotic behaviour of NGP in the long-time limit. Using
the relation between the $n$th moment of ${\bm r}(t)$ and he Fourier-transformed PDF $P({\bm k},t)$,
i.e., Eq.~(\ref{relation between moments and F}), together with 
Eq.~(\ref{Fourier asymptotic}), the second and fourth moments are given by
\begin{equation}
\langle{\bm r}(t)^2\rangle_{\rm st}=2d\mu_t
\end{equation}
and
\begin{equation}
\langle{\bm r}(t)^4\rangle_{\rm st}=4d(d+2)(\sigma_t^2+\mu_t^2).
\end{equation}
Substituting the expressions for $\mu_t$ and $\sigma_t^2$, the asymptotic
behaviour of the NGP becomes
\begin{equation}
A(t)\sim\frac{c_D}{\langle D\rangle_{\rm st}^2}\frac{1}{t}
\end{equation}
for $t\to\infty$. This asymptotic scaling $1/t$ for $t\to\infty$ is a general feature
within the LEFD framework, as demonstrated in Ref.~\cite{Uneyama2015}. 
When the instantaneous diffusivity fluctuates, the PDF always exhibits non-Gaussianity at short times but converges to a Gaussian distribution in the long-time limit. This characteristic behaviour is illustrated in Fig.~\ref{fig: main LEFD}, which shows a representative LEFD trajectory with the corresponding time-dependent diffusivity and the evolution of the propagator from non-Gaussian to Gaussian. This characteristic also holds when the underlying process is not standard Brownian motion but fractional Brownian motion
\cite{Deng2009,wei,diego,wei1}.

\subsection{ Relative standard deviation of the time-averaged square displacement}
\label{subsec: RSD in section III}

We here present the theory of global diffusivity fluctuations as encoded by the
fluctuations of the TSD. For ordinary Brownian motion, the TSD is ergodic,
meaning that in the long-time limit, it converges to the ensemble-averaged MSD,
\begin{equation}
\overline{\delta^2(\Delta;t)}\to  \langle | {\bm r}(\Delta)-{\bm r}(0)  |^2\rangle
_{\rm st},
\end{equation}
for $t\to\infty$. 
When $D(t)$ is a stationary
stochastic process, this equilibrium average is equivalent to the stationary ensemble average. 
 To quantify the TSD fluctuations, we introduce the relative standard deviation (RSD) of the TSD,
\begin{equation}
\Sigma(t;\Delta)=\sqrt{\frac{\left<\overline{\delta^2(\Delta;t)}^2\right>-
\left<\overline{\delta^2(\Delta;t)}\right>^2}{\left<\overline{\delta^2
(\Delta;t)} \right>^2}}.
\end{equation}
Although the RSD depends on both the measurement time $t$ and the lag time
$\Delta$, we primarily focus on its dependence on $t$. In an ergodic system,
the RSD vanishes in the long-time limit in the form $\Sigma(t;\Delta)\to0$
as $t\to\infty$ \cite{Deng2009,Akimoto2016}. Conversely, if the RSD converges
to a nonzero constant, the system is non-ergodic, meaning that individual TSDs
continue to fluctuate even at long times. Thus, the RSD serves as a quantitative
measure of ergodicity. The square of the RSD is  also known as the ergodicity breaking
(EB) parameter \cite{He2008,Schulz2013,Metzler2014}, which has been widely used
to characterise non-ergodic diffusion in heterogeneous environments.

We next derive a general expression for the RSD in the case when the instantaneous
diffusivity $D(t)$ is a stationary stochastic process. Under this assumption,
the ensemble-averaged TSD is given by 
\begin{equation}
\left<\overline{\delta^2(\Delta;t)}\right>_{\rm st}=2d\langle D\rangle_{\rm st}
\Delta.
\label{eq: ETSD LEFD general}
\end{equation}
To quantify the fluctuations of the TSD, we evaluate its second moment, which
was previously obtained in Ref.~\cite{Uneyama2015}. For the case when the
characteristic timescale $\tau_D$ of $D(t)$ is significantly longer than the
lag time $\Delta$, and in the long-time regime $t\gg\Delta$, the squared RSD
is given by
\begin{equation}
\Sigma^2(t;\Delta)\approx\frac{2}{t^2}\int_0^tds(t-s)\psi_1(s),
\label{eq: formula RSD}
\end{equation}
where $\psi_1(t)$ is the correlation function of the instantaneous diffusivity,
defined as
\begin{equation}
\psi_1(t)\equiv\frac{\langle D(t)D(0)\rangle_{\rm st}}{\langle D\rangle_{\rm
st}^2}-1.
\end{equation}
At long times, assuming $\psi_1(t)$ decays as $\psi_1(t)=o(t^{-1})$ for $t\gg\tau_
D$, the RSD simplifies to
\begin{equation}
\Sigma^2(t;\Delta)\approx\left\{\begin{array}{ll}\psi_1(0)&(t\ll\tau_D),\\\\
\dfrac{2}{t}\int_0^\infty ds\psi_1(s)&(t\gg\tau_D).\end{array}\right.
\label{eq: approx formula RSD}
\end{equation}
For short measurement times ($t\ll\tau_D$), the RSD thus remains constant,
indicating that fluctuations in the TSD are intrinsic. This reflects that individual realisations retain trajectory-specific variability over such timescales, an intuitive behaviours for a stochastic process.
In contrast, for long measurement times
($t\gg\tau_D$), the RSD exhibits the power-law decay scaling like $1/t$. This
decay arises from the fluctuations of a sum of independent random variables, a
behaviour consistent with the central limit theorem. In the case of ordinary
Brownian motion, the RSD is known to decay as $\Sigma^2(t;\Delta)\sim4\Delta/
(3dt)$ for $t\to\infty$. Thus, while systems with fluctuating diffusivity
initially exhibit persistent TSD fluctuations, in the long-time limit, their
behaviour gradually approaches that of standard Brownian motion.
This transition reflects how fluctuating diffusivity  gives rise to long-lived but ultimately decaying TSD variability---a hallmark of heterogeneous diffusion dynamics.

The derived expression for the RSD establishes a direct connection between the
characteristic timescale of diffusivity fluctuations and the crossover time at
which the RSD transitions from a plateau to a $1/t$ decay. This crossover time,
$\tau_c$, can be estimated as
\begin{equation}
\tau_c\approx\frac{2}{\psi_1(0)}\int_0^\infty ds\psi_1(s). 
\end{equation}
When $\psi_1(t)$ exhibits single-exponential decay, i.e., $\psi_1(t)=\psi_1(0)
e^{-t/\tau_D}$, the crossover time becomes $\tau_c=2\tau_D$. This result indicates
that the crossover time is proportional to the characteristic timescale of the
diffusivity fluctuations. In other words, longer diffusivity correlations lead
to an extended plateau in the RSD before crossing over to the $1/t$ decay.
However, in more complex systems, such as the case of entangled polymers and
glassy materials, the relationship between the characteristic timescale of the
diffusivity fluctuations and the crossover time is generally nontrivial due to
the presence of multiple relaxation mechanisms. Nevertheless, the presence of a
plateau in the RSD serves as a clear indicator of fluctuating diffusivity,
reflecting the persistence of global diffusivity fluctuations in these systems.


Now we turn to the non-stationary case when the instantaneous diffusivity $D(t)$
evolves over time. In what follows, we assume that the lag time $\Delta$ is
significantly shorter than the characteristic timescale $\tau_D$ of the
diffusivity fluctuations, i.e., $\Delta\ll\tau_D$. Following the approach in
Ref.~\cite{Miyaguchi2016}, the ensemble-averaged TSD is expressed as
\begin{equation}
\left<\overline{\delta^2(\Delta;t)}\right>\approx\frac{2d\Delta}{t-\Delta}\int_
0^{t-\Delta}dt'\langle D(t')\rangle.
\end{equation}
Furthermore, the second moment of the TSD can be decomposed into two distinct
contributions,
\begin{align}
\left\langle \left| \overline{\delta^2(\Delta; t)} \right|^2 \right\rangle 
\approx {} & \, \frac{8 d^2 \Delta^2}{(t - \Delta)^2} \int_0^{t - \Delta} dt' 
\Bigg[ \int_0^{t'} dt'' \langle D(t') D(t'') \rangle \notag \\
& \quad + \frac{2 \Delta}{3d} \langle D(t')^2 \rangle \Bigg].
\end{align}
In the regime where $\Delta\ll\tau_D\ll t$, thus, the RSD can be decomposed
into two parts
\begin{equation}
\Sigma^2(t;\Delta)\approx\Sigma_{\rm id}^2(t;\Delta)+\Sigma_{\rm ex}^2(t;\Delta),
\label{eq: rsd id and ex}
\end{equation}
where
\begin{equation}
\Sigma_{\rm id}^2(t;\Delta)\equiv\frac{4\Delta}{3d}\frac{\int_0^tdt'\langle
D(t')^2\rangle}{\int_0^tdt'\int_0^tdt''\langle D(t')\rangle\langle D(t'')\rangle}
\end{equation}
represents the ideal part, accounting for fluctuations in a single-mode
diffusion process, and
\begin{equation}
\Sigma_{\rm ex}^2(t;\Delta)\equiv\frac{\int_0^tdt'\langle\delta D(t')\delta D
(t'')\rangle}{\int_0^tdt'\int_0^tdt''\langle D(t')\rangle\langle D(t'')\rangle}
\end{equation}
represents the excess part, arising due to the fluctuating diffusivity,
with $\delta D(t')\equiv D(t')-\langle D(t')\rangle$. The asymptotic behaviour
of the ideal part then reads $\Sigma_{\rm id}^2(t;\Delta)\propto1/t$ for $t\to
\infty$. In particular, for a single-mode diffusion process with 
diffusion coefficient $D$, the RSD is given by $\Sigma^2(t;\Delta)\sim4\Delta
/(3dt)$ for $t\to\infty$. However, when diffusive fluctuations are present,
the excess part becomes the dominant contribution to the RSD. This implies
that the presence of a fluctuating diffusivity significantly alters the ergodic
properties of the system, as will be discussed in detail in later sections. 


\subsection{Derivation of Langevin equation with fluctuating diffusivity}
\label{sec: derivation of langevin equation with fluctuating diffusivity}

The LEFD can be interpreted as a coarse-grained dynamic equation that captures
the essential features of the underlying microscopic dynamics. A natural
question then arises: how does the LEFD relate to the microscopic description
of diffusion? Naively, one might expect that the LEFD emerges as an effective
coarse-grained equation obtained by eliminating fast degrees of freedom
from the microscopic dynamics. In standard coarse-graining approaches, 
the generalised Langevin equation (GLE) typically arises as the effective
dynamical equation for a subset of slow variables, incorporating a memory kernel
and non-Markovian effects. The GLE is rigorously derived from microscopic
Hamiltonian dynamics using the projection operator method \cite{Grabert-book}.
However, in many cases---particularly under the widely used Gaussian
noise approximation---the GLE leads to purely Gaussian diffusion, which contrasts with
the non-Gaussian features observed in systems governed by the LEFD. 

In this subsection, we derive the LEFD starting from a Hamiltonian system. This derivation is carried out under the specific set of physical and mathematical assumptions stated explicitly here, to clarify the scope and applicability of the result:
\begin{enumerate}
\item Equilibrium and homogeneous environment---the underlying many-particle system is in a stationary state and statistically homogeneous on the coarse-grained scale.
\item Separation of timescales---the variables of interest (slow variables) evolve on timescales much longer than those associated with the remaining degrees of freedom (fast variables).
\item  Markovian-like coarse-grained dynamics---the evolution of the position variable (a slow variable) can be effectively described by a memoryless stochastic process, provided that the dynamics of auxiliary slow variables are specified and the fast variables have been averaged out.
\end{enumerate}
Under these assumptions, coarse-graining via either the local-equilibrium trap-potential approach or the projection-operator method leads to an LEFD-type description in which the instantaneous diffusivity is a slowly varying stochastic process determined by the underlying microscopic dynamics.

There are several theoretical pathways for deriving the LEFD from
microscopic dynamics. One such approach involves the Langevin equation with
a transient potential (LETP) \cite{Uneyama-2020,Uneyama-2022}, which
introduces a time-dependent fluctuating potential as an auxiliary stochastic
variable. In this framework, microscopic equations of motion, such as the
canonical equations, can be coarse-grained into an effective Langevin
description. The position and momentum of a tagged particle, denoted as
$\bm{r}(t)$ and $\bm{p}(t)$, respectively, serve as the primary coarse-grained
degrees of freedom. In addition, the instantaneous potential
$\phi(\bm{r},t)$---referred to as the transient potential---captures 
the fluctuating energy landscape experienced by the particle. By applying the
projection operator method to these degrees of freedom, we obtain an effective
dynamical equation \cite{Uneyama-2022}. Under suitable approximations, the
resulting LETP takes the form 
\begin{equation}
\label{eq: LETP}
\frac{d\bm{r}(t)}{dt}=-\frac{D_0}{k_BT}\frac{\partial\phi(\bm{r}(t),t)}{
\partial\bm{r}(t)}+\sqrt{2D_0}\bm{\xi}(t).
\end{equation}
Here, $D_0$ is a constant diffusion coefficient and $\phi(\bm{r},t)$ represents
the fluctuating transient potential, which evolves dynamically over time. This
formulation naturally leads to a scenario analogous to the LEFD, in which the
diffusivity varies stochastically due to microscopic interactions. A full
description of the system involves a set of coupled dynamic equations: one
for the particle motion, Eq.~\eqref{eq: LETP}, and another for the evolution
of the transient potential $\phi(\bm{r},t)$. Through the projection operator
formalism, the dynamic equation for $\phi(\bm{r},t)$ reduces to a GLE,
which naturally incorporates memory effects and non-Markovian fluctuations.

In an isotropic and homogeneous system, a tagged particle, on average, does not
experience a net potential force, implying that the mean transient potential
remains constant, $\langle\phi(\bm{r},t)\rangle=\text{const}$. A simplified
yet effective approximation for the transient potential is the harmonic
approximation, for which the potential is expressed as \cite{Uneyama-2020}
\begin{equation}
\label{eq: harmonic transient potential}
\phi(\bm{r},t)\approx\frac{1}{2}\kappa[\bm{r}-\bar{\bm{r}}(t)]^2,
\end{equation}
where $\kappa>0$ is a constant and $\bar{\bm{r}}(t)$ represents the stochastic
centre position of the harmonic potential. The time-dependent fluctuations of
the transient potential arise due to the dynamics of $\bar{\bm{r}}(t)$.
Substituting the harmonic potential into Eq.~\eqref{eq: LETP}, the Langevin
equation with a transient harmonic potential takes the form
\begin{equation}
\label{eq: LETP harmonic transient potential}
\frac{d\bm{r}(t)}{dt}=-\frac{D_0\kappa}{k_BT}[\bm{r}(t)-\bar{\bm{r}}(t)]+
\sqrt{2D_0}\bm{\xi}(t).
\end{equation}
The behaviour of the effective dynamics for $\bm{r}(t)$ depends on the
stochastic model chosen for $\bar{\bm{r}}(t)$. If $\bar{\bm{r}}(t)$ follows a
Langevin equation (as in a coupled oscillator model), then $\bm{r}(t)$
obeys a GLE, exhibiting a Gaussian behaviour. In contrast, if $\bar{\bm{r}}(t)$ follows
a stochastic jump process (e.g., discrete jumps or resampling dynamics),
then $\bm{r}(t)$ exhibits a non-Gaussian behaviour \cite{Uneyama-2020}. On
timescales longer than the characteristic jump time, the MSD exhibits normal
diffusion, $\langle[\bm{r}(t)-\bm{r}(0)]^2\rangle\sim 2dD_{\text{eff}}t$,
where $D_{\text{eff}}$ is the effective long-time diffusion coefficient
(typically much smaller than $D_0$). However, despite the normal MSD growth,
the NGP remains nonzero in this regime, decaying as $A(t)\propto t^{-1}$.
Thus,the harmonic LETP model naturally reproduces the BYNGD behaviour and
can be interpreted as a coarse-grained realisation of LEFD.

By extending the model to allow the particle to escape from the harmonic
transient potential and diffuse freely \cite{hachiya2019unveiling}, the
system can be described by two distinct states: a trapped state (in which the
particle remains confined within the transient potential) and a free state
(in which the particle undergoes unrestricted diffusion). Modelling the
transitions between these states as a stochastic process naturally leads to
BYNGD, similar to the behaviour observed in the LEFD. Thus, combining the
Langevin equation with a transient harmonic potential and stochastic jump
dynamics yields a framework that converges to the LEFD description on sufficiently long timescales as a coarse-graining description. 
From this perspective, the LEFD emerges as a coarse-grained
dynamic equation that effectively captures the interplay between transient
trapping effects and free diffusion. In this coarse-graining pathway, the
fluctuating diffusivity observed in the LEFD arises as a consequence of the
stochastic evolution of the transient potential, rather than an intrinsic
variation in particle's properties.


The second approach to deriving the LEFD begins from the GLE and systematically applies a coarse-graining procedure. As in the first approach, auxiliary degrees of freedom are introduced, and the projection operator method is employed to eliminate fast variables from the microscopic dynamics.
In this framework, the position and momentum of a
tagged particle, $\bm{r}(t)$ and $\bm{p}(t)$, serve as as the primary coarse-grained
variables. Additionally, we introduce the force acting on the tagged particle,
defined as $\bm{f}(t)=-\partial U(\bm{r}(t),t)/\partial\bm{r}(t)$, where
$U(\bm{r}(t),t)$ is a time-dependent microscopic interaction potential reflecting fast environmental degrees of freedom.
 To further
incorporate fluctuations into the interaction potential, we introduce an
auxiliary variable describing the coupling between the particle's momentum
and the local  curvature of the potential. This is defined as
\begin{equation}
\bm{a}(t)=\bm{C}(t)\cdot\bm{p}(t),
\end{equation}
where $\bm{C}(t)$ is the curvature matrix of the microscopic potential, given
by $\bm{C}(t)=\partial^2U(\bm{r}(t),t)/\partial\bm{r}(t)\partial\bm{r}(t)$.
This matrix captures the second-order variations in the potential experienced by the tagged particle due to fast environmental degrees of freedom.
We stress that this form is not assumed arbitrarily but arises naturally when constructing a reduced description via the Mori-type projection operator formalism~\cite{Mori-1965}. In this framework, the fluctuation in curvature---interpreted as a slowly evolving environmental variable---modulates the effective interaction forces experienced by the particle. As a result, the variable $\bm{a}(t)$ encodes the geometric influence of the environment on the particle's dynamics, rather than implying a direct or unphysical coupling between momentum and potential curvature.
 Applying the Mori-type projection operator \cite{Mori-1965}
leads to a GLE for the set of four coupled variables $\bm{r}(t)$ (position),
$\bm{p}(t)$ (momentum), $\bm{f}(t)$ (force), and $\bm{a}(t)$
(curvature-momentum interaction). 
An important feature of this derivation is that the resulting GLE is linear in these coarse-grained variables, which facilitates analytical treatment and preserves physical consistency. This formulation ultimately leads to a fluctuating diffusivity that emerges from systematic coarse-graining of the underlying many-body dynamics, rather than being imposed ad hoc.


Building on this framework, the linearity of the GLE enables further simplification by systematically eliminating additional intermediate variables. 
To express the dynamics more
concisely, we employ the curvature matrix $\bm{C}(t)$ directly. Assuming that
$\bm{C}(t)$ is approximately invertible, we eliminate $\bm{f}(t)$ and $\bm{p}
(t)$, leading to an effective dynamic equation for the tagged particle
position $\bm{r}(t)$ and the fluctuating curvature matrix $\bm{C}(t)$.
Applying the Markov approximation and neglecting memory effects, we arrive at
the coarse-grained Langevin equation: 
\begin{equation}
\label{eq: Langevin equation with curvature matrix}
\frac{d\bm{r}(t)}{dt}=\sqrt{2}\bm{B}(t)\cdot\bm{\xi}(t),
\end{equation}
where the noise coefficient matrix satisfies
\begin{equation}
\label{eq: noise coefficient matrix and curvature matrix}
\bm{B}(t)\cdot\bm{B}^{\mathrm{T}}(t)=D_0\bar{C}\bm{C}^{-1}(t).
\end{equation}
Here, $D_0$ represents the mean diffusion coefficient; and $\bar{C}$ is the
ensemble-averaged curvature matrix, satisfying $\langle\bm{C}\rangle=\bar{C}
\bm{1}$. Since $\bm{C}(t)$ follows a stochastic process, which is generally
non-Markovian, its fluctuations directly influence the tagged particle's
diffusivity. 
As a result, the LEFD form emerges naturally in this coarse-grained description. The fluctuating diffusivity arises from the time evolution of the local curvature $\bm{C}(t)$, linking it directly to microscopic structural variations. This perspective aligns with the earlier derivation based on transient potentials, both highlighting how curvature-related fluctuations generate stochastic diffusivity at coarse-grained scales.

\begin{figure*}[htb]
 \begin{center}
  \includegraphics[width=0.4\linewidth]{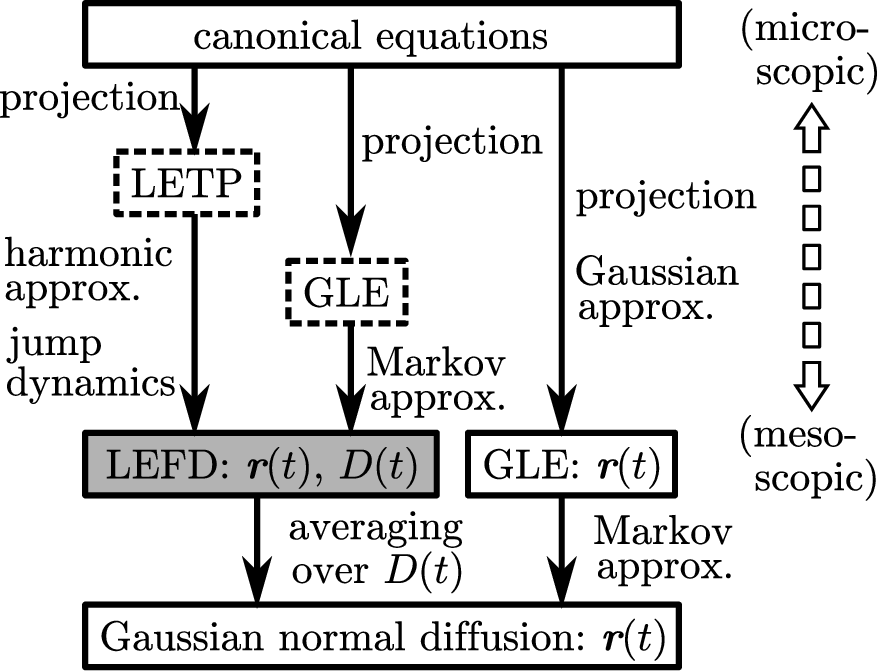}
 \end{center}
\caption{Coarse-graining pathways for deriving the LEFD from the canonical
equations. Boxes represent dynamical equation models, while arrows represent
coarse-graining pathways. The LEFD can be obtained through two primary
approaches: (i) applying a harmonic approximation and introducing jump
dynamics in the LETP, and (ii) employing a Markov approximation for the GLE.
In the conventional coarse-graining framework, the canonical equations are
reduced to the GLE with Gaussian noise, whereas the LEFD emerges when
incorporating a fluctuating diffusivity as an additional stochastic variable. 
If we perform further coarse-graining of the LEFD (or of the GLE), we obtain an effective standard
    Langevin equation with constant diffusivity, which describes Gaussian normal diffusion.
 \label{fig: LEFD derivation}}
\end{figure*}

Figure~\ref{fig: LEFD derivation} summarises the coarse-graining pathways
discussed in this subsection, alongside the conventional coarse-graining
approach that leads to the GLE. In the standard pathway, the GLE serves as
the effective dynamic equation for the tagged particle's position, in which
the memory kernel encapsulates the microscopic dynamics while the diffusivity
remains constant. In contrast, in our proposed coarse-graining pathways, the
tagged particle follows the LEFD, in which the fluctuating diffusivity---rather
than the memory kernel---encodes the influence of the microscopic dynamics. It is important to
recognise that these derivations are not universally applicable and are
constrained to specific systems. For instance, in gaseous systems, the
transient potential acting on a tagged particle is nearly negligible, making the
inversion of the potential curvature infeasible. Consequently, the LEFD for
certain gas systems, as discussed in Section~\ref{sec: binary gas mixture and
lorenz gas}, must be derived through an alternative route. 
Moreover, other yet-to-be-explored coarse-graining pathways may exist for deriving the LEFD in different types of complex systems.


{\color{black}
\subsection{Distinct Mechanisms Leading to BYNGD}

While the LEFD framework captures many experimental features of BYNGD, it is not the only theoretical approach. Several distinct mechanisms have been proposed to explain the emergence of non-Gaussian displacement statistics with linear MSD.

One such framework attributes non-Gaussianity to} the dominance of rare, large displacement events, rather than to fluctuating diffusivity. In this picture, inspired by Laplace's first law of errors, the displacement PDF is assumed to follow a double-exponential (Laplace) distribution \cite{Barkai2020,wang2020large,hamdi2024laplace,Singh2023}.  Unlike LEFD, which links non-Gaussianity to time-dependent or random diffusivity processes, this approach posits that the observed heavy tails arise from statistically rare but physically plausible extreme events within otherwise standard dynamics. This suggests a fundamentally different physical mechanism behind BYNGD in some systems.

{\color{black}Another class of models arises from geometrically induced constraints, such as transient confinement within randomly distributed barriers. These are often encountered in compartmentalized or membrane-like media and can lead to BYNGD characterized by linear MSD and exponential-tailed displacement PDFs at intermediate times} \cite{slkezak2021diffusion}. This behaviour emerges in the intermediate-time regime and can be effectively described by a geometry-induced CTRW, providing a structural---rather than thermal or stochastic---origin for non-Gaussian transport.

{\color{black}
In quenched disordered systems, BYNGD can emerge due to static spatial heterogeneity in local diffusivity. Particles traversing regions of varying mobility sample different effective environments over time, leading to non-Gaussian PDFs at intermediate times. These typically display narrow central peaks due to averaging over diffusivity landscapes \cite{Pacheco-Pozo2021}.

Shot-noise-based models offer yet another explanation \cite{Thiffeault2015}. Here, particle motion is driven by intermittent, impulsive forces \cite{Thiffeault2015}. These models attribute non-Gaussianity to intermittent, impulsive forces or ``kicks," leading to jump-like particle motion. Although conceptually distinct, shot-noise-based models can, under certain conditions, yield statistical behaviour similar to LEFD. 
It has been shown that a fluctuating-diffusivity Langevin model can reproduce shot-noise-like (or CTRW-like) behaviour in the limit of rare, intermittent diffusivity fluctuations  \cite{Uneyama2015}, providing a concrete example of how impulsive transport processes can emerge as a special case within the LEFD framework.
In the overdamped limit, the shot-noise model and the CTRW are formally equivalent descriptions of intermittent transport. Within the LEFD picture, this correspondence can be captured when diffusivity fluctuations produce effectively Gaussian displacements during each dynamical episode. However, the standard two-state LEFD cannot represent completely arbitrary non-Gaussian steps, since the displacement statistics within each state remain Gaussian by construction. Introducing additional diffusivity states could, in principle, approximate a wider range of intermittent behaviours, though such extensions would be primarily phenomenological.
Thus, while the physical origins of LEFD and shot-noise dynamics differ, their statistical behaviour may overlap in certain regimes, particularly when diffusivity fluctuations occur as rare, intermittent bursts.
}

{\color{black}
A particularly illustrative example is found in single-particle tracking experiments involving micron-sized tracers in dilute suspensions of swimming Chlamydomonas algae \cite{Leptos2009}. In the absence of swimmers, the tracer displacements are Gaussian, consistent with thermal Brownian motion, whereas in the presence of swimmers the PDFs develop a Gaussian core with pronounced exponential tails that grow with swimmer concentration. Both the Gaussian width and the exponential decay length scale as $\propto \sqrt{\Delta t}$, indicating linear growth of the MSD despite strongly non-Gaussian statistics. Similar behaviour occurs in quasi-two-dimensional E. coli baths, where tracer trajectories exhibit intermittent ballistic segments embedded in otherwise diffusive motion \cite{Wu2000}. Such dynamics can be captured by a two-state model in which the tracer alternates between (i) rapid advection when entrained in the flow field of a nearby swimmer and (ii) slower Brownian motion when far from swimmers. These alternating modes generate temporal variability in mobility, representing a hydrodynamic realisation of local diffusivity fluctuations at a coarse-grained level.

Tracer diffusion in active suspensions has also been analysed from a non-LEFD perspective. In such systems, passive tracers are advected by the long-range velocity fields generated by swimming microorganisms. Curved or randomly reorienting swimmer trajectories continually deform these flow fields, breaking the symmetry of tracer loops and producing stochastic tracer displacements that accumulate diffusively \cite{Pushkin2013}.  Such tracer diffusion is effectively described by shot-noise dynamics that can emerge from sparse two-body interactions \cite{kanazawa2020loopy, baule2023universal}. Together, these works establish that BYNGD can arise from intermittent active forcing and trajectory curvature, without invoking a fluctuating-diffusivity (LEFD) mechanism.

It is important to distinguish fluctuating-diffusivity dynamics from run-and-tumble motion, which involves alternation between fundamentally different modes of transport---ballistic and diffusive.  
In run-and-tumble dynamics, exemplified by many self-propelled microorganisms such as E. coli, the organism undergoes persistent ballistic runs punctuated by tumbles that randomise its orientation. Although this switching produces intermittent fast and slow phases, it reflects a change in transport mode rather than fluctuations of a diffusivity parameter $D(t)$ \cite{Datta2024}. In cases where run durations follow a broad distribution, the motion exhibits L\'evy-walk-like statistics. Thus, although run-and-tumble and LEFD both generate non-Gaussian displacement statistics, run-and-tumble is better regarded as a coexistence of distinct transport modes, and its description would require either a modified LEFD framework or a separate theoretical treatment.}

Furthermore, a recent addition to this family of models is the telegraphic multifractional Brownian motion, introduced in Ref.~\cite{Balcerek2025}, which describes diffusion with a stochastically varying Hurst exponent. It bridges classical multifractional Brownian motion and stochastic switching processes, offering a framework where the heterogeneity is encoded directly in the time-varying scaling behaviour.


\section{Diffusing diffusivity model}

We investigate how the short-time non-Gaussian propagator arises from a fluctuating
instantaneous diffusivity. Specifically, we analyse the types of fluctuating
diffusivity dynamics that result in an exponential form of the displacement PDF.
Two stochastic models have been identified to exhibit this behaviour. The
first is a discrete random walk model proposed by Chubynsky and Slater
\cite{Chubynsky2014}. The second is the LEFD, in which the diffusivity evolves according to an
Ornstein-Uhlenbeck process \cite{Chechkin2017}. 
In this LEFD framework, the propagator exhibits a Laplace (exponential) core, accompanied by a power-law correction. The exponent of this correction is determined by the dimension of the auxiliary variable used to model the diffusivity, rather than the spatial dimension of the particle's motion.
Both models employ the superstatistical framework to explain the emergence of the non-Gaussian
propagator  \cite{beck2003superstatistics}.

\subsection{Diffusing diffusivity random walk model}

The diffusing diffusivity random walk model provides a framework for 
understanding of  BYNGD, a phenomenon in which
the MSD grows linearly in time (normal diffusion), but the propagator
exhibits non-Gaussian characteristics, such as exponential tails, at short
times \cite{Chubynsky2014}. This behaviour has been observed in crowded
environments, including lipid bilayers, entangled polymer solutions,
lipid membranes, mucus systems, and the cytoplasm of living cells \cite{wang2009anomalous,
wang2012brownian,Bhattacharya2013,Jeon2016,stefanie,lampo2017cytoplasmic}.
The  model is described by an unbiased one-dimensional random walk in which
the displacement $\Delta x_i$ at each step $i$ is drawn from the Gaussian
PDF \cite{Chubynsky2014}
\begin{equation}
P(\Delta x_i|D_i)=\frac{1}{\sqrt{4\pi D_i}}\exp\left(-\frac{\Delta x_i^2}{
4D_i}\right),
\end{equation}
where $D_i$ is the instantaneous diffusivity at step $i$. The instantaneous
diffusivity fluctuates due to an environmental heterogeneity or variations in
the particle shape/size. These fluctuations are modelled as a stochastic
process. Specifically, the diffusivity $D_i$ itself evolves over time through
a biased or unbiased random walk. Assuming that the diffusivity changes only slightly between steps allows 
 the process to be treated in
continuous time $t$. The PDF of the diffusivity, $\pi(D,t)$, evolves according
to the advection-diffusion equation
\begin{equation}
\frac{\partial\pi(D,t)}{\partial t}=-\frac{\partial J}{\partial D},
\end{equation}
where the flux $J$ is defined as
\begin{equation}
J=-d_D\frac{\partial}{\partial D}\pi(D,t)-s\pi(D,t)
\end{equation}
and $d_D$ represents the ``diffusivity of the diffusivity," that is, the diffusion coefficient governing the evolution of  $D$, 
while $-s \pi(D,t)$ is a drift term that drives 
$D$ toward $D=0$. Since $D$ cannot be negative or exceed the free-space
diffusivity $D_{\text{max}}$, reflecting boundary conditions are imposed: $J=0$
at $D=0$ and $D=D_{\text{max}}$. Over long times, the system reaches a stationary
state, in which the diffusivity distribution $\pi(D)$ satisfies $J(D)=0$ for
all $D$.

In the stationary state, the initial distribution of $D$ follows the stationary
distribution, and the PDF of $D$ remains unchanged over time, i.e., $\pi(D,t)=
\pi(D)$. Consequently, the MSD increases linearly with time,
\begin{equation}
\langle x_n^2\rangle=\sum_{i=1}^n\langle\Delta x_i^2\rangle=2\langle D\rangle_{
\rm st}n,
\end{equation}
indicating normal diffusion---even though the instantaneous diffusivity continues to fluctuate. 
In contrast, the fourth moment of the displacement
deviates from its Gaussian value. This deviation is given by
\begin{align}
\langle x_n^4 \rangle - 3 \langle x_n^2 \rangle^2 
= {} & \, 12\left( \langle D^2 \rangle_{\mathrm{st}} 
- \langle D \rangle_{\mathrm{st}}^2 \right) n \notag \\
& + 24 \sum_{i=1}^{n-1} \sum_{j=i+1}^{n} 
\left( \langle D_i D_j \rangle_{\mathrm{st}} 
- \langle D_i \rangle_{\mathrm{st}} 
  \langle D_j \rangle_{\mathrm{st}} \right).
\end{align}
The second term depends on the correlation function of $D$ which itself is a
function of $|i-j|$. Assuming that the relaxation time $\tau_D$ of the
correlation is much longer than the step interval ($\tau_D\gg1$), $D$ changes slowly
over time. Furthermore, we assume that $\langle D^2\rangle_{\rm st}-\langle D
\rangle_{\rm st}^2\sim\langle D\rangle_{\rm st}^2$, which is characteristic of an exponential distribution of diffusivities, where the variance equals the square of the mean. 
Under these assumptions, the NGP can be expressed as
\begin{equation}
A(n)\sim\left\{\begin{array}{ll}1&(n\ll\tau_D),\\\\
\dfrac{\tau_D}{n}&(n\gg\tau_D).\end{array}\right.
\end{equation}
This result demonstrates that significant deviations from Gaussianity occur
when $n\ll\tau_D$, even though the diffusion remains normal (i.e., $\langle
x_n^2\rangle\propto n$). At longer times, $n\gg\tau_D$, the model predicts a
crossover to Gaussian behaviour due to the central limit theorem, with
diffusivity fluctuations averaging out over time.

For $n\ll\tau_D$, the propagator can be expressed using a superstatistical
approach in the form
\begin{equation}
P(x,n)=\int_0^{D_{\max}}\frac{\pi(D)}{\sqrt{4\pi D}}\exp\left(-\frac{\Delta
x_i^2}{4Dn}\right) dD,
\end{equation}
where $\tau_D$ is the characteristic timescale of $D$. For $D_{\max}\to\infty$,
the stationary distribution $\pi(D)$ is given by
\begin{equation}
\pi(D)=\frac{1}{D_0}\exp(-D/D_0),
\end{equation}
where $D_0=d/s$. Substituting this distribution into the PDF yields, for $n\ll
\tau_D$, the Laplace distribution
\begin{equation}
P(x,n)=\frac{1}{\sqrt{4D_0n}}\exp\left(-\frac{|x|}{\sqrt{D_0n}}\right), 
\end{equation}
which highlights the emergence of exponential tails in the PDF due to the
fluctuating diffusivity. Notably, the Laplace distribution persists even when
$D_{\max}$ is finite, provided that $|x|$ is not large. This demonstrates that
a fluctuating diffusivity induces the non-Gaussian behaviour characterised by
exponential tails, even in the presence of a bounded diffusivity. The diffusing
diffusivity model provides a theoretical foundation for the understanding of
experimental observations of non-Gaussian displacement PDFs such as the Laplace
PDF, as seen in single-particle tracking experiments or simulations in lipid
bilayers and crowded polymer solutions.

\subsection{Diffusing diffusivity LEFD model}

The diffusing diffusivity model, as introduced by Chubynsky and Slater
\cite{Chubynsky2014} and extended by Chechkin et al. \cite{Chechkin2017},
provides a minimal yet powerful framework for describing BYNGD. By coupling stochastic processes for particle
position and diffusivity, the model captures key non-Gaussian features,
such as exponential tails in the propagator and their eventual crossover
to Gaussian behaviour at longer times.

The extension proposed by Chechkin et al. reformulates the diffusing
diffusivity model as the LEFD \cite{Chechkin2017}
\begin{equation}
\frac{d{\bm r}(t)}{dt}=\sqrt{2D(t)}{\bm\xi}(t),
\label{eq: DD LEFD}
\end{equation}
where the time-dependent coefficient $D(t)$ represents the fluctuating
diffusivity, governed by $D(t)={\bm Y}^2(t)$, and  the auxiliary variable ${\bm Y}(t)$ evolves
according to an Ornstein-Uhlenbeck process:
\begin{equation}
\frac{d{\bm Y}(t)}{dt}=-\frac{1}{\tau}{\bm Y}(t)+{\bm\eta}(t),
\label{eq: DDFD OU}
\end{equation}
where ${\bm\eta}(t)$ is a $n$-dimensional Gaussian white noise independent of ${\bm\xi}(t)$. 
This squaring of the stationary process ${\bm Y}(t)$ ensures the positivity of
$D(t)$ without requiring reflecting boundary conditions, in contrast to the original random walk model 
 \cite{Chubynsky2014}. In this formulation, ${\bm\xi}(t)$ and ${\bm\eta}(t)$ are independent white Gaussian
noise terms, with ${\bm \xi}(t)$ acting in $d$-dimensional space, and ${\bm
\eta}(t)$ in $n$-dimensional space.\footnote{The choice of $n$ is a priori
free to choose, but in the sense of Occam's razor $n=1$ or $n=d$ appear
reasonable. The values of $d$ and $n$ determine power-law corrections to the
Laplace PDF at short times \cite{Chechkin2017}.} The parameter $\tau$
represents the characteristic timescale over which ${\bm Y}(t)$ relaxes, i.e.,
it is the correlation time of the auxiliary process ${\bm Y(t)}$. Since ${\bm
Y}(t)$ is governed by the Ornstein-Uhlenbeck process, its stationary PDF $f_{
\rm st}({\bm Y})$ follows the Boltzmann distribution,
\begin{equation}
 f_{\rm st}({\bm Y})= \left(\frac{1}{\tau \pi}\right)^{n/2} e^{- \frac{{\bm Y}^2}{\tau}}. 
\label{eq: st Y Botzmann}
\end{equation}
The stationary PDF $P_{\rm st}(D)$ of the fluctuating diffusivity $D(t)$ then
follows in the form
\begin{equation}
P_{\rm st}(D)=\left\{\begin{array}{ll}\dfrac{1}{\sqrt{\pi\tau D}}e^{-D/\tau}
\quad&(n=1),\\\\
\dfrac{1}{\tau} e^{-D/\tau}&(n=2),\\\\
\dfrac{2\sqrt{D}}{\sqrt{\pi\tau^3}}e^{-D/\tau}&(n=3).\end{array}\right.
\label{eq: st FD Botzmann}
\end{equation}
For large $D$, the stationary PDF $P_{\rm st}(D)$ exhibits exponential tails,
which underpin the emergence of the non-Gaussian, Laplace-like behaviour in
the propagator at short times.

Since the model is described by the LEFD framework, the displacement moments
can be calculated as outlined in Section~\ref{sec: Fundamental properties of LEFD}.
From Eq.~(\ref{eq: normal diffusion in LEFD}), it follows that the MSD exhibits
normal diffusion, $\langle\{{\bm r}(t)-{\bm r}(0)\}^2\rangle_{\rm st}=2d\langle
D\rangle_{\rm st}t$, which holds for all times $t$. Here, the stationary average
$\langle D\rangle_{\rm st}$ is determined by the stationary distribution of $D$,
i.e., Eq.~(\ref{eq: st FD Botzmann}). Notably, this stationary average depends
on the dimension $n$ of the auxiliary process ${\bm Y}(t)$, as reflected in the
power-law prefactor of the exponential distribution.

\subsection{Short-time Non-Gaussian behaviours}

To characterise deviations from the Gaussian distribution in the propagator,
we use the NGP $A(t)$ defined in Eq.~(\ref{eq: NGP def}). In the short-time
limit, the NGP is given by Eq.~(\ref{eq: NGP short-time}): $A(t)\sim{\rm Var}
_{\rm st}(D)/\langle D\rangle_{\rm st}^2$. Substituting the stationary variance
and the mean of $D$ for different dimensions $n$, the short-time NGP becomes 
\begin{equation}
A(t)\sim\left\{\begin{array}{ll}2 &(n=1),\\\\
1&(n=2),\\\\\dfrac{2}{3}&(n=3).\end{array}\right.
\label{eq: short-time NGP DD model}
\end{equation}
This result, shown in Eq.~(\ref{eq: short-time NGP DD model}), demonstrates
that the propagator deviates from a Gaussian PDF. In particular, a positive
NGP indicates a leptokurtic shape, i.e., one with heavier tails than a Gaussian.

In the short-time limit ($t\ll \tau$), the explicit form of the PDF can be
derived as shown in \cite{Chechkin2017}. For the case $d=n=1$, it is given
by
\begin{equation}
P(x,t)\sim\frac{1}{\sqrt{2\pi|x|t^{1/2}}}\exp\left(-\frac{|x|}{t^{1/2}}\right).
\label{eq: prop DD model short-time}
\end{equation}
For the case $d=n=2$, the PDF takes on the form
\begin{equation}
P({\bm r},t)\sim\frac{1}{2\sqrt{2\pi|{\bm r}|t^{3/2}}}\exp\left(-\frac{|{\bm
r}|}{t^{1/2}}\right),
\end{equation}
and for $d=n=3$, it becomes
\begin{equation}
P({\bm r},t)\sim\frac{1}{(2\pi)^{3/2}|{\bm r}|^{1/2}t^{5/4}}\exp\left(-\frac{
|{\bm r}|}{t^{1/2}}\right).
\end{equation}
These expressions capture the exponential tail observed in experiments of
BYNGD.  Moreover, the scaling $t^{1/2}$, required to maintain a linear MSD, has been explicitly confirmed in experimental data \cite{wang2009anomalous}.



\subsection{Long-time Gaussian behaviours}

In the long-time limit, the NGP decreases as $A(t)\propto1/t$, indicating that
the PDF converges to the Gaussian form
\begin{equation}
P({\bm r},t)\sim\frac{1}{(4\pi\langle D\rangle_{\rm st}t)^{d/2}}\exp\left(-
\frac{|{\bm r}|^2}{4\langle D\rangle_{\rm st}t}\right),
\label{eq: prop DD model long-time}
\end{equation}
as  diffusivity fluctuations average out over time. The width
of the Gaussian PDF is dominated by the effective diffusivity $\langle D\rangle
_{\mathrm{st}}$. This model provides a unified framework for BYNGD, bridging
the gap between theoretical models and experimental observations, explaining
key features such as short-time non-Gaussian tails and long-time Gaussian
behaviour. It also highlights the role of the dimensionality of the auxiliary
process in shaping the subleading behaviour of the diffusivity distributions.


\section{Two-state model}
\label{sec: two-state model}

In complex systems such as supercooled liquids and intracellular environments,
the instantaneous diffusivity $D(t)$ varies over time due to environmental
heterogeneity and dynamic interactions. To model fluctuating diffusive
states, a dichotomous process is often employed \cite{Richert-2002,yamamoto2015anomalous,Hidalgo-Soria2020,hidalgo2021cusp,Kimura2022,kimura2023non}, where  $D(t)$  switches between two discrete states, 
 $D_+$ and $D_-$. Despite its simplicity,
this model provides exact theoretical insights into both the displacement PDF
and global diffusivity fluctuations, making it a valuable framework for
studying heterogeneous diffusion processes.


A dichotomous process  for the diffusivity can be described by an alternating
renewal process \cite{Cox,Akimoto2023}, where the system switches between two
diffusive states with random sojourn times. In this framework, the sojourn
times spent in each state ($D_+$ or $D_-$) are independent and identically distributed (IID) random variables drawn from 
prescribed PDFs. Let $\rho_+(\tau)$ and $\rho_-(\tau)$ represent the PDFs of the
sojourn times in the $D_+$ and $D_-$ states, respectively. Both Markovian and
non-Markovian cases of $D(t)$  can be formulated within this framework. In the
non-Markovian case, the sojourn-time PDFs typically exhibit power-law PDFs,
leading to long-tailed waiting times. Specifically, $\rho_\pm(\tau)\sim a_\pm\tau^{
-1-\alpha_\pm}/|\Gamma(-\alpha_\pm)|$ for $\tau\to\infty$, where $a_{\pm}$ are positive scale factors that determine the overall amplitude of the power-law tails in the sojourn-time distributions for the fast (+) and slow (-) states, respectively.
 A key mathematical
tool for analysing dichotomous diffusion processes is the Laplace transform of
the sojourn-time PDFs. The Laplace transform gives a compact mathematical representation of how a process behaves over different timescales, and its small-$s$  behaviour is especially useful for understanding long-time dynamics. The asymptotic
form of the Laplace transform for power-law sojourn times, in the limit $s\to0$
and for $\alpha_\pm<1$, is given by\footnote{The Laplace transform of the PDF
$\rho_{\pm}(\tau)$ is defined as
\[
\hat{\rho}_{\pm}(s)=\int_0^{\infty}\rho_{\pm}(t)e^{-st}dt.
\]
}
\begin{equation}
\hat{\rho}_\pm(s)=1-a_\pm s^{\alpha_\pm}+O(s).
\end{equation}
This expression highlights the dominant power-law behaviour at long times,
which is crucial for determining the memory effects and non-Markovian dynamics
in systems with fluctuating diffusivity.

We assume that the first sojourn time distributions in the $D_+$ and $D_-$
states follow the same PDFs as $\rho_\pm(\tau)$. In renewal theory, this
setup corresponds to an ordinary renewal process \cite{Cox}. In fact, different 
types of renewal processes are distinguished by their initial conditions. 
 In ordinary renewal process, the system starts at $t=0$  with a
typical non-equilibrium initial condition, where the first sojourn-time PDF
 is identical to $\rho_\pm(\tau)$. In contrast, an equilibrium renewal process assumes 
the system has been evolving for a long time before the
measurement starts,  resulting in a different PDF for the first sojourn time \cite{Cox}. Throughout this
study, we primarily consider the ordinary renewal process as the initial
condition, i.e., the first sojourn-time PDF is identical to $\rho_\pm(\tau)$.
The impact of the initial ensemble selection on the statistical properties of
the dichotomous diffusivity process was extensively discussed in
Ref.~\cite{Miyaguchi2016}. These effects play an important role in determining
the long-time behaviour of diffusion processes in heterogeneous environments.

\subsection{ Mean squared displacement}
\label{subsec: two-state model: MSD}

As established previously, the MSD for a system with time-dependent diffusivity
follows $\langle |{\bm r}(t)-{\bm r}(0)|^2\rangle=2d\langle D(t)\rangle t$.
When $D(t)$ is a stationary stochastic process, the MSD exhibits normal
diffusion, following $\langle |{\bm r}(t)-{\bm r}(0) |^2\rangle=2d\langle D
\rangle_{\rm st}t$, where $\langle D\rangle_{\rm st}$ represents the stationary
average of the diffusivity---it can be derived from the probability $p_+(t)$ of
finding $D_+$ state at time $t$. Using renewal theory \cite{Cox,Miyaguchi2013,
Akimoto2023}, this probability converges to its long-time equilibrium value
\begin{equation}
p_\pm(t)\to p_\pm^{\rm eq}\equiv\frac{\mu_\pm}{\mu_++\mu_-},
\end{equation}
where $\mu_\pm$ is the mean sojourn times for the states $D_\pm$, respectively.
Consequently, the ensemble-averaged global diffusivity $\langle D\rangle_{\rm
st}$ is given by
\begin{equation}
\langle D\rangle_{\rm st}=\frac{D_+\mu_++D_-\mu_-}{\mu_++\mu_-}.
\end{equation}
This expression provides a fundamental relation between the global diffusivity
and the statistical properties of the fluctuating diffusive states, ensuring a
quantitative characterisation of the diffusive dynamics in systems with a
dichotomous diffusivity.

When $D(t)$ is not a stationary process---that is, when at least one of the mean sojourn times 
diverges---the ensemble-averaged diffusivity becomes explicitly
time-dependent,
\begin{equation}
\langle D(t)\rangle=D_-+(D_+-D_-)p_+(t).
\end{equation}
As a consequence, the MSD no longer exhibits linear time dependence.
Instead, it follows the sublinear growth
\begin{equation}
\label{e.MSD.two-state-lefd.subdiffusion}
\langle |{\bm r}(t)-{\bm r}(0) |^2\rangle\propto t^\alpha.
\end{equation}
This subdiffusive behaviour arises when the mean sojourn time in the $D_+$ state
is finite, while the system spends increasingly long periods in the $D_-$ state.
 A prominent example is the ``freezing" scenario, in which  $D_- = 0$, and the sojourn-time PDF in this state follows a power law, 
$\rho_-(\tau)\propto\tau^{-1-\alpha}$ for $\tau\to\infty$ \cite{Miyaguchi2013,
Akimoto2023}. These conditions lead to a progressive slowdown in the diffusion,
akin to subdiffusion observed in quenched trap models and subdiffusive CTRWs
\cite{bouchaud90,metzler00}.

The CTRW framework describes a random walker that undergoes waiting times
$\tau$ between successive jumps \cite{bouchaud90,metzler00}, whose PDF follows
a power-law, $\rho_-(\tau)\propto\tau^{-1-\alpha}$ for $\tau\to\infty$. This
leads to subdiffusion characterised by the anomalous diffusion exponent $\alpha$, similar
to the two-state diffusivity model with $D_-=0$. However, an important
distinction exists between the two models: in the CTRW model, 
the walker remains immobilised during the waiting period and then performs an instantaneous jump. In contrast, in the two-state diffusivity model, the particle undergoes continuous diffusion with coefficient  $D_+$  during the sojourns in the  $D_+$  state, making the displacement gradual rather than instantaneous.
This fundamental difference implies that the two-state model with $D_-=0$
can be effectively approximated by a CTRW model with non-instantaneous jumps
\cite{Kimura2022}. This approximation offers a more physically realistic
description of subdiffusive processes in heterogeneous environments, akin
to the noisy CTRW framework \cite{jeon2013noisy}, in which the trajectory between jumps is influenced by underlying noise.

\begin{widetext}
\subsection{ Brownian yet non-Gaussian diffusion}

The displacement PDF for the two-state diffusivity model can be derived in a
manner analogous to the CTRW framework, following the Montroll-Weiss equation
\cite{montroll1965random,Shlesinger1982,Miyaguchi2013}. The propagator $P_\pm
({\bm r},t)d{\bm r}$, which represents the probability of finding the particle in the
$D_\pm$ state at position $({\bm r},{\bm r}+d{\bm r})$ at time $t$, satisfies
the following equations:
\begin{equation}
P_\pm({\bm r},t)=G_d({\bm r},t;D_{\pm})I_\pm(t)p_\pm(0)+\int_0^tdt'\int d{\bm
r}'G_d({\bm r}-{\bm r}',t-t';D_{\pm})R_\pm({\bm r}',t')I_\pm (t-t'),
\end{equation}
with
\begin{equation}
R_\pm({\bm r},t)=G_d({\bm r},t;D_{\mp})\rho_\mp(t)p_\mp(0)+\int_0^tdt'\int d{\bm
r}'G_d({\bm r}-{\bm r}',t-t';D_{\mp})R_\mp({\bm r}',t')\rho_\mp(t-t'),
\end{equation}
where $I_\pm(t)=\int_t^{\infty}d\tau\rho_\pm(\tau)$ is the survival probability,
describing the probability that the particle remains in the $D_\pm$ state until
time  $t$, and $R_\pm({\bm r},t)d{\bm r}$ represents the probability of finding
the particle at position $({\bm r},{\bm r}+d{\bm r})$ immediately after a
transition between diffusivity states. Applying the Fourier and Laplace
transforms, we obtain the transformed PDF
\begin{equation}
\tilde{\hat{P}}_\pm({\bm k},s)=\frac{1-\hat{\rho}_\pm(s_\pm)}{s_\pm}p_\pm(0)+
\frac{1-\hat{\rho}_\pm(s_\pm)}{s_\pm}\frac{p_\pm(0)\hat{\rho}_\pm (s_\pm)\hat{
\rho}_\mp(s_\mp)+\hat{\rho}_\mp(s_\mp)p_\mp(0)}{1-\hat{\rho}_+(s_+)\hat{\rho}_
-(s_-)},
\end{equation}
where $s_\pm=s+D_\pm{\bm k}^2$. This Laplace-Fourier representation allows for
a direct analysis of the asymptotic behaviour of the PDF, particularly in
long-time and small-wavenumber limits, revealing how subdiffusion and
non-Gaussianity emerge in two-state diffusivity models.
\end{widetext}
 
 \subsubsection{Non-Gaussian propagator in a short-time limit} 

We first analyse the asymptotic behaviour of the PDF in the short-time limit,
$t\ll\tau_D$. Then, the diffusivity remains approximately constant, meaning that
$D(t)$ takes either the value $D_+$ or $D_-$ over the duration $t$. The
characteristic timescale $\tau_D$ of $D(t)$ is defined here as $\tau_D=\min
(\mu_+,\mu_-)$. If both mean sojourn times diverge, no well-defined
characteristic time exists. However, the following discussion remains valid
for any finite observation time $t$. For $t\ll\tau_D$, the PDF of ${\bm r}(t)$ becomes a
superposition of two Gaussian PDFs corresponding to the diffusion coefficients
$D_+$ and $D_-$:
\begin{equation}
P_\pm({\bm r},t)=p_+(0)G_d({\bm r},t;D_+)+p_-(0)G_d({\bm r},t;D_-).
\end{equation}
This expression clearly demonstrates a non-Gaussian propagator, even though
the MSD remains normal---a hallmark of BYNGD. When the sojourn-time PDF
follows a power law, the mean sojourn time diverges for $\alpha_\pm<1$. If
both mean sojourn times are divergent, the deviation from Gaussianity becomes
particularly pronounced even at finite times \cite{Miyaguchi2016}. This result
highlights that non-Gaussian PDFs are a generic feature of the LEFD in a
two-state model.

 \subsubsection{Gaussian and non-Gaussian propagators in a long-time limit} 

We now investigate the asymptotic behaviour of the PDF in the long-time limit.
In Fourier-Laplace space, the probability of finding the particle in the
$D_\pm$ state at position $({\bm r},{\bm r}+d{\bm r})$ at time $t$, denoted $P_\pm
({\bm r},t)d{\bm r}$, 
can be obtained in the hydrodynamic limit, where $s_\pm\ll1$ and $s\sim{\bm
k}^2\ll1$, as
\begin{equation}
\tilde{\hat{P}}_\pm({\bm k},s)\sim\frac{1-\hat{\rho}_\pm(s_\pm)}{1-\hat{\rho}_+
(s_+)\hat{\rho}_-(s_-)}.
\end{equation}
When the mean sojourn times $\mu_\pm$ are finite, the total PDF in
Fourier-Laplace space, given by the sum $\tilde{\hat{P}}({\bm k},s)=\tilde{
\hat{P}}_+({\bm k},s)+\tilde{\hat{P}}_-({\bm k},s)$, takes the asymptotic
form
\begin{equation}
\tilde{\hat{P}}({\bm k},s)\sim\frac{1}{s+\langle D\rangle_{\rm st}
{\bm k}^2}
\label{eq: propagator Laplace two-state long-time}
\end{equation}
for $s_\pm\to 0$ and $s\sim{\bm k}^2\to0$. This result indicates that in the
long-time limit the PDF converges to a Gaussian shape with variance $2d\langle
D\rangle_{\rm st}t$. Thus, despite the initial non-Gaussian characteristics,
the system ultimately exhibits normal diffusion at long times, with an effective
diffusivity given by the stationary average $\langle D\rangle_{\rm st}$.

Even when both mean sojourn times diverge, the PDF still converges to a
Gaussian form \cite{Miyaguchi2016}. However, the effective variance differs from
that of the stationary case. Instead of having the variance $2d\langle D\rangle
_{\rm st}t$, the PDF exhibits a Gaussian shape with variance $2dD_+t$ if $\alpha_
+<\alpha_-$ \cite{Miyaguchi2016}. This result indicates that in the long-time
limit the $D_+$  state dominates the dynamics, effectively making the system
behave as Brownian motion with diffusion coefficient $D_+$. However, when the
power-law exponents are equal ($\alpha=\alpha_+=\alpha_-$), the system cannot
be described solely by either of the two diffusive states. Instead, the TSD
exhibits trajectory-to-trajectory fluctuations, reflecting strong global
diffusivity fluctuations \cite{Akimoto2016}. In this regime, the PDF of the
diffusion coefficient $D$, as obtained from TSD analysis, follows the
generalised arcsine density $g_{\alpha,b}(D)$ \cite{Akimoto2016},
\begin{equation}
P(D)=\frac{1}{D_+-D_-}g_{\alpha,b}\left(\frac{D-D_-}{D_+-D_-}\right),
\end{equation}
where $b=a_-/a_+$ and the generalised arcsine density is defined as
\begin{equation}
\begin{split}
g_{\alpha,b}(x) \equiv {} & \frac{b \sin(\pi\alpha)}{\pi} \\
& \times \frac{x^{\alpha - 1}(1 - x)^{\alpha - 1}}{b^2 x^{2\alpha} 
+ 2b x^{\alpha}(1 - x^{\alpha}) \cos(\pi\alpha) 
+ (1 - x)^{2\alpha}}.
\end{split}
\end{equation}
This PDF was originally discovered in the context of occupation time statistics
in stochastic processes \cite{Lamperti1958}. Notably, its mean value is given
by $1/(1+b)$, which implies that the distribution becomes asymmetric with
respect to $x=1/2$ when $b\neq1$, reflecting an imbalance in the proportion
of time spent in each diffusive state. As a consequence, the PDF of displacements does not
converge to a Gaussian shape but instead follows a superstatistical form:
\begin{equation}
P_\pm({\bm r},t)\sim\int_{D_-}^{D_+}G_d({\bm r},t;D)P(D)dD.
\label{eq: propagator Laplace two-state long-time non-Gauss}
\end{equation}
This result demonstrates that, even in the long-time limit, the PDF  remains
non-Gaussian, highlighting the fundamental role of diffusivity fluctuations
in shaping anomalous diffusion behaviour.

\subsection{ Relative standard deviation of the time-averaged squared displacement}

The RSDs for both Markovian and non-Markovian cases can be systematically
derived. The ensemble average of the TSD for a two-state process with
switching diffusivity $D(t)$ is given by 
\begin{equation}
\left\langle\overline{\delta^2(\Delta;t)} \right\rangle=\frac{2d}{t-\Delta}\int_0^{t-\Delta}
dt'\int_{t'}^{t'+\Delta}d\tau\langle D(\tau)\rangle.
\end{equation}
When $D(t)$ is a stationary process, the ensemble average simplifies to
form
\begin{equation}
\left\langle\overline{\delta^2(\Delta;t)} \right\rangle_{\rm st}=2d\langle D\rangle_{\rm
st}\Delta.
\label{eq: ETSD two-state}
\end{equation}
This equality between the ensemble-averaged TSD and the stationary MSD is a necessary condition for ergodicity. Whether full ergodicity holds---i.e., whether individual time-averaged trajectories converge to this mean---will be examined the following discussion.

\subsubsection{Markovian case}

We first consider the Markovian case, in which the sojourn-time PDFs for the two
diffusive states follow exponential distributions with mean sojourn times $\mu_\pm$.
For the equilibrium process with $p_\pm(0)=p_\pm^{\rm eq}$ the correlation
function of $D(t)$ is given by
\begin{equation}
\label{eq: psi1 markovian two state}
\psi_1(t)=\frac{p_+^{\rm eq}p_-^{\rm eq}(D_+-D_-)^2}{\langle D\rangle_{\rm
st}^2}e^{-t/\tau},
\end{equation}
where $\tau=\mu_+\mu_-/(\mu_++\mu_-)$ is a characteristic time of $D(t)$. By
substituting this expression into the general formula (\ref{eq: formula RSD}), we obtain 
\begin{align}
\label{eq: rsd markovian two state}
&\Sigma^2(t;\Delta) \nonumber\\
&\approx\frac{p_+^{\rm eq}p_-^{\rm eq}(D_+-D_-)^2}{\langle D
\rangle_{\rm st}^2}\frac{2\tau^2}{t^2}\left(e^{-t/\tau}-1+\frac{t}{\tau}\right).
\end{align}
Asymptotically, the RSD exhibits the behaviours
\begin{equation}
\Sigma(t;\Delta)\approx \left\{\begin{array}{ll}
\frac{\sqrt{p_+^{\rm eq}p_-^{\rm eq}}(D_+-D_-)}{\langle D\rangle_{\rm st}^2}
&(t\ll\tau),\\\\
\frac{\sqrt{p_+^{\rm eq}p_-^{\rm eq}}(D_+-D_-)}{\langle D\rangle_{\rm st}^2}
\sqrt{\frac{2\tau}{t}}&(t\gg\tau) , \end{array}\right.
\label{eq: RSD two-state Markov}
\end{equation}
where we assume that $D_+>D_-$. The crossover time between the short-time plateau and the long-time decay of the RSD
 is estimated as $\tau_c=2\tau$. This result confirms that the fluctuations in the TSD persist at short times
but decay as $t^{-1/2}$ at long times, thereby satisfying a key criterion for ergodicity. In other words, although significant trajectory-to-trajectory fluctuations exist at short times, these vanish asymptotically, and time-averaged quantities converge to their ensemble-averaged counterparts.

\subsubsection{non-Markovian case}

To extend our analysis beyond the Markovian case, we now consider a scenario
 when the sojourn-time PDFs follow a power-law distribution, $\rho_
\pm(\tau)\sim a_\pm\tau^{-1-\alpha_\pm}/|\Gamma(-\alpha_\pm)|$ for $\tau\to\infty$.
The long-tailed nature of these PDFs significantly alters the statistical
properties of the process. The RSD follows from Eq.~(\ref{eq: rsd id and ex}),
revealing distinct asymptotic behaviours depending on whether the mean sojourn
time $\mu$ is finite or not.

When $\alpha_\pm>1$, the mean sojourn time $\mu$ remains finite. For $1<\alpha_+
<\alpha_-< 2$, the excess part in Eq.~(\ref{eq: rsd id and ex}) becomes dominant
in the asymptotic behaviour of the RSD, indicating  that the RSD exhibits a slow
relaxation,
\begin{equation}
\Sigma^2(t;\Delta)\approx\frac{2(\alpha_+-1)(D_+-D_-)^2(p_-^{\rm eq})^2}{\langle D\rangle_{
\rm st}^2\Gamma(4-\alpha_+)}\frac{a_+}{\mu}t^{1-\alpha_+},
\label{eq: RSD two-state non-Markov}
\end{equation}
for $t\to \infty$. Here, it is assumed that the initial sojourn-time PDFs are identical to the origin PDFs, corresponding to an ordinary renewal process.
Other choices of initial conditions such as the equilibrium sojourn-time PDF can lead to different asymptotic behaviour of the RSD. These cases are discussed in detail in Ref.~\cite{Miyaguchi2016}.
Notably, this decay is slower than $t^{-1/2}$, resembling
the slow relaxation observed in CTRWs with a power-law exponent  $\alpha\in
(1,2)$ in the waiting-time PDF \cite{Akimoto2011}. The origin of this behaviour lies in
 the power-law decay of the correlation function of
the instantaneous diffusivity $D(t)$, which takes the asymptotic form:
\begin{equation}
\langle\delta D(t)\delta D(0)\rangle_{\rm st}\sim\frac{(D_+-D_-)^2(p_-^{\rm
eq})^2}{\Gamma(2-\alpha_+)}\frac{a_+}{\mu}t^{1-\alpha_+},
\end{equation} 
for $t\to\infty$. Here, it is assumed that the initial sojourn-time PDFs are given by the equilibrium PDFs, i.e.,  the system has been prepared long before the observation starts, corresponding to an equilibrium renewal process. This result highlights the long memory effects: correlations of diffusivity fluctuations persist over extended timescales, sustaining non-ergodic behaviour at intermediate times, even though the system ultimately becomes ergodic in the long-time limit.

When $\alpha_\pm<1$, the mean sojourn time $\mu$ diverges, and the process
deviates significantly from Markovian dynamics. However, for $\alpha_+<\alpha_-
<1$, the TSD fluctuations decrease over time, and the RSD converges to zero:
\begin{equation}
\Sigma^2(t;\Delta)\approx\left(\frac{D_+-D_-}{D_-}\right)^2\frac{a_+}{a_-}\frac{
2(1-\alpha)t^{\alpha_+-\alpha_-}}{\Gamma(3+\alpha_--\alpha_+)},
\end{equation}
for $t\to\infty$. The negative exponent $\alpha_+-\alpha_-< 0$ ensures that
the RSD decays to zero over time. This result implies that, despite the
heavy-tailed sojourn-time distributions, the global
fluctuations in the TSD vanish asymptotically. Consequently, the system behaves effectively
 as Brownian motion with a constant diffusion coefficient $D_+$ in the long-time limit. In
this regime, trajectory-to-trajectory variability in the diffusivity diminishes,
ultimately resulting in ergodic-like behaviour at long times.

\section{Annealed transit time model}
\label{sec: ATTM}


The annealed transit time model (ATTM), introduced in Ref.~\cite{Massignan2014}, was designed to describe anomalous behaviours observed in single
particle tracking experiments in biological systems \cite{Manzo2015}. The
ATTM is a temporally heterogeneous diffusion model that can be formulated
based on the $d$-dimensional LEFD framework, i.e., Eq.~(\ref{LEFD}). In the
experimental context, $d$ typically takes values of two (e.g., diffusion in
a membrane) or three (e.g., diffusion within the cellular cytoplasm). The model
assumes that the instantaneous diffusivity $D(t)$ is a stochastic process that
remains constant over a sojourn time---defined as the time interval before the
diffusivity transitions to a new value. This diffusivity is coupled to the
sojourn time via the relation
\begin{equation}
D_\tau=\tau^{\sigma-1},
\label{eq: relation between D and tau}
\end{equation}
where $D_\tau$ represents the diffusion coefficient associated with a sojourn
time of duration $\tau$. At any given time $t$, the
instantaneous diffusivity is therefore expressed as $D(t)=\tau_t^{\sigma-1}$, where
$\tau_t$ denotes the sojourn time that straddles $t$---that is, the time the particle remains in the current state before switching. Since we assume $0<
\sigma<1$, longer sojourn times correspond to a lower instantaneous diffusivity,
effectively capturing the interplay between trapping and slow transport.

To characterise anomalous statistical properties in ATTM, we consider the
power-law sojourn-time PDF
\begin{equation}
\psi(\tau)\sim\frac{c}{|\Gamma(-\alpha)|}\tau^{-1-\alpha},\quad(\tau\to\infty),
\label{eq: sojourn-time PDF ATTM}
\end{equation}
where $c>0$ is a scale factor. When $\alpha\leq1$, the mean sojourn time
diverges, leading to intrinsically non-equilibrium dynamics. In this regime,
the system exhibits anomalous features such as subdiffusion, ageing, and weak
ergodicity breaking. Conversely, when $\alpha>1$, the system can reach
equilibrium, provided it has evolved over a sufficiently long period. However,
for $\alpha\leq1$, true equilibrium is never attained. In the following analysis,
we focus on the non-equilibrium case, assuming that the sojourn-time PDF at
$t=0$ follows Eq.~(\ref{eq: sojourn-time PDF ATTM}).

\subsection{ Mean squared displacement}

To analyse the MSD, we consider the displacement PDF $P({\bm r},t)$. This PDF
can be derived using the Montroll-Weiss equation \cite{AkimotoYamamoto2016a},
a fundamental relation in CTRW theory \cite{montroll1965random}. Let $Q({\bm r},
t)$ represent the PDF of the position ${\bm r}$ under the condition that the
instantaneous diffusivity $D(t)$ changes precisely at time $t$. Assuming the
initial condition $P({\bm r},0)=\delta({\bm r})$, the PDF $Q$ satisfies the
renewal equation
\begin{equation}
Q({\bm r},t)=\int_{-\infty}^\infty d{\bm r}'\int_0^tdt'Q({\bm r}-{\bm r}',t-t')
\phi({\bm r}',t')+\delta({\bm r}),
\end{equation}
where $\phi({\bm r},t')$ is the joint PDF of the displacement ${\bm r}$ and
sojourn time $t'$ immediately preceding time $t$. That is, $t'$ represents the duration for which the particle has remained in its current diffusive state prior to a possible transition occurring at time $t$. This joint PDF is given by
\begin{equation}
\phi({\bm r},t)=G_d({\bm r},t;D_t)\rho(t).
\end{equation}
Here, $G_d({\bm r},t;D_t)$ denotes the propagator for Brownian motion having diffusivity
$D_t$, and $\rho(t)$ is the sojourn-time PDF. The conditional PDF
$P({\bm r},t;\tau)$ for the displacement ${\bm r}$  given the sojourn time
$\tau$ is expressed as
\begin{equation}
P({\bm r},t;\tau)=\int_{-\infty}^\infty d{\bm r}'\int_0^t dt'Q({\bm r}-{\bm
r}',t-t')\Phi({\bm r}',t';\tau),
\end{equation}
where $\Phi({\bm r},t;\tau)$ is the joint PDF of the displacement ${\bm r}$
and sojourn time $\tau$, given that no renewal (i.e., no transition in the
diffusivity) occurs within the time $t$. It is expressed as $\Phi({\bm r},t;
\tau)=G_d({\bm r},t;D_\tau)\rho(\tau)\theta(\tau-t)$, where $\theta(t)$ is
the Heaviside step function---ensuring that the process remains in the same
sojourn state for $t<\tau$. Integrating $P({\bm r},t;\tau)$ over all possible
sojourn times $\tau$ leads to the PDF $P({\bm r},t)$,
\begin{equation}
P({\bm r},t)=\int_0^\infty P({\bm r},t;\tau)d\tau.
\end{equation}
Taking the Fourier-Laplace transform of this expression yields
\begin{equation}
\tilde{\hat{P}}({\bm k},s)=\frac{1}{1-\tilde{\hat{\phi}}({\bm k},s)}\int_0^
\infty\tilde{\hat{\Phi}}({\bm k},s;\tau)d\tau.
\end{equation}
This provides an exact representation of the propagator in Fourier-Laplace
space, generalising the random walk framework developed in earlier studies
\cite{montroll1965random,Shlesinger1982,Akimoto2013a,Akimoto2014}.

For $\alpha<1$, the system reaches a stationary state. As expected in Section~\ref{sec: III BYNGD}, 
the MSD exhibits normal diffusion,
\begin{equation}
\langle |{\bm r}(t)-{\bm r}(0) |^2\rangle\sim2d\langle D\rangle_{\rm st}t,
\end{equation}
where the stationary average $\langle D\rangle_{\rm st}$ is given by
\begin{equation}
\langle D\rangle_{\rm st}=\int_0^\infty\frac{\tau\rho(\tau)D_\tau}{\langle
\tau\rangle} d\tau.
\end{equation} 
We note that $\tau\rho(\tau)/\langle\tau\rangle$ is the PDF of the sojourn
time that straddles time $t$ for $t\to\infty$, corresponding to an equilibrium
renewal process \cite{Cox,God2001,Barkai2014,AkimotoYamamoto2016a}. Since the
system reaches equilibrium in this regime, the stationary distribution of the
instantaneous diffusivity can be directly related to the equilibrium
distribution of the sojourn time straddling $t$.

In contrast, for $\alpha\leq1$, the MSD exhibits subdiffusion. Using the
relation n Eq.~(\ref{relation between moments and F}) between the second moment
and $\tilde{\hat{P}}({\bm k},s)$, inverse Laplace transform yields the
asymptotic behaviour of the MSD in the long-time limit,
\begin{align}
&\langle |{\bm r}(t)-{\bm r}(0) |^2\rangle \nonumber\\
&\sim\left\{\begin{array}{ll}\dfrac{
2d\Gamma(\sigma-\alpha)}{(1+\alpha-\sigma)|\Gamma(-\alpha)|\Gamma(1+\sigma)}
t^\sigma\quad&(\sigma>\alpha),\\\\
\dfrac{2d}{|\Gamma(-\alpha)|\Gamma(1+\alpha)}t^\alpha\ln t&(\sigma=\alpha),\\\\
\dfrac{2d\langle D_\tau\tau\rangle}{c\Gamma(1+\alpha)}t^\alpha&(\sigma<
\alpha).\end{array}\right.
\label{eq: attm msd}
\end{align}
where the expectation value $\langle D_\tau\tau\rangle$ is defined as $\langle
D_\tau\tau\rangle=\int_0^\infty d\tau\rho(\tau)D_\tau \tau$. These results
demonstrate that the MSD exhibits subdiffusion, and the subdiffusive anomalous
diffusion exponent depends on both the power-law exponent $\alpha$ of the
sojourn-time PDF and the coupling exponent $\sigma$ between the instantaneous
diffusivity and the sojourn time. A transition in the power-law exponent of
the MSD occurs at $\alpha=\sigma$, at which the integral $\langle D_\tau\tau
\rangle$ diverges. This divergence marks a fundamental change in the diffusion
behaviour, highlighting the crucial interplay between the statistics of the
sojourn-time PDF and the fluctuating diffusivity in the ATTM.

\subsection{ Non-Gaussian propagator}

The PDF in the ATTM is inherently non-Gaussian due to the underlying
fluctuations in the instantaneous diffusivity. For $\alpha>1$, a
characteristic timescale $\tau_D$ exists in the dynamics of $D(t)$. In the
short-time limit $t\ll\tau_D$, the PDF can be approximated using the
superstatistical approach [Eq.~(\ref{eq: superstatistics approach})], yielding
\begin{equation}
P({\bm r},t)\sim\int_0^\infty G_d({\bm r},t;D)P(D)dD.
\end{equation}
The stationary distribution $P(D)$ can be derived as
\begin{equation}
P(D)=\frac{\rho(D^{\frac{1}{\sigma-1}})D^{\frac{1}{\sigma-1}-1}}{1-\sigma}.
\end{equation}
The asymptotic behaviour for $D\to0$ becomes
\begin{equation}
P(D)\sim\frac{c}{(1-\sigma)\Gamma(|-\alpha|)\langle\tau\rangle}D^{-1+\frac{
\alpha-1}{1-\sigma}}.
\end{equation}
As established in Section~\ref{sec: III BYNGD}, the NGP in in the short-time regime $t\ll\tau_D$ is determined by
the variance of $P(D)$, confirming that the NGP remains positive in the ATTM,
i.e., reflecting non-Gaussian displacement PDFs. 
 
 For $\alpha<1$, the characteristic time $\tau_D$  diverges, indicating that the system does not reach equilibrium.
 Consequently, the PDF remains non-Gaussian at
all times, even as $t\to\infty$. This behaviour is reminiscent of
heterogeneous random walk models \cite{Akimoto2013a,Akimoto2014}. The
persistent non-Gaussianity in the ATTM for $\alpha<1$  arises due to its
anomalous diffusion properties, similar to the CTRW model \cite{metzler00}.
This highlights a fundamental distinction between ATTM and normal Brownian
motion, reinforcing that heterogeneous diffusion environments
inherently lead to non-Gaussian transport phenomena. 

\subsection{ Relative standard deviation of the time-averaged squared displacement}

To analyse the RSD of the TSD, we again employ renewal theory \cite{Cox,
God2001,Akimoto2023} in the case when $D(t)$ is a non-stationary stochastic
process. Specifically, we assume that the instantaneous diffusivity follows
a semi-Markov process. In such a semi-Markov process, the diffusivity $D(t)$
remains constant over random time intervals. That is, for each interval
$t\in(t_k,t_{k+1})$, the diffusivity takes a random value $D_k$, and
transitions occur at random times $ t_k$. A simple example of this framework
is the two-state process above, in which $D(t)$  switches between two discrete
values. For $\Delta \ll t$, we approximate the TSD as
\begin{equation}
\overline{\delta^2(\Delta;t)}\sim\frac{1}{t}\left\{\sum_{k=0}^{N_t-1}\int_{t_k}
^{t_{k+1}}\delta{\bm r}^2(\Delta;t')dt'+ \Delta I\right\},
\label{eq: TSD approx semi-Markov}
\end{equation}
where $\delta{\bm r}(\Delta;t)\equiv{\bm r}(t+\Delta)-{\bm r}(t)$, $t_k$ is
the $k$th transition time of the diffusive state with $t_0=0$,  $N_t$ is
the number of diffusive-state changes until time $t$, and 
$
\Delta I=\int_{t_{N_t}}^t\delta {\bm r}^2(\Delta;t')dt'.
$
Taking the ensemble
average of the TSD for a fixed realisation of $D(t)$, we obtain
\begin{equation}
\left\langle\overline{\delta^2(\Delta;t)}\right\rangle\sim\frac{\Delta}{t}\left\{\sum_{k=0}^{
N_t-1}2dD_k\tau_k+2dD_{N_t}(t-N_t)\right\},
\label{eq: ETSD semi-Markov}
\end{equation}
where $\tau_k\equiv t_{k+1}-t_k$ represents the sojourn time in the $k$-th
state. Notably, Eq.~\eqref{eq: ETSD semi-Markov} remains random since both
$D_k$ and $\tau_k$  are stochastic variables. To facilitate the computation
of the second moment of the TSD, we approximate the TSD by its
ensemble-averaged form for a given realisation of $D(t)$. This yields
$\overline{\delta^2(\Delta;t)}\approx2dZ(t)\Delta/t$, where we define
\begin{equation}
Z(t)=\sum_{k=0}^{N_t-1}D_k\tau_k+D_{N_t}(t-N_t).
\label{eq: Z(t) semi-Markov}
\end{equation}
Thus, the RSD of the TSD can be determined by analysing the RSD of $Z(t)$.
This formulation provides a systematic approach to quantifying diffusivity
fluctuations in semi-Markovian systems, offering insights into non-ergodic
behaviour and anomalous diffusion.

 The moments of $Z(t)$ can be evaluated using a Montroll-Weiss-like equation
\cite{AkimotoYamamoto2016a}, since the stochastic process $Z(t)$ shares key
similarities with CTRWs \cite{Shlesinger1982,metzler00}. In this framework,
the sojourn times $\tau_k$ correspond to waiting times, while the diffusivity
values $D_k$ play the role of jumps in the CTRW analogy. To systematically
analyse the moments, we define the propagator $P_D(z,t)$, which represents the PDF of  $Z(t)$. Let $Q_D(
z,t)$ represent the PDF of displacement $z$, given that the instantaneous
diffusivity $D(t)$ undergoes a transition exactly at time $t$. These PDFs
satisfy the renewal equations:
\begin{equation}
Q_D(z,t)=\int_0^\infty dz'\int_0^tdt'Q_D(z-z',t-t')\phi_D(z',t')+\delta(z),
\end{equation}
where $\phi(z,t)$ is the joint PDF of displacement $z$ and sojourn time $t$,
described by $\phi_D(z,t)=\rho(t)\delta(z-t^\sigma)$, as follows from the relation between diffusivity and sojourn time in Eq.~(\ref{eq: relation between D and tau}). The PDF $P_D(z,t;\tau)$
of displacement $z$ at time $t$, given that the sojourn time straddling $t$
is $\tau $, is given by
\begin{equation}
P_D(z,t;\tau)=\int_0^\infty dz'\int_0^tdt'Q_D(z-z',t-t')\Phi_D(z',t';\tau),
\end{equation}
where $\Phi_D (z,t;\tau)$ is the joint PDF of displacement $z$ at time $t$
and sojourn time $\tau$, given that the diffusivity remains constant over the
interval $[0,t]$ and the sojourn time satisfies $t<\tau$. This function is
expressed as $\Phi_D(z,t;\tau)=\delta(z-\tau^{\sigma-1}t)\rho(\tau)\theta
(\tau-t)$, where $\theta(t)$ is the Heaviside step function. By integrating
$P_D(z,t;\tau)$ over all possible sojourn times $\tau$, we obtain the PDF
$P_D(z,t)$. Applying a double Laplace transform to the renewal equations
yields the transformed propagator
\begin{equation}
\hat{P}_D(k,s)=\frac{1}{1-\hat{\phi}_D(k,s)}\int_0^\infty\hat{\Phi}_D(k,s;
\tau).
\end{equation}
This formulation generalises the classic Montroll-Weiss equation, enabling
the analysis of anomalous diffusion behaviour in fluctuating diffusivity
models.

For $\alpha>1$, diffusion remains normal, and the process is ergodic in the sense 
that the TSD converges to the MSD: $\overline{\delta^2(\Delta;t)}\to2d\langle D
\rangle_{\rm st}\Delta$ for $t\to\infty$. Consequently, the RSD vanishes as $t
\to\infty$, similar to ordinary Brownian motion. However, due to the fluctuating
diffusivity, the RSD exhibits a different behaviour from that of standard
Brownian motion at intermediate timescales. By calculating the first and second
moments of the TSD, we obtain the associated RSD, which undergoes a transition
between two distinct regimes. For $t\ll\tau_D$, the RSD remains constant,
implying persistent trajectory-to-trajectory fluctuations in the TSD. In
contrast, for $t\gg\tau_D$, the RSD decays as $1/\sqrt{t}$, approaching the
behaviour observed in ordinary Brownian motion. When the sojourn-time PDF
follows an exponential distribution, the squared RSD in the asymptotic limit
$t\gg\tau_D$ and $\Delta\ll t$ decays as 
\begin{equation}
\Sigma^2(t;\Delta)\sim\left(\frac{\Gamma(2\sigma+1)}{\Gamma^2(\sigma+1)}-2
\sigma\right)\frac{\langle\tau\rangle}{t}.
\end{equation}
Conversely, in the short-time regime $t\ll\tau_D$, the RSD remains constant, as
predicted in Section~\ref{subsec: RSD in section III},
\begin{equation}
\Sigma(t;\Delta)\sim\sqrt{\frac{\Gamma(2\sigma)}{\Gamma^2(\sigma+1)}-1}.
\end{equation}
This transition---characterised by an initial plateau followed by a decay at
long times---is a universal feature of the LEFD framework.

For $\alpha<1$, as discussed in previous subsections, the system in this regime exhibits
anomalous diffusion, ageing, and weak ergodicity breaking. The first moment
of the TSD in the long-time limit can be expressed as $\langle\overline{\delta^2
(\Delta;t)}\rangle\sim2d\langle Z(t)\rangle\Delta/t$. Thus, the ensemble-averaged
TSD follows the asymptotic behaviour, depending on $\sigma$ and $\alpha$:
\begin{align}
&\left\langle\overline{\delta^2(\Delta;t)} \right\rangle \nonumber\\
&\sim\left\{\begin{array}{ll}
\dfrac{2d\Gamma(\sigma-\alpha)\Delta}{(1+\alpha-\sigma)|\Gamma(-\alpha)|
\Gamma(1+\sigma)}t^{\sigma-1}&(\sigma>\alpha),\\\\
\dfrac{2d\Delta}{|\Gamma(-\alpha)|\Gamma(1+\alpha)}t^{\alpha-1}\ln t &(
\sigma=\alpha),\\\\
\dfrac{2d\langle D_\tau\tau\rangle\Delta}{c\Gamma(1+\alpha)}t^{\alpha-1}
&(\sigma<\alpha),\end{array}\right.
\label{eq: tsd attm}
\end{align}
where $\langle D_\tau\tau\rangle$ remains finite when $\sigma<\alpha$.
While the ensemble-averaged MSD exhibits anomalous diffusion, the scaling of
the ensemble-averaged TSD indicates normal diffusion. This discrepancy 
highlights the presence of weak ergodicity breaking \cite{pt,Metzler2014}.
Furthermore, the explicit dependence of the ensemble-averaged TSD on the
measurement time $t$ implies ageing behaviour: the TSD systematically decreases
over time. Such ageing effects are well known in the CTRW framework \cite{He2008,
Miyaguchi2011a,Miyaguchi2013,Metzler2014,Schulz2014}, as well as in the quenched
trap model \cite{Miyaguchi2011, Miyaguchi2015}, and have been observed in
biological experiments \cite{Weigel2011,Manzo2015}.
Additional evidence for ageing and anomalous diffusion was reported in a study of protein internal dynamics using high-resolution molecular simulations combined with neutron scattering measurements \cite{jeremy}. The analysis revealed  that the TSD of the inter-domain centre-of-mass distance continued to grow in time without saturation and showed ageing over 13 decades. Notably, the observed subdiffusion was attributed to exploration of a rugged energy landscape, consistent with a  CTRW framework.

The second moment of $Z(t)$ yields the squared RSD, which remains finite even
in the long-time limit,
\begin{widetext}
\begin{equation}
\Sigma^2(t;\Delta)\sim\left\{\begin{array}{ll}\dfrac{2(1+\alpha-\sigma)\Gamma^2
(1+\sigma)}{\Gamma(1+2\sigma)}\left(\dfrac{(1+\alpha-\sigma)\Gamma(2\sigma-
\alpha)|\Gamma(-\alpha)|}{(2+\alpha-2\sigma)\Gamma^2(\sigma-\alpha)}+1\right)
-1&(\sigma>\alpha),\\\\
\dfrac{2\Gamma^2(1+\alpha)}{\Gamma(2\alpha+1)}-1 &(\sigma\leq\alpha).\end{array}
\right.
\label{eq: rsd attm}
\end{equation}
\end{widetext}
For $\sigma\leq\alpha$, the RSD is independent of $\sigma$ and identical to the
RSD of the CTRW \cite{He2008}. However, for $\sigma>\alpha$, the RSD explicitly
depends on $\sigma$, indicating a transition in the TSD distribution at $\sigma
=\alpha$. Similar transitions arise in the
time-averaged kinetic energy of subrecoil laser cooling models \cite{Barkai2021,
barkai2022gas,Akimoto2022infinite}. For $\sigma\leq\alpha$, the distribution
of the TSD is well described by the Mittag-Leffler distribution with order
$\alpha$ \cite{AkimotoYamamoto2016a}, a key result in infinite ergodic theory
\cite{Aaronson1981,Aaronson1997,Akimoto2010} and CTRW \cite{He2008,
Miyaguchi2011a,Miyaguchi2013,Metzler2014}. However, for $\sigma>\alpha$, the
TSD distribution deviates from the Mittag-Leffler form and depends explicitly
on the observable \cite{AkimotoYamamoto2016a,Barkai2021,barkai2022gas,
Akimoto2022infinite}. These non-universal distributions are relevant in the
infinite ergodic theory of non-integrable observables with respect to the
infinite density \cite{Akimoto2015}.
These results illustrate that time-averaged observables in non-equilibrium systems can show normal-like scaling (e.g., linear TSD), yet retain signatures of ageing and weak ergodicity breaking in their fluctuations.

\section {Generalised Langevin equation with fluctuating diffusivity }
Many experimental and simulation studies have reported both a nonlinear
dependence of the MSD on time $t$ and deviations from Gaussian displacement
PDFs (e.g., \cite{He2008,wang2012brownian,parry2014bacterial,Manzo2015,
lampo2017cytoplasmic,sabri20,janczura21,amanda,Jeon2016,stefanie,beta,beta1}).
However, in the standard LEFD framework, if $D(t)$ is a stationary process,
the MSD remains linear in $t$. This follows directly from the fact that the
noise in the LEFD is delta-correlated. Specifically, it satisfies $\langle\bm{\xi}(t)\cdot\bm{\xi}(t')
\rangle=2dD(t)\delta(t-t')$, which leads to the normal diffusion behaviour
as described by Eq.~(\ref{eq: normal diffusion in LEFD}).

For the LEFD to exhibit a nonlinear MSD, such as the subdiffusion dynamics
in Eq.~(\ref{e.MSD.two-state-lefd.subdiffusion}), the fluctuating diffusivity
$D(t)$ must be non-stationary. This implies that the standard LEFD
with stationary diffusivity cannot simultaneously account for both subdiffusion
and non-Gaussian statistics. To capture both features within an
equilibrium and stationary framework, it is necessary to extend the LEFD by
incorporating a temporally correlated noise instead of delta-correlated noise.
Such modifications introduce memory effects in the stochastic process,
potentially capturing the long-range correlations observed in experimental
systems.\footnote{We note that for intrinsically non-equilibrium systems such
as living biological cells or tissues, anomalous diffusion in the presence of
long-range dependence is also achieved by combining a diffusing diffusivity
or a two-state diffusivity dynamics with fractional Brownian motion
\cite{wang20,diego,wang20b,wei1}, as also discussed below.}
In the following section, we explore such extensions and their ability to reconcile anomalous scaling with non-Gaussian statistics.


\subsection {Description of model}

Recently, several studies have incorporated correlated noise into the LEFD
\cite{slezak18,wang20,wang20b,sabri20,janczura21,dieball22}. For
instance, Ref.~\cite{wang20} examined the Langevin equation
\begin{equation}
\label{e.glefd_wang20}
\frac{dx}{dt}=\sqrt{2D(t)}\xi_H(t)
\end{equation}
where $\xi_H(t)$ represents fractional Gaussian noise, generating fractional
Brownian motion (FBM). This is a simple model capable of describing both
subdiffusion and non-Gaussianity. 
Interestingly, the model exhibits a crossover: when the noise correlation shows
anti-persistence, the MSD displays subdiffusive behaviour at short times but
crosses over to normal diffusion at long times.
 A similar model with a different mathematical construction of FBM does not show
this crossover \cite{diego}. In fact, a recent study demonstrates that, in the
presence of a stochastic diffusivity different definitions of FBM lead to
different dynamic behaviours of the MSD \cite{wei1}. We also note that in the case of FBM described by 
Eq.~(\ref{e.glefd_wang20}), the fractional Gaussian noise is interpreted as
``external noise" in the sense of Klimontovich \cite{klimontovich} and the dynamics do not satisfy
 the fluctuation-dissipation relation. This is justified
in intrinsically non-equilibrium systems such as living biological cells or
tissues. 
These observations raise fundamental questions about the universality of subdiffusion in correlated-noise systems and the role of model-specific memory kernels.

An alternative approach involves using the overdamped GLE \cite{miyaguchi22,miyaguchi24}
\begin{equation}
\label{e.gle.wo.fd.simple}
\frac{d\bm{r}}{dt}+\int_0^t\phi(t-t')\dot{\bm{r}}(t')dt'=\sqrt{2D}\bm{\xi}+
\sqrt{D}\bm{\xi}_c^0,
\end{equation}
where $\bm{r}(t)$ is the position of the tracer, and $\bm{\xi}(t)$ and $\bm{\xi}
_c^0(t)$ are Gaussian white noise and coloured noise vectors
in three dimensions, respectively. These noise terms satisfy the fluctuation-dissipation relations:
\begin{align}
\label{e.<xi-xi>.simple}
\langle\bm{\xi}(t)\bm{\xi}(t')\rangle&=\delta(t-t')\bm{1},\\[0.1cm]
\label{e.<xi_c-xi_c>.simple}
\langle\bm{\xi}_c^0(t)\bm{\xi}_c^0(t')\rangle&=\phi(t-t')\bm{1}.
\end{align}
The overdamped GLE in Eq.~\eqref{e.gle.wo.fd.simple} also describes the dynamics of a single bead
in the Rouse model \cite{panja10b} and in elastic network models \cite{miyaguchi24}, where memory effects play a key role.

Since the overdamped GLE in Eq.~(\ref{e.gle.wo.fd.simple}) is a Gaussian process,
it does not produce non-Gaussian displacement statistics. To describe non-Gaussianity, a
fluctuating diffusivity is incorporated into the GLE 
via a Markov embedding, in which the kernel function is approximated by a
superposition of exponential functions as
\begin{equation}
\label{e.phi(t)_approx}
\phi(t-t')\approx\beta D\sum_{i=0}^{N-1}k_ie^{-\beta k_i D_i |t-t'|},
\end{equation}
where $k_i$ and $D_i$ are fitting parameters and $\beta k_0D_0$ is a high
frequency cutoff. Here, each  $D_i $ represents the effective diffusivity of an auxiliary variable
 $\bm{r}_i$, and $k_i$ represents the strength of the harmonic coupling  between the
auxiliary variable  $\bm{r}_i$ and the tracer's position $\bm{r}$. Using these auxiliary variables, 
the non-Markovian equation of motion in Eq.~(\ref{e.gle.wo.fd.simple})
can be recast as a set of Markovian equations.
It is important to clarify that these $D_i$ are not stochastic at this
stage; rather, they are constant parameters employed to approximate a
memory kernel of arbitrary shape [e.g., a power-law memory kernel in
Eq.~(\ref{e.phi(t).power})] \cite{goychuk09}. Consequently, no fluctuating
diffusivity is involved in Eq.~(\ref{e.phi(t)_approx}). The fluctuating
diffusivity is introduced later in Eq.~(\ref{e.glefd1.markov.2}), which
describes the stochastic dynamics of the auxiliary variables $\bm{r}_i$ with
fluctuating diffusivity $D_i(t)$, thereby rendering an effective
diffusivity $D(t)$ of the tracer a stochastic time-dependent quantity.

Replacing the diffusivities $D_i$ in the Markovian equations with stationary stochastic
processes $D_i(t)$, which  satisfy $\langle D_i
(t)\rangle=D_i$, we obtain
\begin{align}
\label{e.glefd1.markov.1}
\frac{d\bm{r}(t)}{dt}&=\sqrt{2D}\bm{\xi}(t)-\beta D\sum_{i=0}^{N-1}k_i(\bm{r}
-\bm{r}_i),\\[0.1cm]
\label{e.glefd1.markov.2}
\frac{d\bm{r}_i(t)}{dt}&=-\beta D_i(t)k_i(\bm{r}_i-\bm{r})+\sqrt{2D_i(t)}
\bm{\eta}_i(t), 
\end{align}
where $\bm{\eta}_i(t)$ is a white Gaussian noise satisfying
\begin{equation}
\label{e.<eta_i(t)eta_j(t')>}
\langle\bm{\eta}_i(t)\bm{\eta}_j(t')\rangle=\delta_{ij}\delta(t-t')\bm{I}.
\end{equation}
Note that Eq.~(\ref{e.glefd1.markov.2}) is an LEFD with a harmonic force,
closely related to the Ornstein-Uhlenbeck process with a fluctuating
diffusivity discussed in Section~\ref{s.ou-proc}.

Equations (\ref{e.glefd1.markov.1}) and (\ref{e.glefd1.markov.2}) can 
be reformulated as an integral equation:
\begin{equation}
\label{e.glefd1}
\frac{d\bm{r}}{dt}+\int_0^t\phi(t,t')\dot{\bm{r}}(t')dt'=\sqrt{2D}\bm{\xi}
+\sqrt{D}\bm{\xi}_c.
\end{equation}
This integral equation is referred to as a GLE with fluctuating diffusivity
(GLEFD). More precisely, the memory kernel $\phi(t,t')$ is
  actually a functional of the diffusivities $D_i(t)$ [see
  Eq.~(\ref{e.phi(t,t')})]. However, for notational simplicity, we denote it as 
  $\phi(t,t')$ by assuming a fixed realization of the diffusivities ${D_i(t)}$.
An important feature of the GLEFD is that it satisfies the (generalised)
fluctuation-dissipation relation
\begin{align}
\label{e.phi(t,t')}
\phi(t,t')\bm{I}&=\langle\bm{\xi}_c(t)\bm{\xi}_c(t')\rangle\nonumber\\
&=\bm{I}\beta D\sum_{
i=0}^{N-1}k_i\exp\left(-\beta k_i\left|\int_{t'}^t D_i(u)du\right|\right). 
\end{align}
Notably, the memory
  kernel in Eq.~\eqref{e.phi(t,t')} depends on both $t$ and $t'$ individually, and
  cannot be written as a function of the lag time $t - t'$ alone
  \cite{miyaguchi22}.

 Interestingly, a memory kernel with separate dependence on $t$ and $t'$ has been derived using the projection operator method for Hamiltonian systems with explicit time dependence \cite{kawai11}. However,
  whether the GLEFD itself can be derived from microscopic dynamics remains an open
  question. This is because the derivation in Ref. \cite{kawai11} pertains to non-stationary
  (non-equilibrium) processes, whereas the GLEFD is intended to describe systems in equilibrium. 
  It is worth noting that Eqs.~(\ref{e.glefd1.markov.1}) and
(\ref{e.glefd1.markov.2}) can be utilised for numerical integration, providing a
practical approach to simulate systems with memory and fluctuating transport
properties.

\subsection {Fractional Brownian motion with fluctuating diffusivity}

Due to the non-convolutional form of the kernel in Eq.~(\ref{e.phi(t,t')}),
general analytical results for the GLEFD remain limited. Two simple examples
have been investigated in Ref.~\cite{miyaguchi22}: a dimer model and an FBM
with fluctuating diffusivity (FBMFD); see also the discussion in \cite{wang20,
wang20b,diego}. In the following, we present the definition and selected numerical results
for FBMFD.

As a canonical example of a long-memory kernel, we consider the power-law form:
\begin{equation}
\label{e.phi(t).power}
\phi(t)=\frac{D}{D_{\alpha}}\frac{t^{-\alpha}}{\Gamma(1+\alpha)\Gamma(1-\alpha)}
=At^{-\alpha},
\end{equation}
where $\alpha$ is the power-law exponent with $0<\alpha<1$, 
$D_{\alpha}$ is the generalised diffusion
coefficient for subdiffusion, and $A$ is the prefactor given by $A=D/D_{\alpha}\Gamma(1+
\alpha)\Gamma(1-\alpha)$.

With the power-law memory kernel given in Eq.~(\ref{e.phi(t).power}), the GLE
(\ref{e.gle.wo.fd.simple}) without the fluctuating diffusivity displays
normal diffusion at short times and subdiffusion at long times 
\cite{miyaguchi22}:
\begin{align}
\label{e.gle.msd}
\langle\delta\bm{r}^2(t)\rangle\sim \begin{cases}2nDt,&t\ll t_c,\\
2nD_{\alpha}t^{\alpha},&t\gg t_c,\end{cases}
\end{align}
where $\delta\bm{r}(t) = \bm{r}(t)-\bm{r}(0)$ is a displacement vector and $t_c$
is a crossover time defined by $t_c=[\Gamma(1+\alpha)D_\alpha/D]^{1/(1-\alpha
)}$. More precisely, the displacement correlation $\langle \delta\bm{r}(t)
\delta\bm{r}(t')\rangle$ is given at long times by $\langle\delta\bm{r}(t)\delta\bm{r}(t')
\rangle\sim D_{\alpha}(t^{\alpha}+t'^{\alpha}-|t-t'|^{\alpha})$, 
which corresponds to the correlation function of FBM
with Hurst exponent \( H = \alpha/2 \). 
This equivalence holds at the level of second-order statistics, and---due to the Gaussian nature of the processes---extends to all higher-order moments. Nonetheless, the underlying stochastic processes remain distinct in their construction.
Despite differences in the short-time
dynamics, we refer to the GLE with the power-law kernel
in Eq.~(\ref{e.phi(t).power}) as FBM for simplicity.

\begin{figure}[t!]
  \centerline{\includegraphics[width=0.8\linewidth]{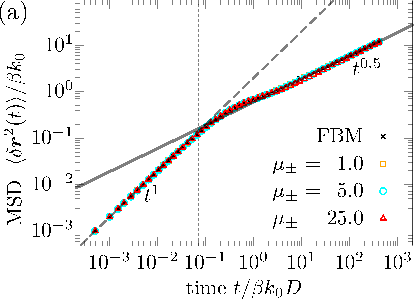}}
  \centerline{\includegraphics[width=0.8\linewidth]{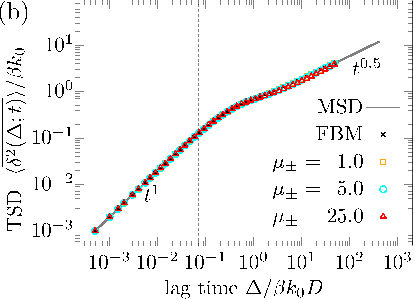}}
  \caption{\label{f.fbm} MSD and TSD for FBMFD at three different
    values of $\mu_{\pm}$: $\mu_{ \pm}=1.0$, $5.0$, and $25.0$. The symbols are
    data obtained from numerical integration of Eqs.~(\ref{e.glefd1.markov.1})
    and (\ref{e.glefd1.markov.2}) with the memory kernel
    (\ref{e.phi(t,t').two-state.power}). The power-law index $\alpha$ and the
    generalised diffusivity $D_\alpha$ are fixed as $\alpha =0.5$ and
    $D_\alpha/D(\beta k_0 D)^{1-\alpha}=0.3$, respectively. The fast and slow
    state diffusivities $\kappa_\pm$ are set as
    $(\kappa_+,\kappa_-)=(1.95, 0.05)$. Simulations for FBM with the same
    parameter values of $\alpha$ and $D_{\alpha}$ are also displayed
    (crosses). (a) MSD vs time. The dashed and full lines are the short-time and
    long-time predictions in Eq.~(\ref{e.fbmfd.msd}), respectively. The vertical
    dotted line indicates the crossover time $t_c'$.  (b)
      Ensemble-averaged TSD vs lag time $\Delta$. The solid line represents the
      MSD for $\mu_{\pm} = 1.0$, which is the same data shown as squares in
      panel.~(a). Thus, the MSD and TSD are consistent, indicating that the FBMFD
      can be considered ergodic.} 
\end{figure}

To approximate the power-law function in Eq.~(\ref{e.phi(t).power}) using the exponential
form in Eq.~(\ref{e.phi(t)_approx}), we must specify the fitting parameters $k_i$
and $D_i$. First, we set $\beta k_iD_i=\beta k_0D_0/b^i$, where $b>1$ is
a constant. Second, the coefficients  $k_i$ are determined by $\beta Dk_i=A'(\beta D_ik_i)^{
\alpha}=A'(\beta k_0D_0/b^i)^{\alpha}$, where $A'=A(b-1)/b^{1/2}\Gamma(\alpha)$ is a prefactor chosen to match the amplitude of the power-law kernel.
Under this choice, the power-law kernel (\ref{e.phi(t).power}) is approximated as
\begin{equation}
\label{e.phi(t).power.markov}
\phi(t)\approx A'\sum_{i=0}^{N-1}(\beta D_ik_i)^{\alpha}e^{-\beta D_ik_i|t|}.
\end{equation}
This approximation becomes exact at $b\to1$ \cite{goychuk09}, whereas we set
$b=2$ for numerical simulations shown in Fig.~\ref{f.fbm}. In these simulations, we
also use $k_0$ as a unit of $k_i$, thereby specifying all fitting parameters
$k_i$ and $D_i$.
Because of the correspondence between Eqs.~(\ref{e.phi(t)_approx}) and
(\ref{e.phi(t,t')}), the fluctuating memory kernel $\phi(t,t')$ is obtained
by modifying Eq.~(\ref{e.phi(t).power.markov}) as
\begin{equation}
\label{e.phi(t,t').power}
\phi(t,t')=A'\sum_{i=0}^{N-1}(\beta D_ik_i)^{\alpha}\exp\left(-\beta k_i\left|
\int_{t'}^tD_i(u)du\right|\right).
\end{equation}
This model is referred to as FBMFD \cite{miyaguchi22}. 

As a simple case study, let us focus on the FBMFD with a two-state diffusivity.
As shown in Section~\ref{sec: two-state model}, the two-state diffusivity has
been frequently used  to explain statistical properties of experimental
single-particle-tracking data \cite{sabri20,janczura21,dieball22}. Here, the
two-state diffusivity is defined as
\begin{equation}
\label{e.glefd.two-state}
D_i(t)=D_i\kappa(t)=\begin{cases}D_i\kappa_+,\\[0.1cm]D_i\kappa_-,\end{cases}
\end{equation}
where $\kappa(t)$ is a dimensionless two-state process. More precisely, the
diffusivity $D_i(t)$ is assumed to switch between the two states at random
times $t_1,t_2,\dots$. Let $\tau_k=t_k-t_{k-1}$, $(k=1,2,\dots)$ be sojourn
times of the two states, where we define $t_0 = 0$ for convenience. The $+$
($-$) state is referred to as the fast (slow) state, and the sojourn time
PDF of the fast (slow) state, denoted as $\rho^+(\tau)$ [$\rho^-(\tau)$], is
assumed to follow an exponential distribution with mean $\mu_+$ ($\mu_-$),
$\rho^{\pm}(\tau)=\exp(-\tau/\mu_{\pm})/\mu_{\pm}$.

By substituting Eq.~\eqref{e.glefd.two-state} into Eq.~\eqref{e.phi(t,t').power},
the memory kernel becomes 
\begin{equation}
\label{e.phi(t,t').two-state.power}
\phi(t,t')=A'\sum_{i=0}^{N-1}(\beta D_ik_i)^{\alpha}\exp\left(-\beta D_ik_i
\left|\int_{t'}^t\kappa(u)du\right|\right).
\end{equation}
If there is no switching, the diffusive state remains constant:
$\nu_i(t) \equiv \nu_i\kappa_+$ or $\nu_i(t) \equiv \nu_i\kappa_-$ for all $t$, 
depending on the initial state. Consequently, if the initial state is $\pm$, the
MSD grows as $\langle\delta\bm{r}^2(t)\rangle\sim 2nD_{\alpha}^{\pm}t^{\alpha}$
with $D^{\pm}_{\alpha}=D_{\alpha}\kappa_{\pm}^{\alpha}$ at long times. If the
diffusivity switches very slowly between the two states, 
the dynamics can be viewed as alternating between two subdiffusive modes, each characterised by an effective diffusion coefficient  $D_\alpha^\pm$  and exponent  $\alpha $.

In general, however, the situation is more subtle. If the diffusive state
switches from one state to the other, information from the previous states is not
lost immediately due to the memory effect encoded in the integral of
Eq.~(\ref{e.phi(t,t').two-state.power}). 
This highlights the subtlety of memory effects: the system does not adapt instantaneously to a new diffusive state, but instead exhibits a transient regime in which the influence of past states persists, due to the non-local nature of the memory kernel.
By setting $\beta k_iD_i=\beta k_0D_0/
b^i$ and applying the approximation discussed above, Eq.~\eqref{e.phi(t,t').two-state.power}
can be expressed as
\begin{equation}
\label{e.phi(t,t').two-state.power.approx}
\phi(t,t')\approx\frac{D}{D_{\alpha}}\frac{\left|\int_{t'}^t\kappa(u)du\right|
^{-\alpha}}{\Gamma(1+\alpha)\Gamma(1-\alpha)}.
\end{equation}
Note that, if $\kappa(t)$ is independent of time $t$,
Eq.~(\ref{e.phi(t,t').two-state.power.approx}) represents the algebraic decay
$\sim |t-t'|^{-\alpha}$, thus recovering the stationary kernel
in Eq.~\eqref{e.phi(t).power} as a special case.

\subsection {Subdiffusion and non-Gaussianity}

If $t-t'$ is much larger than the mean sojourn times $\mu_{\pm}$, and $\kappa(
t)$ is a stationary process with mean $\langle\kappa\rangle$, the integral in
Eq.~(\ref{e.phi(t,t').two-state.power.approx}) can be approximated as $\int_{t'}
^t\kappa(s)ds\approx\langle\kappa\rangle (t-t')$. Thus, for large $t-t'$, we
obtain 
\begin{equation}
\label{e.phi(t,t').two-state.power.approx.long-time}
\phi(t,t')\approx\frac{D}{D_{\alpha}^{\mathrm{eq}}}\frac{|t-t'|^{-\alpha}}
{\Gamma(1+\alpha)\Gamma(1-\alpha)},
\end{equation}
where $D_{\alpha}^{\mathrm{eq}}=D_{\alpha}\langle\kappa\rangle^{\alpha}$. Since
this memory kernel is the same form as that in Eq.~(\ref{e.phi(t).power}), 
the long-time behaviour is also equivalent to that described in Eq.~\eqref{e.gle.msd}, with
the effective subdiffusion coefficient \( D_{\alpha}^{\mathrm{eq}} \).
Moreover, since 
the memory kernel does not affect the short-time dynamics,  the
short-time behaviour remains identical to that in  Eq.~(\ref{e.gle.msd}).
It follows that the MSD has the form:
\begin{align}
\label{e.fbmfd.msd}
\langle\delta\bm{r}^2(t)\rangle\sim\begin{cases}2nDt,&t\ll t_c',\\[.15cm]
2nD_{\alpha}^{\mathrm{eq}}t^{\alpha},&t\gg t_c',\end{cases}
\end{align}
where $t_c'$ is defined by $t_c'=[\Gamma(1+\alpha)D_{\alpha}^{\mathrm{eq}}/D]
^{1/(1-\alpha)}$ \cite{miyaguchi22}. Therefore, the MSD exhibits subdiffusion
at long times $t\to\infty$.
This is consistent with the model studied in \cite{diego}, and contrasts with the models in \cite{wang20, wang20b}, where a crossover to normal diffusion is observed at long times (see also the discussion in \cite{wei1}).
The prefactor  $D_{\alpha}^{\mathrm{eq}}$  encodes the time-averaged effect of the diffusivity fluctuations in the stationary regime.

\begin{figure}[t!]
  \centerline{\includegraphics[width=0.8\linewidth]{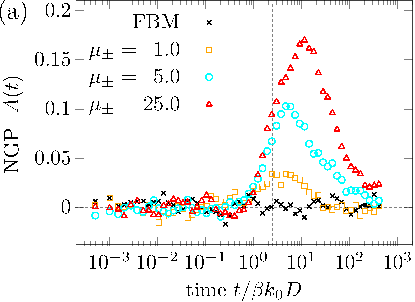}}
  \centerline{\includegraphics[width=0.8\linewidth]{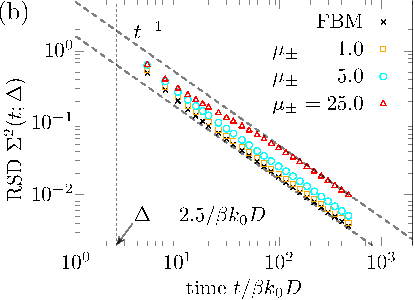}}
  \caption{\label{f.fbm.ngp} NGP and RSD for FBMFD at three
      different values of $\mu_{\pm}$: $\mu_{ \pm}=1.0$, $5.0$, and $25.0$. The
      other parameters are the same as those in Fig.~\ref{f.fbm}. (a) NGP vs
      time. (b) RSD vs time. The lag time $\Delta$ is set as
      $\Delta = 2.5/\beta k_0D$, which is indicated by the vertical dotted lines
      in both panels.  The dashed lines, representing $1/t$ decay, are shown as
      guides for the eye.}  
\end{figure}

In Fig.~\ref{f.fbm}, numerical results for (a) the MSD
  $\langle\delta\bm{r}^2\rangle$ and (b) the ensemble-averaged TSD
  $\langle \overline{\delta(\Delta; t)} \rangle$ are shown for FBM (without
  fluctuating diffusivity) and the FBMFD at three different values of
  $\mu_{\pm}$. In Fig.~\ref{f.fbm.ngp}, (a) the NGP $A(t)$ [Eq.~(\ref{eq: NGP
    def})], and (b) the RSD $\Sigma^2(t;\Delta)$ are displayed.  In the
  numerical simulations, we assume that the sojourn times $\tau$ for the fast
  and slow states follow exponential PDFs with the same mean, $\mu_+=\mu_-$.
  Note that Figs.~\ref{f.fbm} and \ref{f.fbm.ngp} use units different from those employed in Ref.~\cite{miyaguchi22}.

As shown in Fig.~\ref{f.fbm}(a), the MSD exhibits no qualitative difference
between systems with and without fluctuating diffusivity. As a result, 
to extract information about the fluctuating diffusivity in the present model,
higher-order moments, such as the NGP, must be examined.
The fact that the
fluctuating diffusivity cannot be inferred from the second-order moment is a
notable feature of the LEFD, in which the memory kernel is the delta function
\cite{Uneyama2015, miyaguchi17}. This characteristic property appears to remain
valid for the GLEFD framework as well.
As shown in Fig.~\ref{f.fbm}(b), the MSD and the ensemble-averaged
  TSD are in good agreement, indicating that the FBMFD is ergodic. Note,
  however, that if the stochastic process $D_i(t)$ are not stationary, FBMFD
  becomes a non-equilibrium process, and the ergodicity is consequently violated
  as in the case of the LEFD [Section~\ref{sec: two-state model}].

As shown in Fig.~\ref{f.fbm.ngp}(a), the NGP exhibits a clear difference between
FBM and  FBMFD. For FBM, the NGP remains zero because it is a Gaussian
process. In contrast, for the FBMFD, the NGP increases from zero around the
crossover time $t_c'$, reflecting the onset of non-Gaussianity due to diffusivity fluctuations. 
Moreover, the NGP $A(t)$ displays  a unimodal shape,
with its peak located near the mean sojourn times $\mu_{\pm}$. This suggests
that the NGP could serve as a proxy for estimating  the mean sojourn time
$\mu_{\pm}$. The peak height of $A(t)$ is larger for slower switching
dynamics, where $\mu_{\pm}$ are large, whereas for fast switching, when
$\mu_{\pm}\lesssim t_c'$, the NGP $A(t)$ vanishes, making any deviations
from FBM effectively undetectable.
This highlights the usefulness of the NGP as a diagnostic tool for distinguishing between FBM---i.e., a stationary Gaussian process---and FBMFD, which exhibits non-Gaussian fractional Brownian diffusion due to fluctuating diffusivity.

 As shown in Fig.~\ref{f.fbm.ngp}(b), the RSD $\Sigma^2(t;\Delta)$
  for FBMFD also displays marked deviations from that of standard FBM. This is expected, as the RSD is
  closely related to the NGP $A(t)$ \cite{Uneyama2015}. Notably, the RSD can be
  computed from fewer samples than the NGP, making it a more practical and accessible diagnostic tool in experimental and simulation settings.
   Importantly, the RSD strongly
  depends on the lag time $\Delta$; in particular, $\Delta$ should be chosen
  close to the peak time of the NGP $A(t)$ to obtain a clear distinction from
  the FBM case. In the simulations shown in Fig.~\ref{f.fbm.ngp}(b), the lag
  time $\Delta$ was set to $\Delta = 2.5/\beta k_0 D$, which is close to the
  peak times of the NGPs in Fig.~\ref{f.fbm.ngp}(a).  
  

\section{Ornstein-Uhlenbeck process with fluctuating diffusivity}
\label{s.ou-proc}
{\color{black}
The LEFD framework can capture a wide range of intriguing non-Gaussian diffusion phenomena, as discussed above. 
In the preceding sections we have examined the statistical properties of the LEFD for free particles. It is instructive to extend this framework to the case of confined diffusion, where the particle experiences a harmonic restoring force. The resulting dynamics, known as the Ornstein-Uhlenbeck process with fluctuating diffusivity (OUFD), provides a minimal model that connects stochastic diffusivity with confined Brownian motion \cite{Uneyama2019,Miyaguchi2019}.}

\subsection{Model and Governing Equations}

Assuming the underdamped Langevin equation with fluctuating viscosity (see Eq.~\eqref{eq: LE fluctuating viscosity}) in the presence of a harmonic potential $U(x) = \frac{1}{2}k x^2$, where $k > 0$ is a spring constant, the equation of motion becomes
\begin{equation}
m \dot{v}(t) = -\gamma(t) v(t) - k x(t) + \sqrt{2 \gamma(t) k_{\rm B} T} \, \eta(t).
\end{equation}
Introducing the time-dependent diffusivity $D(t)=k_{\rm B} T/\gamma(t) $, and considering the overdamped limit $m \to 0$, the equation reduces to
\begin{equation}
\dot{x}(t) =  - \frac{k D(t)}{k_{\rm B} T} x(t) + \sqrt{2 D(t)} \, \eta(t).
\end{equation}
Thus, in the presence of a harmonic potential $U(\bm{r}) = u k_{B} T \bm{r}^{2} / 2$, where $u > 0$ is a dimensionless constant, the LEFD equation [Eq.~\eqref{LEFD}] is modified to
\begin{equation}
 \label{eq: OUFD}
 \frac{d\bm{r}(t)}{dt} = - u D(t) \bm{r}(t) + \sqrt{2 D(t)} \bm{\xi}(t).
\end{equation}
The first term on the right hand side of Eq.~\eqref{eq: OUFD} represents a
 linear restoring force, causing $\bm{r}(t)$ to 
fluctuate around the origin. When $D(t)$ is constant, Eq.~\eqref{eq: OUFD} reduces to
the classical Ornstein-Uhlenbeck process. Accordingly, we refer to  Eq.~\eqref{eq: OUFD} as
the Ornstein-Uhlenbeck process with the fluctuating diffusivity (OUFD).
The OUFD model describes the motion of a particle subject to both a harmonic potential and time-dependent diffusivity, and it can serve as a minimal model for systems such as a colloidal particle confined in an optical trap embedded within a dynamically heterogeneous medium. In contrast, systems like supercooled liquids---where no external trapping force is present---are more appropriately modelled by the LEFD without a restoring force.
Alternatively, if we interpret $\bm{r}(t)$ as 
the relative position between two atoms in a molecule, the OUFD model provides a useful description of conformational fluctuations within a dynamically heterogeneous environment.

\subsection{Relaxation function and Dynamical Properties}
To characterise dynamical properties of OUFD, we introduce the relaxation
function, defined as
\begin{equation}
 \label{eq: OUFD relaxation function phi}
 \Phi(t) = \frac{\langle \bm{r}(t) \cdot \bm{r}(0) \rangle_{\text{st}}}{\langle \bm{r}^{2} \rangle_{\text{st}}}.
\end{equation}
This function also provides a compact expression for the MSD:
\begin{equation}
 \langle \lbrace \bm{r}(t) - \bm{r}(0)  \rbrace^{2} \rangle_{\text{st}}
  = 2 \langle \bm{r}^{2} \rangle_{\text{st}} [1 - \Phi(t)].
\end{equation}
By combining Eqs.~\eqref{eq: OUFD} and \eqref{eq: OUFD relaxation function phi},
the relaxation function $\Phi$(t) can be expressed in a particularly simple form \cite{Uneyama2019}:
\begin{equation}
 \label{eq: OUFD relaxation function phi with diffusivity}
 \Phi(t) = 
  \left\langle
   \exp\left[ - u \int_{0}^{t} D(t') dt' \right]
\right\rangle_{\text{st}}.
\end{equation}
To highlight the impact of a fluctuating diffusivity, we consider the limiting case where the diffusivity is constant, $D(t) = D$. In this case, Eq.~\eqref{eq: OUFD relaxation function phi with diffusivity}
reduces to a simple exponential decay: $\exp(-u D t)$. 
However, the presence of diffusivity fluctuations qualitatively alters the relaxation dynamics.
Unlike the MSD in the LEFD model---which depends linearly on  $D(t)$---the relaxation function in OUFD depends  nonlinearly on  $D(t)$. 
This nonlinearity enables a wide range of relaxation behaviours, including deviations from exponential decay. In particular, it can give rise to dynamics resembling stretched exponential forms, which are commonly observed in experiments on complex and heterogeneous systems.

In the case where $D(t)$ evolves according to Markovian dynamics \cite{Uneyama2019}, Eq.~\eqref{eq: OUFD relaxation function phi with diffusivity}
can be formally solved using a transfer operator approach. The relaxation time
distribution can be directly obtained from  the eigenvalue spectrum of
the transfer operator. 
If the diffusivity can only take two values (as in the two-state model), the relaxation function is
expressed as the sum of two exponential decays with distinct 
relaxation times. In general, these relaxation times do not coincide with the
inverse transition rates. 
This discrepancy arises because relaxation in the OUFD model results from a combination of two effects: the linear restoring force and the stochastic switching between diffusivity states.
When the diffusivity follows the diffusing
diffusivity model in one dimension, the relaxation function involves infinitely many relaxation modes.
The  superposition of these modes leads to power-law relaxation
$\Phi(t) \propto t^{-1/2}$ in the short-time region. 
This type of relaxation behaviour is reminiscent of that observed in complex systems such as polymers (e.g., the Rouse model) and critical gels.

If $D(t)$ follows non-Markovian dynamics \cite{Miyaguchi2019}, the relaxation function exhibits
significantly more complex behaviour. For the non-Markovian two-state model with power-law
 sojourn time distributions discussed in Section~\ref{sec: two-state model}, the asymptotic
forms of the relaxation function can be analytically evaluated.
We consider the case where  $1 < \alpha_{+} \le 2$, so that the mean sojourn time
for the fast state exists. We also assume that the diffusive
motion is strongly suppressed in the slow state ($D_{-} \ll D_{+}$).
The long-time behaviour of $\Phi(t)$ depends on the value of
$\alpha_{-}$. If $1 < \alpha_{-} < 2$, the system reaches equilibrium. 
In this case, a power-law decay followed by exponential relaxation is observed in the long-time regime:
\begin{equation}
 \label{eq: OUFD relaxation function phi nonmarkov equilibrium}
 \Phi(t) \sim t^{1 - \alpha_{-}} \exp(-u D_{-} t).
\end{equation}
If $0 < \alpha_{-} < 1$, the mean sojourn time in the slow state diverges, rendering the system intrinsically non-equilibrium. In this case, the asymptotic relaxation function becomes
\begin{equation}
 \label{eq: OUFD relaxation function phi nonmarkov no mean sojourn time}
 \Phi(t) \sim \exp
  \left[  - u D_{-} t
  - \frac{\mu_{+} u (D_{+} - D_{-})}{a_{-} \Gamma(1 + \alpha_{-})} t^{\alpha_{-}} \right],
\end{equation}
where $a_{-}$ is  the coefficient from the small-$s$ expansion of the Laplace-transformed sojourn
time distribution, $\hat{\rho}_{-}(s) = 1 - a_{-} s + \dots$.
By setting $D_{-} = 0$ in Eq.~\eqref{eq: OUFD relaxation function phi nonmarkov no mean sojourn time},
one obtains a stretched exponential form:
$\Phi(t) \sim \exp[-(t / \bar{\tau})^{\alpha_{-}}]$, where $\bar{\tau}$ is a characteristic time constant. 
This stretched exponential relaxation has been observed in various glassy systems. Within the OUFD framework using the non-Markovian two-state model, such relaxation emerges when the particle remains immobilised for extended periods in the slow state, reflecting intrinsic non-equilibrium dynamics. The power-law exponent \( \alpha_{-} \) is directly connected to the tail of the sojourn time distribution. 
Thus, the OUFD model provides a valuable framework for analysing and interpreting complex relaxation phenomena in glassy systems.
{\color{black}These results demonstrate that the OUFD model serves as a versatile theoretical platform for studying relaxation dynamics in heterogeneous or glassy systems, bridging stochastic diffusivity models and experimentally observed non-exponential relaxation.}


\section{Some applications}

In this section, we explore how the theoretical framework of the LEFD has been applied to a range of experimental systems and molecular dynamics simulations. These examples demonstrate the versatility of the LEFD approach in capturing the rich and often non-intuitive behaviour of diffusion in heterogeneous environments.

We begin with the reptation model, where LEFD describes the motion of entangled polymer chains subject to topological constraints. We then examine glassy systems, highlighting how diffusivity fluctuations emerge from dynamic heterogeneity and result in long-time tails in correlation functions. Next, we consider binary gas mixtures and Lorenz gas models, which serve as minimal systems exhibiting fluctuating diffusivity even in the absence of complex internal degrees of freedom.

The LEFD framework is also shown to be effective in modelling more abstract and theoretical systems such as the Ornstein-Uhlenbeck process with fluctuating diffusivity, and in addressing practical problems like the diffusive search problem, where diffusivity fluctuations can influence search efficiency.

Biological relevance is highlighted through applications to protein diffusion, both in the ATTM and in crowded cellular environments, where molecular crowding and phase separation lead to subdiffusion and non-Gaussian dynamics. We also discuss the case of macromolecular diffusion in motile amoebae, where intracellular transport is influenced by both active motion and fluctuating diffusivity, and membrane systems, where phase heterogeneity dynamically regulates protein mobility.

Through these diverse examples, we illustrate how LEFD provides a unified and powerful theoretical tool for interpreting complex diffusion phenomena across physics, chemistry, and biology.

\subsection{Reptation model}

\label{sec: reptation model}

Polymers exhibit characteristic diffusion behaviour   arising from their internal conformational
degrees of freedom. A polymer molecule is a long, chain-like (or thread-like) entity that can be modelled as a series of beads connected linearly by springs.
Various theoretical models have been proposed to describe the dynamics of such polymer chains. 
When a polymer chain is dissolved in a solvent, its dynamics 
 can be described by a set of  Langevin equations for beads.
However,  the situation becomes significantly more complex when the polymer is embedded in a melt, where it is surrounded by many other chains. 
In this environment, the polymer cannot cross through other chains, resulting in kinetically constrained dynamics for a single, tagged polymer. This topological constraint is referred to as the entanglement effect, and gives rise to highly nontrivial transport behaviour.

\begin{figure}[htb]
 \begin{center}
  \includegraphics[width=0.8\linewidth]{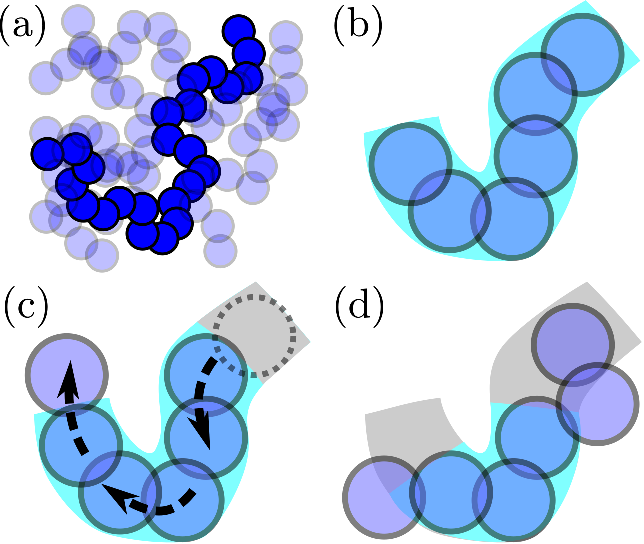}
 \end{center}
\caption{Schematic illustration of the reptation model. 
 (a) A single tagged chain (blue) and surrounding chains (light blue)
 in  a polymer melt. (b) The tagged chain is modelled as a chain
 confined within a tube. (c), (d) The tagged chain moves along its contour via reptation
 and escapes from the tube. Dark blue and light blue
 regions indicate the remaining and relaxed tube segments, respectively.
 \label{fig: reptation image}}
\end{figure}


The reptation model provides a theoretical framework for describing the dynamics of entangled polymers \cite{doi1978dynamics, Doi-Edwards-book}. 
Figure~\ref{fig: reptation image} shows a schematic illustration of this model.
Within the reptation picture, the motion of a tagged polymer chain is conceptualised as that of a chain diffusing through a tube-like constraint imposed by surrounding polymers. The dynamics of the chain's centre of mass can then be described using the LEFD framework, incorporating a time-dependent noise coefficient matrix as defined in Eq.~\eqref{eq: noise matrix in reptation}.
In this context, the instantaneous diffusion matrix ${\bm D}(t)$ becomes 
 \begin{equation}
{\bm D} (t)   = 3D_{\rm com}\frac{{\bm R}_e(t){\bm R}_e(t)}{\langle {\bm R}_e^2\rangle},
\label{eq: noise matrix in reptation model}
\end{equation} 
where $ {\bm R}_e(t)$  represents the end-to-end vector of the chain.
The MSD shows normal diffusion, and the effective diffusion coefficient $D_{\rm eff}$ is given by
\begin{equation}
D_{\rm eff} = \frac{1}{3} {\rm tr} \langle {\bm D} \rangle_{\rm st},
\end{equation} 
which evaluates to  $D_{\rm eff}=D_{\rm com}$. 

\begin{figure}[htb]
 \begin{center}
  \includegraphics[width=0.9\linewidth]{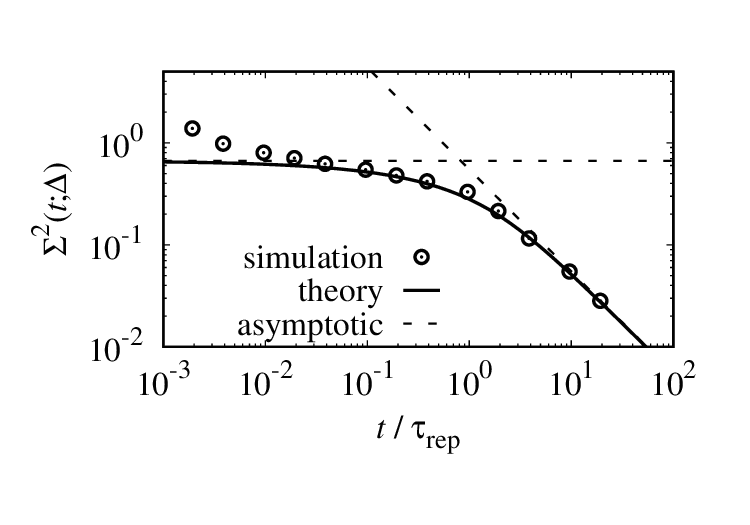}
 \end{center}
\caption{RSD of the reptation model. In numerical simulations, the number of tube segments is $Z = 80$ and the lag time is $\Delta=10\tau_{\rm rep}$, where
$\tau_{\rm rep}$  is the characteristic time of the longitudinal motion of a segment along the tube. Symbols represent kinetic Monte Carlo simulation data.
 The solid curve shows the full analytical solution, i.e., Eq.~(\ref{eq: reptation rsd}), while the dashed lines indicate its asymptotic forms.
 (Reprinted with permission from Ref.~\cite{Uneyama2015}.)}
 \label{fig: rsd reptation}
\end{figure}

To explore the RSD of the TSDs, one must compute the correlation function $\psi_1(t)$ of ${\bm D}(t)$, defined by 
$\psi_1(t) \equiv \langle  {\rm tr} {\bm D}(t) {\rm tr} {\bm D}(0) \rangle - ({\rm tr} \langle {\bm D} \rangle_{\rm st} )^2$. 
When applying the decoupling approximation, this correlation function simplifies to
\begin{equation}
\psi_1(t)  = \frac{ \langle   {\bm R}_e(t)  {\bm R}_e(0) \rangle_{\rm st}}{\langle {\bm R}_e^2 \rangle_{\rm st}^2} -1.
\end{equation} 
An analytical evaluation of the four-body two-time correlation function $\psi_1(t)$  has been derived in  in Ref.~\cite{Uneyama2015}, taking the form:
 \begin{equation}
\psi_1(t)  = \frac{16}{3\pi^2} \sum_{k: {\rm odd}} \frac{1}{k^2} E_2(k^2 t/\tau_d) ,
\end{equation} 
where $E_m(z)$ denotes the generalised exponential integral of the $m$-th order \cite{olver2010nist}, and 
$\tau_d$ denotes the disengagement time in the reptation model, characterising the longest relaxation time of the polymer chain. This timescale corresponds to the characteristic decorrelation time $\tau_D$ of the fluctuating diffusivity ${\bm D}(t)$; while these are distinct concepts---$\tau_d$ arising from chain dynamics and $\tau_D$ from mobility fluctuations---they are quantitatively related in this context.
At $t=0$, this function becomes 
 \begin{equation}
\psi_1(0)  = \frac{16}{3\pi^2} \sum_{k: {\rm odd}} \frac{1}{k^2} = \frac{2}{3}. 
\end{equation} 
The integral of $\psi_1(t) $ with respect to $t$ from 0 to $\infty$ evaluates to 
 \begin{equation}
\int_0^\infty \psi_1(t) dt = \frac{\pi^2 \tau_d}{36} .
\end{equation} 
Using Eq.~(\ref{eq: formula RSD}), the squared RSD is given by  
\begin{equation}
\Sigma^2(t; \Delta) =
\frac{\pi^2 \tau_d}{18t}
- \frac{\pi^4 \tau_d^2}{270t^2}
+ \frac{32 \tau_d^2}{3 \pi^2 t^2} 
 \sum_{k\,\text{odd}}  \frac{1}{k^6} E_4\left( \frac{k^2 t}{\tau_d} \right) .
 \label{eq: reptation rsd}
\end{equation}
The asymptotic form of the RSD becomes
\begin{equation}
\Sigma^2 (t; \Delta) \approx \left\{
\begin{array}{ll}
\dfrac{2}{3} &(t\ll \tau_d),\\
\\
\dfrac{\pi^2 \tau_d}{18t} 
&(t\gg \tau_d) .
\end{array}
\right.
\label{eq: asymptotic RSD reptation}
\end{equation}
For short times, the squared RSD exhibits plateau, reflecting persistent fluctuations in diffusivity. In the long-time regime, the squared RSD decays as $1/t$, 
considered with the fact that the coarse-grained dynamics resemble standard Brownian motion. 
Figure~\ref{fig: rsd reptation} shows 
the RSD of the reptation model as obtained from simulations, compared with the analytic prediction above.
At short times, the simulation data deviate slightly from the plateau. This deviation arises from intrinsic statistical fluctuations of the TSD in the Brownian motion, even when the diffusion coefficient remains constant.
The crossover time $\tau_c$ of the RSD is estimated as
\begin{equation}
\tau_c  = \dfrac{\pi^2 \tau_d}{12} \approx 0.822 \tau_d.
\end{equation} 
Therefore, $\tau_c$ is proportional and nearly equal to $\tau_d$. This relation between $\tau_c$ and $\tau_d$ was originally found in 
molecular simulations \cite{Uneyama2012}. The RSD analysis particularly useful for identifying the longest characteristic time of 
${\bm D}(t)$, in paticular cases where the dynamics of the instantaneous diffusivity are not known a priori. 

The deviation from the $1/t$ scaling of the squared RSD at short times indicates that the displacement PDF is non-Gaussian in this regime \cite{Uneyama2015}. This non-Gaussianity, despite a linear MSD, is a defining characteristic of BYNGD. Accordingly, the system exhibits BYNGD. Notably, similar non-Gaussian behaviours have also been reported in other polymer dynamics contexts, including self-avoiding walks, polymerization processes, and critical phenomena in polymer systems \cite{marcone2022brownian, baldovin2019polymerization,nampoothiri2022brownian,Nampoothir2021}.

\subsection{Glassy systems}
\label{sec: glassy systems}

Even in the absence of conformational degrees of freedom,
non-Gaussian diffusion behaviour and the fluctuating diffusivity can emerge.
In a normal liquid above the melting temperature, the motion of a molecule 
can be effectively described by the underdamped Langevin equation \cite{huang2011direct}.
On a short timescale, the MSD shows ballistic ($ \propto t^2$) behaviour, while at longer times, it transitions to normal diffusion.
As the temperature decreases below the melting point, the liquid enters a metastable supercooled state. If crystallisation is avoided through rapid cooling, the system instead evolves into a disordered solid known as a glass \cite{Kob1995,Berthier2011}. In supercooled liquids and glasses, the dynamics of individual molecules differ markedly from those in equilibrium liquids. One key feature is the emergence of dynamic heterogeneity---strong spatiotemporal correlations in particle mobility---observed in simulations and experiments
 \cite{Hurley1995, Kob1997,
Yamamoto-Onuki-1998,Yamamoto-Onuki-1998a,sillescu1999heterogeneity,weeks2000three,Doliwa2000,
Richert-2002,Berthier2011}.
Such heterogeneity arises even in systems composed of simple particles without internal structure. For instance, even spherical particles interacting via purely central force potentials can exhibit glassy dynamics~\cite{Kob1995}. Theoretical approaches such as mode-coupling theory \cite{gotze2009complex,gotze1999recent,Das2004,Reichman2005} predict a sharp slowing-down of relaxation due to the build-up of caging constraints and collective rearrangements. In contrast, dynamical facilitation theory \cite{Garrahan2002,garrahan2003coarse} models glassy dynamics as arising from sparse, kinetically constrained mobility excitations that propagate through facilitation rather than thermodynamic transitions.
A complementary perspective is provided by the random energy landscape framework \cite{bouchaud1992,debenedetti2001supercooled,Doliwa2003}, in which the system's configurational coordinates evolve within a complex potential landscape punctuated by energy traps of varying depths. In this view, the dynamics are governed by thermally activated hopping between traps, leading to broad distributions of waiting times and long-lived metastable states. 
Furthermore, diffusion on a heterogeneous energy landscape, designed to mimic the dynamics of atoms near the glass transition, has also been investigated in Refs. \cite{Odagaki1988,Odagaki1990}. 
More recent work has investigated diffusion in quenched random environments and demonstrated that the displacement PDF becomes non-Gaussian, while the local diffusivities exhibit pronounced trajectory-to-trajectory fluctuations \cite{Akimoto2016a, Akimoto2018,Luo2018,Luo2019}. These findings highlight the role of static disorder and rugged energy landscapes in producing fluctuating diffusivity and non-Gaussian transport statistics, consistent with the LEFD framework. 
These complementary approaches consistently show that glassy systems deviate from the assumptions of classical Brownian motion and necessitate a more nuanced description---such as that offered by the LEFD framework---especially when describing effective, time-dependent diffusivity and its statistical signatures.

\begin{figure}
 \begin{center}
  \includegraphics[width=0.7\linewidth]{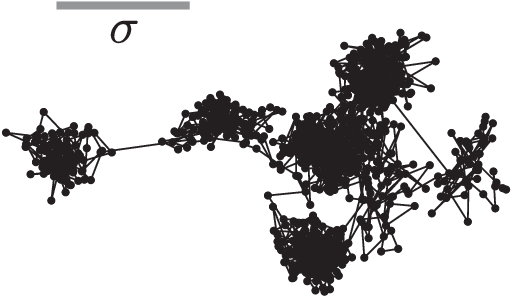}
 \end{center}
\caption{A typical trajectory of a single tagged particle in a
 supercooled binary Lennard-Jones mixture.
 The grey bar indicates the diameter of the tagged particle.
 \label{fig: binary lj glass trajectory}}
\end{figure}

Due to the dynamic heterogeneity, the dynamics of a single tagged molecule
in a supercooled liquid cannot  be accurately captured by a simple underdamped Langevin
equation. While spatial inhomogeneity may not be directly observable when observing a single tagged particle, 
temporal inhomogeneity can still be detected \cite{Hurley1995, Kob1997, Yamamoto-Onuki-1998,Yamamoto-Onuki-1998a,sillescu1999heterogeneity,
Richert-2002,Chaudhuri2007,Berthier2011}.
By tracking the trajectory of a tagged particle, it becomes clear that the particle exhibits varying diffusivity over time \cite{Richert-2002,pastore2015dynamic}. Some intervals display rapid motion, while others reflect a temporarily localised state. 
Figure~\ref{fig: binary lj glass trajectory} illustrates a typical trajectory of a
single tagged particle in a binary Lennard-Jones mixture within a supercooled liquid state.
The trajectory reveals that, in certain regions, the particle is effectively trapped and exhibits confined diffusion, whereas in others, it diffuses more freely.
These localised trapping regions arise from the cage effect \cite{Yamamoto-Onuki-1998,Yamamoto-Onuki-1998a,weeks2000three}, whereby neighbouring particles temporarily confine the tagged particle. 
This phenomenon can be modelled by the LETP
(see Section~\ref{sec: derivation of langevin equation with fluctuating diffusivity}).
The particle intermittently escapes from one cage and diffuses until it becomes trapped again in a newly formed one---reflecting the dynamic reconstruction of the trap landscape \cite{Hurley1995, Kob1997,Yamamoto-Onuki-1998,Yamamoto-Onuki-1998a}.
Such intermittent dynamics can be effectively modelled using two-state frameworks that alternate between a trapped state and a mobile diffusive state \cite{Fodor2016,hachiya2019unveiling}.
Numerical simulations have explicitly demonstrated the presence of fluctuating instantaneous diffusivity in supercooled liquids \cite{kaneko2018phase,hachiya2019unveiling}.
Over longer timescales, the MSD of the tagged particle exhibits normal diffusion, yet the displacement distribution remains non-Gaussian owing to the intermittent nature of the trapping and escape events \cite{Chaudhuri2007}.

A simple yet insightful model for glassy dynamics of a single tagged particle is an LEFD framework combined with a two-state switching process
 \cite{Uneyama2015}, as discussed in Section~\ref{sec: two-state model}. 
 In this model, the particle exists in one of two diffusive states---slow or fast---representing, for example, whether the particle is temporarily caged or free. This formulation provides a minimal yet physically meaningful representation of dynamic heterogeneity in glassy systems.
 In the slow state, the particle has low diffusivity, whereas in the fast state, its diffusivity is significantly higher. Transitions between these two states
 are governed by a stochastic process that is independent
of the particle position.
The dynamics of a single tagged particle can thus be described by LEFD with a time-dependent diffusion coefficient,
\begin{equation}
 \label{eq: diffusion coefficient two state model glass}
 D(t) = 
 \begin{cases}
  D_{+} & (\text{at the fast state}), \\
  D_{-} & (\text{at the slow state}),
 \end{cases}
\end{equation}
where $D_{+}$ and $D_{-}$ ($D_{+} > D_{-}$) are the diffusion coefficients
for the fast and slow states, respectively. 
In a supercooled liquid, molecular motion  typically exhibits strong correlations and memory effects. Accordingly, the stochastic process governing the switching between states is generally non-Markovian in nature.

Before turning to the general non-Markovian case, we first consider the simpler Markovian scenario. Here, the transition between the two states is modelled as a reversible first-order reaction, governed by rate equations. The statistical properties of the switching process are fully characterised by two transition rates:  $1/\mu_{-} $, representing the transition rate from the fast to the slow state, and  $1/\mu_{+} $, the rate from the slow to the fast state.
In this case, the correlation function $\psi_{1}(t)$ and the squared RSD $\Sigma^{2}(t;\Delta)$
are given by Eqs.~\eqref{eq: psi1 markovian two state} and \eqref{eq: rsd markovian two state}, respectively. The crossover time is estimated as  $\tau_{c} = 2 \tau $, where  $\tau$  is the characteristic switching time. This estimate is physically reasonable, since the fluctuations in diffusivity are directly tied to transitions between states. 
In MD simulations of glassy systems, the instantaneous diffusivity of a tagged particle is typically difficult to measure unambiguously. In contrast, the RSD analysis offers a robust means of accessing information about diffusivity fluctuations without requiring direct measurement of the instantaneous diffusivity. As such, RSD serves as a valuable tool for uncovering the non-Gaussian nature of dynamics in a wide range of glassy systems.

Here, we consider the non-Markovian case. 
In deeply supercooled systems, once a particle enters the slow state, it may remain trapped there for an extended period---potentially longer than the experimental timescale. In such regimes, some particles become effectively immobilised, and the mean sojourn times may grow very large or even diverge.
This suggests that
the sojourn time distributions may have power-law-type tails for  longer times: $\rho_{\pm}(\tau) \propto \tau^{-1 - \alpha_{\pm}}$ for large $\tau$.
If the mean sojourn times remain finite, the asymptotic behaviour of the RSD is similar to that in the Markovian case. The crossover time is then given by
\begin{equation}
 \tau_{c} = 
  \left(\frac{\langle \tau_{-}^{2} \rangle_{\text{eq}} - \langle \tau_{-} \rangle_{\text{eq}}^{2}}{\langle \tau_{-} \rangle_{\text{eq}}^{2}}
   + \frac{\langle \tau_{+}^{2} \rangle_{\text{eq}} - \langle \tau_{+} \rangle_{\text{eq}}^{2}}{\langle \tau_{+} \rangle_{\text{eq}}^{2}}
  \right) \mu,
\end{equation}
where $\tau_{\pm}$ represents the sojourn times at the fast and slow states,
respectively, and $\mu^{-1}= \langle \tau_{+} \rangle_{\text{eq}}^{-1} + \langle \tau_{-} \rangle_{\text{eq}}^{-1}$.
This expression indicates that  $\tau_c$  depends not only on the mean sojourn times but also on their fluctuations; i.e., the variances of the sojourn time distributions. In particular, large variances amplify $ \tau_c $, meaning that diffusivity fluctuations persist over longer timescales even when the mean sojourn times are finite.

If, however, the mean sojourn times diverge, the asymptotic behaviour of the RSD changes qualitatively. In this case, the correlation function of the instantaneous diffusivity exhibits a long-time tail, and the relaxation time associated with diffusivity fluctuations becomes ill-defined. Such extended relaxation reflects the strongly non-equilibrium nature of glassy systems, as observed in experiments and simulations \cite{debenedetti2001supercooled,Berthier2011}.
Importantly, when the mean sojourn time is infinite, the RSD no longer shows the typical Gaussian decay $\Sigma^{2}(t;\Delta) \propto t^{-1/2}$ \cite{Miyaguchi2016, Akimoto2016}. 
Instead, its decay becomes slower and highly sensitive to the specific properties of the sojourn time distributions and state switching dynamics. In some regimes, slower power-law or even logarithmic decay may emerge. 
For instance, in idealised cases discussed in Section~\ref{sec: two-state model}, the RSD may exhibit a scaling of the form
 $\Sigma^{2}(t;\Delta) \propto t^{1 - \alpha_{+}}$. 
Interestingly, such a scaling behaviour has also been reported in numerical simulations of supercooled liquids \cite{kaneko2018phase}, providing further evidence for the presence of fluctuating diffusivity dynamics in glassy systems.



\subsection{Binary gas mixture and Lorenz gas}
\label{sec: binary gas mixture and lorenz gas}

From the results discussed in Sections~\ref{sec: reptation model} and \ref{sec: glassy systems},
one might be led to believe that the fluctuating diffusivity can only arise in systems
with conformational degrees of freedom or spatial heterogeneity---such as polymers or supercooled liquids. 
However, in this subsection, we present a contrasting example where a fluctuating diffusivity emerges effectively in much simpler systems,
 including monoatomic gases without internal structure or spatial constraints \cite{Nakai-Masubuchi-Doi-Ishida-Uneyama-2023,Nakai-Uneyama-2023}.

\begin{figure}[htb]
 \begin{center}
  \includegraphics[width=0.8\linewidth]{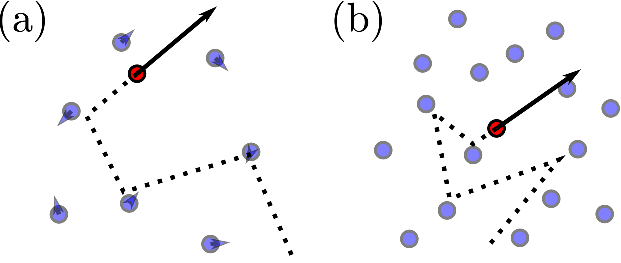}
 \end{center}
\caption{Schematic illustration  of a binary gas mixture and the Lorenz gas.
 Red and blue circles represent particles in gases $A$ and $B$, respectively.
 Arrows  indicate the velocity and dotted line segments represent the trajectory of a particle in $A$.
 (a) A binary gas mixture comprising a single $A$ particle
 and multiple $B$ particles.
 (b) The Lorentz gas. This system resembles the binary gas mixture, but with 
 $B$ particles fixed in space, acting as immobile obstacles. Only the
 particle in $A$ is mobile in this configuration. 
 \label{fig: binary gas mixture and lorenz gas image}}
\end{figure}

Here, we consider a mixture of two monoatomic gases, denoted as gases  $A$ and $B$.
Gases $A$ and $B$ consist of particles with different sizes and masses, denoted by $\sigma_{A}$ and $\sigma_{B}$ for size, and $m_{A}$ and $m_{B}$ for mass, respectively.
Assuming gas $A$ is highly dilute, we effectively follow the motion of a single $A$ particle moving through a background of $B$ particles, as illustrated in
 Fig.~\ref{fig: binary gas mixture and lorenz gas image}(a).
When tracking the position of a particle in gas $A$, we observe diffusive behaviour over long timescales. 
 Intuitively, we expect that a particle in gas $A$ moves ballistically at short timescales. 
 After many collisions with the more abundant $B$ particles, its momentum becomes randomised,
resulting in normal, Gaussian diffusion at long times.
This crossover from ballistic to diffusive behaviour is well-described by kinetic theory (e.g., Enskog theory) and is not in itself surprising.

\begin{figure}[htb]
 \begin{center}
  \includegraphics[width=0.8\linewidth]{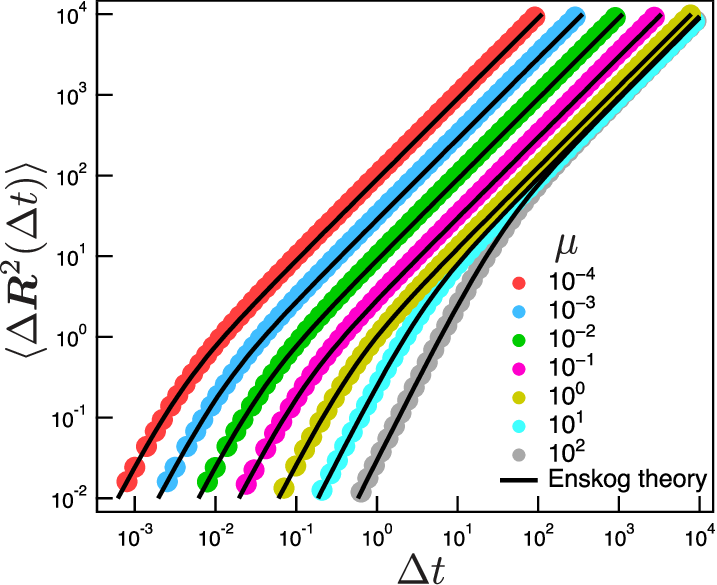}
 \end{center}
\caption{Mean squared displacement of the $A$ particle in the binary gas mixture
 for different mass ratios $\chi$.
 Symbols denote simulation results, and solid curves indicate predictions based on
  Enskog theory.
 (Reprinted with permission from Ref.~\cite{Nakai-Masubuchi-Doi-Ishida-Uneyama-2023}.)
 \label{fig: msd binary gas}}
\end{figure}

\begin{figure}[htb]
 \begin{center}
  \includegraphics[width=0.8\linewidth]{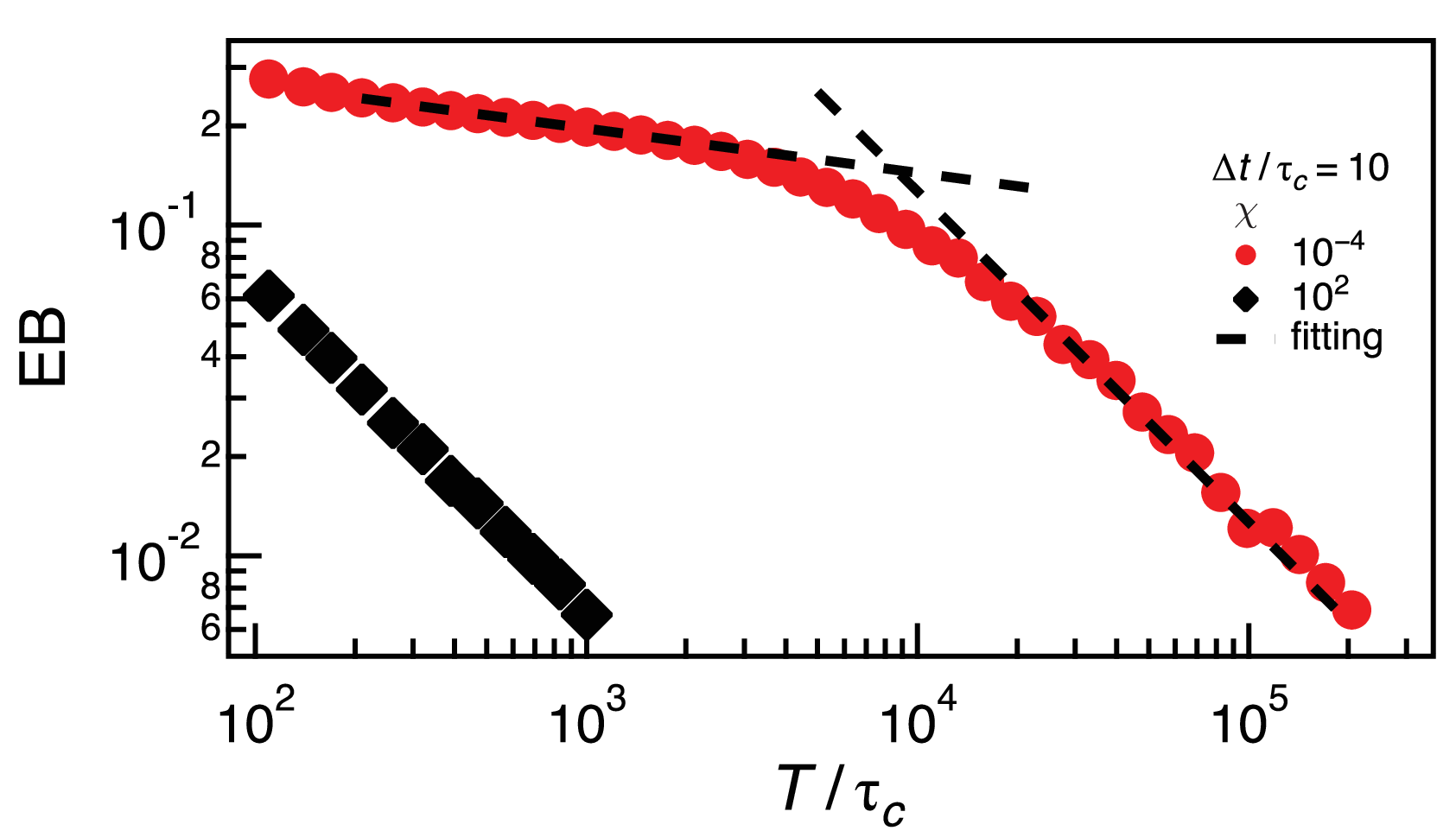}
 \end{center}
\caption{Squared RSD of  the TSDs---equivalent to the EB parameter---for the $A$ particle in the binary gas mixture
 for two mass ratios, $\chi = 10^{2}$ and $10^{-4}$.
 Dashed lines indicate fits to the power-law form 
 $\Sigma^{2}(t;\Delta) \propto t^{-\beta}$. At short measurement times $T \ll t_c$, the Lorenz gas model predicts $\Sigma^2(t;\Delta) \approx
    0.178$, as given by Eq.~(\ref{eq: rsd lorenz gas short time limit}). Note that $T$ here refers to the measurement time, not temperature.
 (Reprinted with permission from Ref.~\cite{Nakai-Masubuchi-Doi-Ishida-Uneyama-2023}.)
 \label{fig: rsd binary gas}}
\end{figure}

Recent simulations~\cite{Nakai-Masubuchi-Doi-Ishida-Uneyama-2023} have shown, however, that the rate and character of this crossover depend sensitively on the mass ratio $\chi = m_A / m_B$. For example, the MSD of the $A$ particle exhibits an explicit dependence on $\chi$.
Figure~\ref{fig: msd binary gas} shows the MSD for various mass ratios, which are accurately predicted 
 by the gas kinetic theory---specifically, Enskog theory \cite{Chapma-Cowling-book}.
For large $\chi$ (i.e., a heavy $A$ particle), the  motion of the $A$ particle
follows the underdamped Langevin equation. 
 For small values of $\chi$, the MSDs in the long-time regime do not collapse onto a single curve. This indicates that both the effective diffusion coefficient and the corresponding friction coefficient depend on $\chi$. The emergence of a $\chi$-dependent diffusion coefficient reflects a qualitative change in the underlying collision dynamics. Specifically, for small $\chi$, a single collision can significantly alter the direction of the A particle. This pronounced reorientation leads to a fundamental change in the nature of diffusion. Even in the regime where the MSD grows linearly with time---indicative of diffusive behavior---the diffusion coefficient itself exhibits temporal fluctuations, reflecting persistent dynamical heterogeneity. 
Figure~\ref{fig: rsd binary gas} displays the RSD of the $A$ particle for
two characteristic cases: $\chi = 10^{2}$ and $10^{-4}$. For $\chi = 10^{-4}$, the RSD 
$\Sigma^{2}(t; \Delta)$ exhibits a slow power-law  decay at 
short times: $\Sigma^{2}(t; \Delta) \propto t^{-\beta}$ with $\beta < 1$.
The characteristic crossover time $\tau_{c}$ exceeds the momentum relaxation time---that is, the time marking the transition from ballistic to diffusive motion, suggestive of persistent effective diffusivity fluctuations beyond the kinetic scale.

We stress that this system does not represent a canonical example of LEFD, where a stochastic $D(t)$ is explicitly defined. Rather, the fluctuating diffusivity in this case emerges, due to a separation in the relaxation timescales of the particle's velocity magnitude and direction. For $\chi \ll 1$, the $A$ particle's direction decorrelates rapidly upon collision with heavier $B$ particles, while its speed $|\bm{v}_A(t)|$ retains memory. Because the instantaneous diffusivity is effectively proportional to speed ($D(t) \propto |\bm{v}_A(t)|$) \cite{Nakai-Masubuchi-Doi-Ishida-Uneyama-2023, Nakai-Uneyama-2023, dorfman2021contemporary}, these speed fluctuations manifest as a time-dependent diffusivity. The crossover timescale seen in the RSD corresponds well to the speed relaxation time.
This mechanism is qualitatively different from those in Sections~\ref{sec: reptation model} and~\ref{sec: glassy systems}, where diffusivity fluctuations arise from internal or collective structural dynamics. Here, fluctuations emerge from single-particle kinetics—a form of emergent time-dependent diffusivity.

In the limit $\chi \to 0$, $B$ particles act as immobile obstacles and the system reduces to the Lorentz gas
 [Fig.~\ref{fig: binary gas mixture and lorenz gas image}(b)].
In the Lorenz gas model, the evolution of the particle's velocity can be described as a point process, 
 where collisions with the fixed  $B$  particles are treated as stochastic events \cite{Nakai-Uneyama-2023}. When gas $B$ is dilute, 
 spatial and temporal correlations between collisions become negligible,
 simplifying the dynamics 
into a Markovian point process. This approach enables the analytical calculation of various statistical quantities for the motion of the $A$ particle.
Despite the randomness of scattering, the motion becomes fully Gaussian after velocity decorrelation. This is consistent with standard diffusion theory, as the speed remains constant, and hence, the diffusivity does not fluctuate.

This distinction is key: in the binary gas system with $\chi \ll 1$, the apparent non-Gaussianity and RSD behaviour result from the slow relaxation of particle speeds. In contrast, in the Lorentz gas model, the particle's speed remains constant. The speed relaxation time diverges as $\chi \to 0$, effectively eliminating speed relaxation. Thus, the LEFD framework becomes applicable only when the speed fluctuates, i.e., over longer timescales where an effective diffusivity emerges.
In this regime,  the binary gas system with sufficiently
small $\chi$ can be interpreted as an superstatistical ensemble of
Lorentz gas molecules, each  averaged over a distribution of particle speed. 
Non-Gaussian diffusion behaviour emerges when observing such ensemble-averaged quantities.
In the long-time limit, the correlation function $\psi_{1}(t)$ can be approximated as
\begin{equation}
 \psi_{1}(t) 
  \approx \frac{\langle |\bm{v}_{A}|^{2} \rangle_{\text{eq}}}{\langle |\bm{v}_{A}| \rangle_{\text{eq}}^{2}} - 1
  = \frac{3 \pi}{8} - 1,
\end{equation}
which leads to the following expression for the RSD:
\begin{equation}
 \label{eq: rsd lorenz gas short time limit}
 \Sigma^{2}(t;\Delta) \approx \frac{3 \pi}{8} - 1 \approx 0.178.
\end{equation}
(The non-Gaussian parameter also takes the same value in the long-time limit: $A(t) \approx 3 \pi / 8 - 1$.)
The RSD data for $\chi = 10^{-4}$ shown in Fig.~\ref{fig: rsd binary gas}
are consistent with the prediction from Eq.~\eqref{eq: rsd lorenz gas short time limit}.
These results suggest that statistical features resembling LEFD behaviour can emerge even in minimal systems, provided there exists sufficient separation of relaxation timescales. While we do not claim that these systems obey LEFD in the strict sense, they illustrate how effective diffusivity fluctuations may arise from deterministic dynamics, broadening the contexts in which LEFD-type models may serve as useful approximations.

\subsection{Diffusive search problem}

Determining the time it takes for a diffusing particle to reach a target is a fundamental question at both microscopic scales---such as in molecular and cellular biology---and macroscopic scales  \cite{Benichou2011}. In this context, the focus is placed on how fluctuating diffusivity influences the first-passage time (FPT) for diffusing particles to locate a target. 

Analytical expressions for the mean first-passage time (MFPT) have been derived under conditions in which both the searcher and the targets—located either at the boundary or within the domain—undergo diffusion with fluctuating diffusivities \cite{lawley2019diffusive}. 
In the model, the diffusivities of both the searcher and targets are randomly fluctuating, and the boundary and interior targets 
exhibit stochastic gating behaviour, transitioning between absorbing and reflecting the searcher. 
A target that absorbs the searcher is referred to as ``open," while one that reflects the searcher is termed ``closed." 
Diffusive and gate states are described by an irreducible continuous-time Markov jump process on the finite state space. 
In this general setting, the mean first-passage time (MFPT) was analytically derived. The study examined how the correlation between the diffusivity of the searcher and that of the target affects the MFPT. In particular, a scenario was considered in which the boundary target diffusivity randomly switches between slow $D^-$ and fast $D^+$ states, while the searcher diffusivity alternates between $D_0^-$ and $D_0^+$.
The MFPT is minimised when diffusivities of both the searcher and the target are positively correlated---that is,  the searcher's diffusivity 
is $D_0^-$ or $D_0^+$ when the target diffusivity is  $D^-$ or $D^+$, respectively. 
In cases where all the targets are immobile ($D=0$), and always open, the MFPT can be characterised by the average searcher 
diffusivity. Moreover, when the interior targets diffuse with fluctuating diffusivity, the MFPT is also determined by the average diffusivities
 of the searcher and the interior targets. However, in situations where the boundary targets diffuse with fluctuating diffusivity, the MFPT 
 cannot be simply calculated by substituting the fluctuating boundary target diffusivity with the average diffusivity. 
 Therefore, the effects of fluctuating diffusivities become non-trivial and the minimization of the MFPT is achieved through a non-trivial structure in the jump process governing the diffusive states. 

Redundancy plays a crucial role in biological reactions, particularly in cases where out of a group of $N$ particles, only one particle is capable of contributing to a diffusion-limited reaction. 
In such cases, the particle with the shortest FPT initiates the reaction. A classic example is human reproduction, in which only a single sperm cell---out of approximately 100 million---is successful in locating and fertilising the egg.

\begin{figure}[]
\begin{center}
\includegraphics[width=0.8\linewidth]{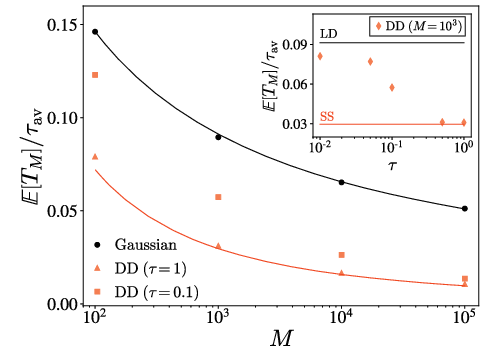}
\end{center}
\caption{Extreme MFPT $E(T_M)/ \tau_{\mathrm{av}}$ as a function of the number of searchers $M$.
The results compare the diffusing-diffusivity  model (orange, or gray in grayscale) with standard Brownian motion (black). The ratio $E(T_M) / \tau_{\mathrm{av}}$ is independent of both the target position and the average diffusivity. Solid lines represent theoretical predictions, while symbols denote results from numerical simulations performed with $d = 1$. Inset: Crossover between the superstatistics and large deviation regimes is shown by varying the diffusivity correlation time $\tau$ at fixed $M$, illustrating the influence of diffusivity dynamics on search efficiency.
}
\label{fig: sposini2024}
\end{figure}

The effect of fluctuating diffusivity on extreme FPT statistics---specifically, the behaviour of the fastest searcher---has been analysed using the diffusing-diffusivity model \cite{Sposini2024PRL, *Sposini2024PRE}. This analysis revealed that the mean of the fastest FPT, also known as the extreme MFPT, is reduced in comparison to standard Brownian motion.
 Specifically, for $M$ non-interacting particles searching for a target, the ratio of the extreme MFPT 
 $E(T_M)$ to the MFPT $\tau_{\mathrm{av}}$ scales as $E(T_M)/\tau_{\mathrm{av}} \propto 1/(\ln M)^2$ in the diffusing-diffusivity model, 
whereas for standard Brownian motion, it follows  $E(T_M)/\tau_{\mathrm{av}} \propto 1/(\ln M)$. Consequently, fluctuating diffusivities lead to a reduction 
in the extreme MFPT in systems with multiple searching particles. 
Figure~\ref{fig: sposini2024} shows how the MFPT of the fastest searcher, normalised by the MFPT  $\tau_{\mathrm{av}}$, varies as a function of the number $M$ of searchers. The diffusing-diffusivity model consistently predicts shorter extreme MFPTs as compared to standard Brownian motion, especially in regimes where the diffusivity exhibits strong temporal correlations. Notably, this acceleration of the search process arises despite the average diffusivity being fixed, highlighting how temporal fluctuations in diffusivity can optimise search efficiency. The inset further reveals a crossover between two dynamical regimes---superstatistics and large deviation---as the correlation time $\tau$ is varied, emphasising the nontrivial impact of diffusivity dynamics on extreme statistics.

Such a reduction is not always observed in general search processes.
For a single searcher following a non-Gaussian diffusion process like the LEFD, 
the characteristic time to reach the target is generally longer than in standard Brownian motion \cite{lanoiselee2018diffusion, sposini2018first,  lanoiselee2019non}. 
This implies that fluctuating diffusivity can be either advantageous or disadvantageous, depending on the context.
In particular, the presence of a higher probability for a few fast-moving particles in a non-Gaussian distribution plays a key role in reducing the extreme MFPT---
that is, the time for the fastest among many searchers to reach the target.
This effect becomes especially beneficial in redundant search scenarios, where only a few successful binding events are needed (e.g., fertilisation or rare signal transduction).
Conversely, fluctuating diffusivity can be disadvantageous when a large fraction of searchers must succeed, such as in biological systems requiring a high level of receptor activation.

\subsection{ Protein diffusion (ATTM) }

Extensive SPT experiments on DC-SIGN---a receptor with unique pathogen-recognition capabilities---diffusing on living-cell membranes have revealed that fluctuating diffusivity plays a crucial role in generating anomalous features such as subdiffusion and weak ergodicity breaking \cite{Manzo2015}. Here, we explore the relevance of the ATTM, which is described by the LEFD with 
instantaneous diffusivity being coupled with sojourn time,  to the experimental observations. As detailed in Section~\ref{sec: ATTM}, 
the ATTM is characterised by the LEFD with instantaneous diffusivity $D(t)$ given by $D(t)=\tau_t^{\sigma -1}$ \cite{Massignan2014}, where 
$\tau_t$ is the sojourn time straddling time $t$. Sojourn times are randomly distributed according to a PDF denoted by $\rho (\tau)$. 
When the sojourn-time PDF follows a power law, i.e., $\rho(\tau) \propto \tau^{-1-\alpha}$, as shown in Section~\ref{sec: ATTM}, 
the MSD exhibits subdiffusion, ageing of the TSD, and trajectory-to-trajectory fluctuations of the TSDs. These anomalous features 
are in quantitative agreement with those in the CTRW when $\alpha > \sigma$. 

A notable distinction between the two models lies in the concept of a trapped state. In the ATTM, particles never come to a complete stop due to Brownian motion with a non-zero diffusion coefficient. Conversely, in the CTRW model, particles can come to a halt during a trapped state.
In the SPT experiments, a localisation of a particle was observed. A detailed analysis suggests that durations of these trappings 
are distributed according to a power law with exponent $\alpha \approx 0.83$, suggesting  $\rho(\tau) \propto \tau^{-1-\alpha}$ for 
$\tau\to\infty$ because the localisation indicates a lower diffusion coefficient 
such that $D<D_{\rm th} = 6\times 10^{-4}$ $\mu$m${}^2$s${}^{-1}$ \cite{Manzo2015}. 
The CTRW with a waiting-time PDF following $\rho(\tau) \propto \tau^{-1-\alpha}$ for $\tau\to\infty$ yields 
$\langle | {\bm r}(t) - {\bm r}(0) |^2 \rangle \propto t^\alpha$ for $t\to\infty$, which is consistent with the experimental  MSD result. 
However, such a localisation rarely occurs, indicating that single-particle trajectories obtained by the SPT experiments do not 
follow the CTRW model. To confirm whether the CTRW describes the diffusion process, they calculated the MSD with SPT data 
excluding localisation trajectories. The excluded SPT data showed the MSD still exhibits subdiffusion with exponent $\alpha \approx 0.84$, 
implying that the CTRW fails to explain the subdiffusion observed in the experiments. 

\begin{figure}[]
\begin{center}
\includegraphics[height=10.5cm]{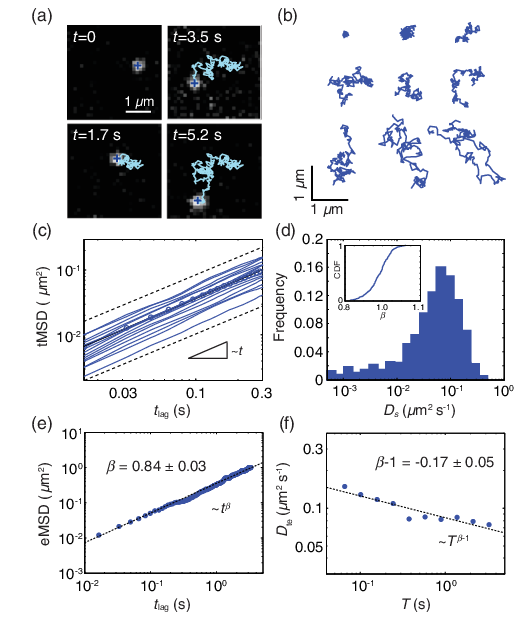}
\end{center}
\caption{DC-SIGN diffusion reveals weak ergodicity breaking and aging \cite{Manzo2015}.
(a) Representative video frames showing a quantum dot-labelled DC-SIGN molecule diffusing on the dorsal membrane of a CHO cell. The centroid of the fluorescent spot ($+$), corresponding to the quantum dot, is tracked and reconstructed into a trajectory (cyan line).
(b) Sample trajectories recorded over the same observation time (3.2 s), highlighting trajectory-to-trajectory variability.
(c) Log-log plot of the TSD for individual trajectories (blue lines). Dashed lines indicate linear scaling with time, consistent with Brownian motion ($\beta=1$). The averaged TSD (circles) is fitted by a power law (black line), yielding $\beta \approx 0.95 \pm 0.05$.
(d) Distribution of short-time diffusion coefficients extracted from linear fits of the TSDs. Inset: Cumulative distribution function (CDF) of the scaling exponent $\beta$ obtained from nonlinear fits of individual trajectories.
(e) Log-log plot of the  MSD, showing subdiffusive scaling with exponent  $\beta\approx 0.84$ (dashed line).
(f) Log-log plot of the ensemble-averaged diffusion coefficients obtained by the TSDs versus observation time $T$. A power-law fit yields an exponent of -0.17, indicating aging dynamics consistent with the subdiffusive behaviour observed in (e).
Reproduced from Ref.~\cite{Manzo2015}, licensed under CC BY 4.0.
}
\label{fig: manzo2015}
\end{figure}

As shown in Section~\ref{sec: ATTM}, the MSD increases as $\langle | {\bm r}(t) - {\bm r}(0) |^2 \rangle\propto t^\alpha$ in the ATTM with  $\alpha > \sigma$. 
The MSD in the SPT data indicates that $\langle | {\bm r}(t) - {\bm r}(0) |^2 \rangle \propto t^{0.84}$, suggesting  $\alpha \approx 0.84$ in the ATTM (see Fig.~\ref{fig: manzo2015}). 
Additionally, the ensemble average of the TSD demonstrates ageing behaviour, 
i.e., $\langle\overline{\delta^2 (\Delta; t)} \rangle \propto t^{-0.17}$, where $t$ represents the 
measurement time (see Fig.~\ref{fig: manzo2015}). The ATTM theory predicts $\langle\overline{\delta^2 (\Delta; t)} \rangle \propto t^{-1+\alpha}$ when $\alpha > \sigma$. 
Therefore, parameter $\alpha$ estimated from the ensemble average of the TSD is consistent with that determined from the MSD. Furthermore, 
a comparison between experimental observations and numerical simulations of the 2D ATTM suggests that parameter $\sigma$ in the ATTM 
is approximately $\sigma \approx 0.275$, implying $\alpha > \sigma$. 
Therefore, the ATTM effectively elucidates the experimental findings, indicating that 
fluctuating diffusivity plays a crucial role in anomalous diffusion of DC-SIGN on living-cell membranes

\subsection{ Protein diffusion in crowded environments }

Biological membranes are highly crowded environments in which the embedded proteins significantly impact the membrane dynamics, including the diffusion of lipid molecules and proteins. A number of experimental and computational studies have reported anomalous lateral diffusion of phospholipids and membrane proteins, particularly, at the protein-rich condition~\cite{horton2010development, Weigel2011, javanainen2013anomalous, goose2013reduced}. However, the precise stochastic mechanisms governing such diffusion remain poorly understood.

To provide new insights into membrane dynamics, extensive molecular dynamics simulations combined with stochastic modelling have been employed to elucidate how protein crowding affects the lateral diffusion of lipids and proteins on membranes~\cite{Jeon2016}. Their study revealed that in non-crowded bilayers, lipid diffusion follows a Gaussian process described by fractional Langevin equation (FLE) in which the memory kernel in the generalised Langevin equation is of a power-law form. However, in protein-crowded membranes, such Gaussian homogeneous diffusion dynamics are disrupted by the presence of proteins. Instead, the diffusion dynamics of lipids become multifractal, non-Gaussian, and spatial-temporally heterogeneous.

\begin{figure}[]
\begin{center}
\includegraphics[height=4.5cm]{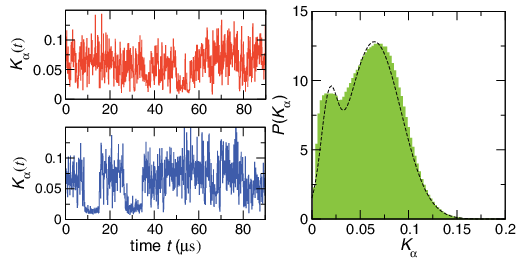}
\end{center}
\caption{Temporal fluctuations of the local diffusivity $K_{\alpha}(t)$ for two representative DPPC lipid molecules in a protein-crowded membrane \cite{Jeon2016}. The diffusivity $K_{\alpha}$ is expressed in units of $\mu\text{m}^2/\text{ns}^{\alpha}$, with a temporal resolution of 100 ns. (Right): PDF $P(K_{\alpha})$ of diffusivity $K_{\alpha}(t)$ constructed from the instantaneous diffusivity values of all DPPC lipid molecules. The dashed line indicates a double-Gaussian fit to the distribution, highlighting the presence of distinct dynamical states. 
Reproduced from Ref.~\cite{Jeon2016}, licensed under CC BY 4.0.
}
\label{fig: protein-crowding}
\end{figure}

\begin{figure}[]
\begin{center}
\includegraphics[height=7cm]{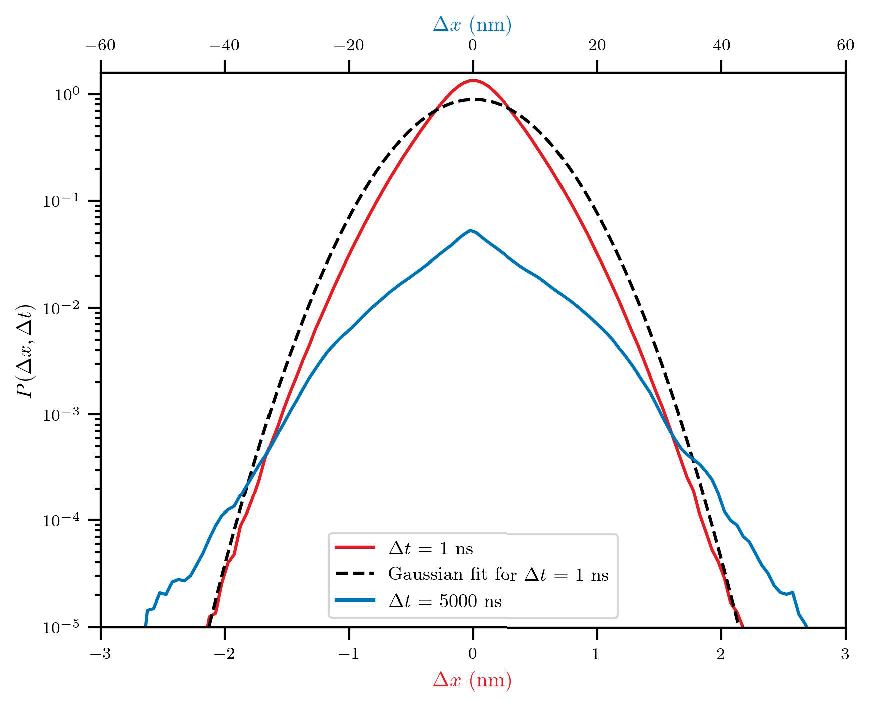}
\end{center}
\caption{Displacement probability density functions (propagators) of DPPC lipid molecules in a protein-crowded membrane.
The red and blue lines shows the PDFs of displacements for two lag times, $\Delta = 1 \, \text{ns}$ and 5000 $\, \text{ns}$, respectively. The dashed line indicates a Gaussian fit  to the data of $\Delta t=1$~ns.
}
\label{fig: protein non-Gaussian}
\end{figure}

These findings highlight the critical role of protein crowding in shaping lateral membrane diffusion and suggest that homogeneous stochastic models such as FLE are inadequate for describing lipid and protein motion in crowded environments. The high concentration of membrane proteins and their slow lateral dynamics introduce pronounced spatiotemporal heterogeneity in the local membrane environment. As a result, lipid diffusion exhibits a fluctuating diffusivity and non-Gaussian displacement statistics, as illustrated in Figs.~\ref{fig: protein-crowding} and \ref{fig: protein non-Gaussian}.  The observed stochastic dynamics of a single protein in such a spatiotemporally heterogeneous medium is in line with the theoretical prediction of the LEFD, demonstrating the importance of stochastic approaches in modelling complex molecular transports within a membrane. This study underscores the necessity of incorporating fluctuating diffusivity models into the theoretical framework of lateral macromolecular diffusions. It represents a significant step towards quantitatively describing diffusion-controlled reactions in biological membranes, with broad implications for cellular signalling, trafficking, and drug targeting.

\subsection{Macromolecular diffusion in motile amoebae}

\begin{figure*}[]
\begin{center}
\includegraphics[height=4.5cm]{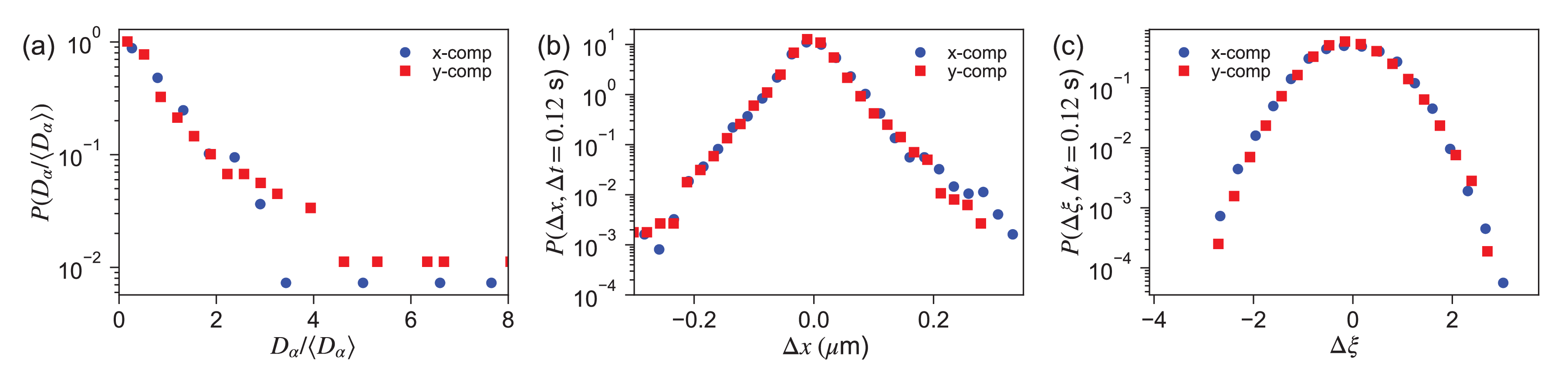}
\end{center}
\caption{Heterogeneous superdiffusion of intracellular particles in {\it Acanthamoeba castellanii}. (a)  Distribution of generalised diffusivities $D_\alpha$, extracted from the TSD of individual trajectories. 
(b) The total propagator  $P(\Delta x,\Delta t)=\langle \delta(\Delta x-[x_i(t+\Delta t)-x_i(t)])\rangle_t$, calculated over all particle trajectories at a given lag time $\Delta t=0.12$~s. 
(c) Rescaled propagator from (b), obtained by dividing each trajectory into shorter segments of equal duration (``chopped trajectories") and plotting displacements against the dimensionless variable $\xi=x/\sqrt{ \overline{\delta^2(\Delta t; t)}}$. Here, the chopped trajectories have the same length of 200 data points (i.e., $t=2.4$~s).
}
\label{fig:amoeba}
\end{figure*}

Macromolecular transport is essential for numerous biological processes within living cells. The intracellular environment is densely packed with biomolecules of varying sizes, while networks of cytoskeletal filaments introduce viscoelastic memory effects and mechanical caging. Under such conditions, intracellular particles frequently display subdiffusive motion \cite{Golding2006, wirtz2009particle, Jeon2011, Tabei2013}. However, active processes, such as ATP-driven molecular motors and cytoplasmic streaming, can induce superdiffusive transport. Notably, such superdiffusion appears to dominate in the cytoplasm of motile cells~\cite{Reverey2015,Krapf2019,Samu2019}. 

A prominent example of such behaviour was reported by Reverey et al.~\cite{Reverey2015}, who employed single-particle tracking techniques to analyse the motion of intracellular particles in {\it Acanthamoeba castellanii}, a motile unicellular organism. The tracked particles---comprising vesicles and granules ranging from several hundred nanometres to a few micrometres---exhibited collective motion and clear superdiffusive behaviour, with an ensemble-averaged anomalous exponent of $\alpha \approx 1.79   \pm0.12$, in stark contrast to the typical subdiffusive dynamics observed in immotile cells.
Disruption of actin filaments or microtubules led to a partial suppression of the superdiffusive motion, with the anomalous exponent decreasing to $\alpha \approx 1.5-1.6$. However, this suppression was not complete, suggesting the involvement of additional active components. Subsequent experiments identified myosin II as a major contributor to the observed transport dynamics. Inhibition of myosin II activity using blebbistatin rendered the intracellular particles nearly immobile, with the anomalous exponent dropping to $\alpha \approx 0.15$.

Further analysis revealed that the observed superdiffusion exhibited significant heterogeneity. This was quantified by examining the distribution of generalised diffusivities $D_\alpha$, extracted from the TSDs of individual trajectories via the relation $\overline{\delta^2 (\Delta; t)} = D_\alpha \Delta^\alpha$, where $\Delta$ is the lag time in units of the time resolution 0.012~s and the measurement times  $t$  vary across experimental realizations. The resulting distribution of $D_\alpha$ followed an exponentially decaying profile [Fig.~\ref{fig:amoeba}(a)], reflecting broad variability in local particle dynamics. In addition, the total propagator (or the Van Hove self-correlation function) $P(\Delta x, \Delta t)$, computed across all particle trajectories at a given lag time $\Delta t$, displayed a Laplace shape with a sharp central peak [Fig.~\ref{fig:amoeba}(b)], further highlighting non-Gaussian displacement statistics.

To investigate the origin of the observed non-Gaussian propagator, Reverey et al. applied a trajectory segmentation method, often referred to as ``trajectory chopping" \cite{Reverey2015}. In this approach, each long particle trajectory is divided into shorter segments of equal duration, and the displacement statistics are evaluated within these segments. This procedure allows one to probe the short-time, local dynamics of individual particles. For each segment $\Delta t$, the displacement distribution is rescaled by the square root of the TSD, yielding a dimensionless variable  $\xi = x / \sqrt{ \overline{ \delta^2(\Delta t; t)} } $. The resulting rescaled propagators collapse onto a Gaussian profile, as shown in Fig.~\ref{fig:amoeba}(c), indicating that the local dynamics within each segment are Gaussian. Thus, the non-Gaussian shape of the full propagator arises not from intrinsically non-Gaussian noise, but from trajectory-to-trajectory variations in diffusivity---a hallmark of superstatistical behaviour. 
This dynamic heterogeneity is attributed to two key factors: (i) spatial variability in the intracellular environment due to cytosolic supercrowding and (ii) intrinsic polydispersity in the tracked particles. Under these conditions, intracellular particles become mobile due to the activity of myosin II motors but experience transient caging and viscoelastic memory effects from surrounding structures. Therefore, the heterogeneous intracellular superdiffusion observed in motile {\it Acanthamoeba} cells can be interpreted within the superstatistical framework: a mixture of local Gaussian superdiffusive dynamics with particle-to-particle variability in diffusivity.


\subsection{Diffusion in phase-separated heterogeneous membranes }

Biological membranes are composed of various proteins and lipids, whose molecular composition governs intricate patterns of phase separation \cite{fan2010formation, levental2011raft, sezgin2017mystery}. In particular, mixtures of saturated and unsaturated lipids promote phase separation into liquid-ordered (Lo) and liquid-disordered (Ld) domains \cite{pike2006rafts, heberle2010comparison, de2015dynamic}, which play a critical role in regulating protein diffusion, partitioning, and localisation. These domains serve as functional platforms for essential cellular processes such as signalling, trafficking, and membrane organisation. Despite extensive experimental investigations, the molecular mechanisms underlying how membrane heterogeneity influences protein mobility and dynamic interactions remain poorly understood.

Recent experimental and computational studies have provided new insights into how phase-separated biological membranes regulate molecular diffusion. SPT experiments have enabled direct visualisation of nanoscopic structures in reconstituted supported bilayer membranes~\cite{WuLinYenHsieh2016}.
Wu et al. utilised iSCAT-based SPT with 20 nm gold nanoparticles (GNPs) as labels to monitor the lateral diffusion of single lipids with nanometer spatial precision and microsecond temporal resolution \cite{WuLinYenHsieh2016}.
Their study revealed that in raft-containing membranes, the Lo domains manifest as distinct nanoscopic regions that transiently trap lipid molecules. Specifically, while lipids diffuse freely and exhibit nearly Brownian motion in the liquid-disordered (Ld) phase, those in the Lo domains display subdiffusion at the microsecond timescale. This behaviour is attributed to transient confinements within nano-subdomains of the Lo phase.
Furthermore, an MD simulation study~\cite{KumarDaschakraborty2023} reported that lipid diffusion during the fluid/gel phase transition exhibits highly non-Gaussian over long timescales, indicating that phase coexistence is sufficient to induce dynamic heterogeneity.
Similar diffusivity fluctuations have been reported in raft-mimetic lipid bilayers, where Brownian yet non-Gaussian diffusion emerges due to nanoscale phase separation \cite{ErimbanDaschakraborty2023}.

\begin{figure}[]
\begin{center}
\includegraphics[width=0.8\linewidth]{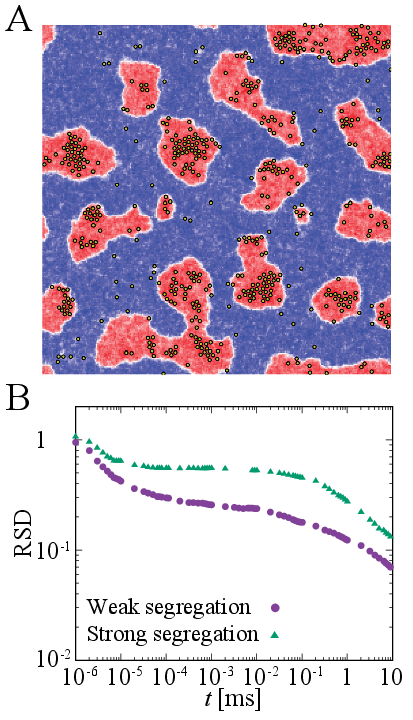}
\end{center}
\caption{Diffusion in heterogeneous media.
(A) Snapshot of particles diffusing in a spatially heterogeneous environment composed of two domains with distinct diffusivities.
(B) Relative standard deviation of TSDs in weakly and strongly segregated environments.
Stronger segregation leads to greater diffusivity fluctuations, leading to broader variability in TSDs.
Figure adapted from Ref.~\cite{sakamoto2023heterogeneous} with modifications.
}
\label{fig: phase separation}
\end{figure}

To further address this complexity, Sakamoto {\it et al.} performed Langevin dynamics simulations coupled with the phase-field method to investigate molecular partitioning and diffusion in phase-separated membranes \cite{sakamoto2023heterogeneous} (see Fig.~\ref{fig: phase separation}A). Their findings reveal that proteins diffuse more rapidly in Ld domains but become intermittently confined in Lo domains, resulting in significant diffusivity fluctuations between slow and fast diffusive states. These fluctuations arise from phase heterogeneity, molecular crowding, and transient trapping, highlighting the role of membrane structure in dynamically regulating protein mobility.

To quantify these effects, they analysed the TSD and its RSD to investigate the effects of fluctuating diffusivity in phase-separated membranes. While the TSD exhibits normal diffusion---consistent with predictions from LEFD theory---the RSD reveals a persistent plateau, indicating sustained diffusivity fluctuations. This plateau reflects the characteristic timescales of protein residence in Lo and Ld phases, capturing their intermittent confinement and transitions between slow and fast diffusive states. 
The degree of phase segregation plays a crucial role in shaping these dynamics.

Importantly, these fluctuations are not solely determined by the presence of phase separation, but are also shaped by its degree of segregation.
In weakly segregated membranes, the contrast between Lo and Ld domains is small, leading to weaker diffusivity fluctuations.
In contrast, strong segregation generates a more pronounced difference in diffusivity between Lo and Ld domains.
This distinction is reflected in the RSD of the TSD: stronger segregation leads to higher RSD values due to enhanced diffusivity heterogeneity, as shown in Fig.~\ref{fig: phase separation}B.
Their results align well with the theoretical framework of RSD in LEFD, further supporting the role of heterogeneous diffusivity in biological membranes.

Moreover, their study demonstrates that molecular crowding induces subdiffusion, characterised by a power-law dependence of the TSD: $\overline{\delta^2} (\Delta; t) \propto \Delta^\alpha$, with the exponent $\alpha$ decreasing from  $\alpha = 1.0$  (normal diffusion) to  $\alpha \approx 0.85$ as protein concentration increases. This subdiffusive behaviour is reminiscent of dynamics observed in glassy systems, soft matter, and biological environments \cite{cipelletti2005slow, weeks2000three, Golding2006, Jeon2011, parry2014bacterial, lampo2017cytoplasmic}, where crowding and interactions significantly impact molecular motion.

While the study does not explicitly analyse non-Gaussian propagators, the presence of diffusivity fluctuations and subdiffusion suggests that heterogeneous diffusion models such as LEFD may provide useful frameworks for describing protein motion. Their results align with previous experimental and theoretical studies that highlight the importance of time-dependent diffusivity in biological transport processes \cite{mashanov2021heterogeneity}.

Beyond its relevance to biological membranes, this study provides a generalizable approach for investigating heterogeneous and out-of-equilibrium systems, offering valuable insights into diffusion in complex environments. The methodologies employed by Sakamoto et al. can be extended to other biological and soft matter systems, reinforcing the role of fluctuating diffusivity as a key factor governing molecular transport in living systems.

\section{Conclusion}

Fluctuating diffusivity is a common phenomenon in nature, with its origins often attributed to the heterogeneous nature of the medium and the 
conformational fluctuations of diffusing particles. 
Specific examples of the former include diffusion in living cells, supercooled liquids and glass-forming liquids, where dynamic heterogeneity leads to 
spatial and temporal variations in diffusivity. Typical examples of the latter are diffusion of a protein in solutions and diffusion of entangled polymers, 
where the diffusing particles have many degrees of freedom and change their conformations over time. 

Recent advances in
high-resolution single-particle-tracking (SPT) techniques and molecular dynamics simulations have enabled the detailed investigation of both local and global diffusivity fluctuations.
From single-particle trajectories, one can examine not only global diffusivity characterised by the mean squared displacement (MSD) and time-averaged squared displacement (TSD) 
but also the local diffusivity. These analyses provide to identify the anomalous diffusion, ageing and ergodic properties.

In this review, we have discussed  statistical properties of the fundamental stochastic models of heterogeneous diffusion with fluctuating diffusivity, particularly the Langevin equation with fluctuating diffusivity (LEFD).
We have examined observables such as the MSD, the propagator, and RSD of the TSDs. 
In particular, we highlighted that a universal crossover behaviour is observed in the relative standard deviation of the LEFD, 
with the crossover time reflecting the characteristic timescale of the underlying diffusivity fluctuations $D(t)$. 
This result suggests that information about $D(t)$ can be inferred from the analysis of single-particle trajectories, thereby offering a robust theoretical framework for identifying local diffusivity fluctuations.

Fluctuating diffusivity  typically results in a non-Gaussian propagator in short-time behaviour, 
a phenomenon commonly observed in glassy systems, soft materials, and biological experiments. 
The diffusing-diffusivity model, in particular,  predicts a Laplace-type propagator that is consistent with experimental observations of Brownian yet non-Gaussian diffusion, 
reinforcing the connection between observed non-Gaussianity and underlying diffusivity fluctuations.

Furthermore, fluctuating diffusivity leads to several important effects such as anomalous diffusion and weak ergodicity breaking. 
Models such as the two-state model and the annealed transit time model (ATTM) capture key signatures such as non-Gaussian propagators, slow relaxation dynamics, and pronounced trajectory-to-trajectory fluctuations in the TSD.
These findings have implications for glassy systems and diffusion on living-cell membranes. 
Although these characteristics are also present in continuous-time random walk (CTRW) models, SPT experiments involving DC-SIGN on living-cell membranes suggest that the ATTM better aligns with the experimental observations compared to the CTRW. 
These results highlight the central role of fluctuating diffusivity in heterogeneous systems.

 To synthesise the diverse range of models discussed throughout this work, we have compiled a comparative summary (Table~\ref{tab:model-comparison}) highlighting their key dynamical features, including mean-square displacement (MSD), short- and long-time behaviour of displacement probability distribution functions (PDFs), time-averaged squared displacement (TSD), relative standard deviation (RSD), and ergodicity. This overview illustrates how different mechanisms---such as switching dynamics, correlated diffusivity, and structural constraints---yield distinct or overlapping dynamical signatures. For example, while the LEFD and diffusing diffusivity models both produce Brownian yet non-Gaussian diffusion (BYNGD), the two-state LEFD and ATTM are capable of generating subdiffusion and weak ergodicity breaking. Other models, such as the reptation model and binary mixture systems, show BYNGD behaviour due to complex underlying dynamics, even in equilibrium. This unified perspective enables a clearer understanding of how microscopic mechanisms shape macroscopic diffusion characteristics and provides a useful reference for selecting appropriate models to describe anomalous diffusion in various physical, chemical, and biological systems.
 
Taken together, these insights underscore the essential role of fluctuating diffusivity in the dynamics of heterogeneous systems.
Looking ahead, further theoretical and experimental developments will be key to uncovering the microscopic origins of diffusivity fluctuations across diverse materials.
In particular, combining stochastic models with high-resolution single-particle data promises to deepen our understanding of complex environments, from intracellular processes to soft and disordered materials.
Fluctuating diffusivity is not merely a correction to standard diffusion, but a fundamental feature of transport in nature.

\section*{Acknowledgement}
TA was supported by JSPS Grant-in-Aid for Scientific Research (No.~C 21K033920). 
JJH was supported by the National Research Foundation (Korea), no. RS-2024-00343900.
RM acknowledges funding from the German Research Foundation (DFG, grants ME 1513/13-1 and ME 1535/20-1).
EY was supported by JST PRESTO Grant Number JPMJPR22EE, Japan.

\begin{table*}[htbp]
\centering
\caption{Summary of dynamical behaviours for various models discussed in Sections III-VIII. Key observables include mean squared displacement (MSD), displacement probability distribution function (PDF), time-averaged squared displacement (TSD), relative standard deviation (RSD), and ergodicity. Note: The asterisk $*$ denotes that the displacement PDF is Gaussian in the short-time regime.}
\label{tab:model-comparison}
\resizebox{\textwidth}{!}{%
\begin{tabular}{|p{2.0cm}|p{2.2cm}|p{2.3cm}|p{2.3cm}|p{2.2cm}|p{2.5cm}|p{2.5cm}|}
\hline
\textbf{Model} & \textbf{MSD} & \textbf{PDF (short-time)} & \textbf{PDF (long-time)} & \textbf{TSD} & \textbf{RSD} & \textbf{Ergodicity} \\
\hline
LEFD (basic) & $\propto t$ [Eq.~\eqref{eq: normal diffusion in LEFD}] & Non-Gaussian [Eq.~\eqref{eq: NGP short-time}] & Gaussian [Eq.~\eqref{eq: prop LEFD long-time}] & $\propto \Delta$ [Eq.~\eqref{eq: ETSD LEFD general}, ergodic] & Crossover [Eq.~\eqref{eq: approx formula RSD}] & Ergodic (via~RSD) \\
\hline
Diffusing diffusivity model & $\propto t$ [Eq.~\eqref{eq: normal diffusion in LEFD}] & Exponential [Eq.~\eqref{eq: prop DD model short-time}] & Gaussian [Eq.~\eqref{eq: prop DD model long-time}] & $\propto \Delta$ (ergodic) & Crossover (captured by LEFD framework) & Ergodic (via~RSD) \\
\hline
Two-state LEFD & $\propto t$ or $t^\alpha$ ($\alpha < 1$) [Eq.~\eqref{e.MSD.two-state-lefd.subdiffusion}] & Non-Gaussian & Gaussian or non-Gaussian [Eqs.~\eqref{eq: propagator Laplace two-state long-time}] & $\propto \Delta$ [Eq.~\eqref{eq: ETSD two-state}, ergodic] & Crossover or slow decay & Usually ergodic (via~RSD) \\
\hline
ATTM & $\propto t^\alpha$ ($\alpha < 1$) [Eq.~\eqref{eq: attm msd}] & Non-Gaussian & Non-Gaussian & $\propto \Delta$ [Eq.~\eqref{eq: tsd attm}] & Non-decaying [Eq.~\eqref{eq: rsd attm}] & Non-ergodic (via~RSD) \\
\hline
CTRW & $\propto t^\alpha$ ($\alpha < 1$) & Non-Gaussian & Non-Gaussian & $\propto \Delta$ [Ref.~\cite{He2008}] & Non-decaying [Ref.~\cite{He2008}] & Non-ergodic [Ref.~\cite{He2008}] \\
\hline
FBMFD & Crossover from $t$ to $t^\alpha$ ($\alpha < 1$) [Eq.~(\ref{e.fbmfd.msd})] & Non-Gaussian${}^*$ {\color{black}(intermediate)} & Gaussian & Same as MSD (ergodic) & Crossover & Usually ergodic \\
\hline
Reptation model & $\propto t$ & Non-Gaussian (via RSD) & Gaussian (via~RSD) & $\propto \Delta$ (ergodic) & Crossover [Eq.~(\ref{eq: asymptotic RSD reptation})] & Ergodic (via~RSD) \\
\hline
Binary mixture & $\propto t$ & Non-Gaussian (via RSD) & Gaussian (via~RSD) & $\propto \Delta$ (ergodic) & Crossover & Ergodic (via~RSD) \\
\hline
\end{tabular}
}
\end{table*}

\clearpage

\appendix

\section{Central limit theorem for correlated random variables}

In many applications, we are interested not only in the statistics of independent random variables, but also in understanding how correlations between them affect their cumulative behaviour. A particularly important question is how the central limit theorem (CLT), which governs the sum of independent random variables, generalises to correlated variables and continuous-time processes. Below, we outline the generalisation of the CLT to such correlated systems.

We consider the PDF of a sum of correlated random variables, $S_n = X_1 + \cdots + X_n$. 
When $X_1, \cdots, X_n$ are IID random variables with a finite variance, the CLT ensures that 
the PDF of $S_n$ converges to a Gaussian with mean $n \langle X \rangle $ and variance $n(\langle X^2 \rangle - \langle X \rangle^2)$
in the large-$n$ limit. 

We now examine the case where random variables  $X_k$  are correlated. When the correlation decays exponentially fast,
 the PDF of  $S_n$  still converges to a Gaussian distribution with mean  $n \langle X \rangle $, as expected. 
 However, the variance generally deviates from the IID expression
$n (\langle X^2 \rangle - \langle X \rangle^2)$ due to the correlations among variables \cite{bouchaud90}. 
The correlation function $C(n)$ of $X_k$ is defined by $C(n)= \langle X_{k} X_{k+n} \rangle - \langle X_k \rangle \langle 
X_{k+n} \rangle$, where we  assume the stationarity.   For simplicity, we assume $\langle X_k \rangle =0$. The second moment of 
$S_n$ becomes
\begin{equation}
\langle S_n^2 \rangle = n \langle X^2 \rangle + 2 \sum_{k=1}^n (n-k)C(k),
\end{equation}
where we assume that the sum of the correlation function $\sum_{k=1}^n (n-k)C(k)$ is finite.
In the long-$n$ limit, this becomes
\begin{equation}
\langle S_n^2 \rangle \sim  \left(2 \sum_{k=1}^{\infty} C(k) + C(0) \right) n.
\label{second moment of Sn Cn}
\end{equation}
To relate this to the classical CLT, we coarse-grain the sequence  $\{ X_k \} $ by introducing $Y_k \equiv X_{(k-1) c + 1} + \cdots + X_{kc}$, where $c$ is a coarse-graining interval. 
When $c$ is sufficiently large, the coarse-grained variables $ \{ Y_k \} $ are approximately uncorrelated and hence can be treated as IID. Then we can write
\begin{equation}
S_n \approx \sum_{k=1}^{\lfloor n/c \rfloor} Y_k, 
\end{equation}
where $\lfloor \cdot \rfloor$ denotes the floor function. By the CLT, the distribution of $S_n$ converges to a Gaussian with mean 0 and 
variance $\langle Y^2 \rangle  n/c$ in the large-$n$ limit. Because  $\langle Y^2 \rangle  = 
\langle X^2 \rangle  c^2$, the second of $S_n$ becomes $\langle S_n^2 \rangle  
= c \langle X^2 \rangle  n$. Comparing the second moment with Eq.~(\ref{second moment of Sn Cn}), we have 
\begin{equation}
c \langle X^2 \rangle = 2 \sum_{k=1}^{\infty} C(k) + C(0). 
\end{equation}
This relation shows that the effective variance of the coarse-grained variables encodes the cumulative effect of correlations across all time lags, illustrating how temporal correlations renormalise the apparent variance in the long-time limit.

The result can be generalised to continuous-time stochastic processes. Let  $X(t)$  be a stationary process with correlation function
$ C(t) = \langle X(0) X(t) \rangle - \langle X \rangle^2 $, and assume that $\int_0^\infty C(t) dt$ is finite. In this case, the time integral $ \int_0^t X(t')\,dt' $ behaves analogously to a sum of correlated random variables.
In the long-time limit, the distribution of this integral converges to a Gaussian with mean  $\langle X \rangle t $ and variance
\begin{equation}
\tilde{c} = 2 \int_0^\infty C(t) dt .
\end{equation}
This result holds even when $ \langle X \rangle \neq 0 $.
This central limit theorem for correlated random variables provides a foundation for understanding long-time Gaussian behaviour in systems with memory. 
It offers a powerful framework for analysing the statistics of time-integrated observables in non-Markovian stochastic processes, where conventional techniques for uncorrelated systems no longer apply.
 When the integral of the correlation function diverges---such as in the case of power-law correlations---standard Gaussian central limit theorems no longer hold, and non-Gaussian limiting distributions are expected~\cite{bouchaud90}.

\end{document}